\documentclass[]{emulateapj}
\usepackage{graphicx, amsmath, amsthm, amssymb}
\usepackage{rotating}
\usepackage{verbatim} 
\def\ba{\begin{eqnarray}}
\def\ea{\end{eqnarray}}
\shorttitle{ Steady-state distributions from Kozai-Lidov migration}
\shortauthors{Petrovich}
\altaffiltext{1}{Department of Astrophysical Sciences, Princeton University, Ivy Lane, Princeton, NJ 08544, USA; cpetrovi@princeton.edu}

\begin{document}

\title{Steady-state planet migration by the Kozai-Lidov 
mechanism in stellar binaries }
\author{Cristobal Petrovich\altaffilmark{1}}
\begin{abstract}
We study the steady-state orbital distributions of
giant planets migrating through the combination of the  
Kozai-Lidov (KL) mechanism due to a 
stellar companion and 
friction due to tides raised on the planet by the host star.
We run a large set of Monte Carlo simulations that describe the 
secular evolution of a star-planet-star triple system including
the effects from general relativistic precession, stellar 
and planetary spin evolution, and tides.  
Our simulations show that KL migration produces Hot Jupiters (HJs)
with semi-major axes that are generally 
smaller than in the observations
and they can only explain the observations if the following are both 
true: (i) tidal dissipation at high eccentricities
is at least $\sim 150$ times more efficient than the upper limit
inferred from the Jupiter-Io interaction;
(ii) highly eccentric planets get tidally disrupted at distances 
$\gtrsim 0.015$ AU. 
Based on the occurrence rate and semi-major axis distribution 
of HJs, we find that KL migration in stellar binaries can produce 
at most  $\sim 20\%$ of the observed HJs.
Almost no intermediate-period  (semi-major axis $\sim0.1-2$ AU)
planets are formed by this mechanism---migrating 
planets spend most of their lifetimes undergoing KL oscillations
at large orbital separations  ($>2$ AU) or as Hot Jupiters.
\end{abstract}

\section{Introduction}\label{sec:intro}

\subsection{Planets in binary systems}

Current observations show that $\sim20\%$ of exoplanet host stars 
are in binary (or even higher order) stellar systems
\citep{DB07,MN09,ragha11}.
The observed gas giant planets in such binary systems seem to
reside preferentially in close-in orbits ($<0.1$ AU) and 
have large masses relative to the planets in single stars
 \citep{ZM02,US07,Law14}.
Also, the binary systems harboring giant planets
have preferentially wide orbital separations 
($>100$ AU) \citep{egg11}.
These findings suggest that wide stellar companions 
may play a significant role at shaping the planetary 
orbits of giant planets.

It might be expected that the protostellar accretion disks in binary stars 
form in alignment with the orbit of the binary if its semi-major
axis is not too large. 
Moreover, since the stars accrete high angular momentum 
gas from the protostellar disks 
their spins can also be brought into alignment with the binary 
orbit and the accretion disk. 
By measuring the inclination to the line of sight of 
the spin of the stars in binaries \citet{hale94} inferred 
that binaries with semi-major axes
$\lesssim30-40$ AU are spin-aligned, while at larger 
separations the spin vectors become randomly oriented relative
to each other.
This result suggests that protostellar disks, in which planets will
ultimately form, have angular momentum vectors that
are not correlated with the angular momentum vector of the
binary for wide enough binary separations.
Note, however, that polarimetry studies of protostellar disks 
find that the disks in individual stars in binaries 
are preferentially aligned with each other for binary 
separations up to a few hundred AU
\citep{jensen04,monin06}.
More recently, ALMA observations have revealed protoplanetary 
disks in wide (separations of 
$\sim400$ AU)  binaries are misaligned by 
$\sim60-80$ degrees \citep{jensen14,williams14}.

An important assumption we make throughout this work is that 
the angular momentum of the
binary and that of the planetary system are uncorrelated
for semi-major axis $>100$ AU.

\subsection{The Kozai-Lidov mechanism}
\label{sec:KL}
The long-term stability of star-planet-star system requires that 
the system is hierarchical  ($a_{\rm in}\ll a_{\rm out}$) and that
the eccentricity of the outer binary $e_{\rm out}$ is small enough so
the inner and outer orbits do not experience close approaches
(e.g., \citealt{HW99}).
In such systems, a large mutual inclination between the inner and 
outer orbits can produce large-amplitude oscillations of the eccentricity 
and inclination; this is the so-called Kozai-Lidov (KL) mechanism 
\citep{kozai,lidov}.
Such oscillations have a characteristic timescale \citep{HTT97}
\ba
\tau_{\mbox{\tiny{KL}}}=\frac{2P_{\rm out}^2}{3\pi P_{\rm in}}\frac{m_1+m_2+m_3}{m_3}
\left(1-e_{\rm out}^2\right)^{3/2},
\label{eq:tau_KL} 
\ea
where $m_1$ and $m_2$ are the masses of the inner binary 
(host star and planet),
while $m_3$ is the mass of the perturber (stellar companion). 
The inner binary has a period $P_{\rm in}$, while the
outer binary 
has a period and eccentricity $P_{\rm out}$ and $e_{\rm out}$.

KL cycles can be studied analytically by averaging over
the orbital phases of the inner and outer binaries (usually
called the secular approximation)  \citep{kozai,ford2000}, which is 
generally a good approximation because the precession
time is much longer than the orbital period of either binary
(although see \citealt{anto13} and discussion in \S\ref{sec:secular}).
Under the secular approximation the semi-major axis of the
inner binary $a_{\rm in }$ and outer binary $a_{\rm out }$ are both 
conserved.

The perturbing potential of the outer companion 
can be written in the quadrupole approximation (expansion
up to $a_{\rm{in}}^2/a_{\rm{out}}^3$) in the limiting case when 
$a_{\rm in }\ll (1-e_{\rm out})a_{\rm out }$, which implies the following
important results:
(i) this averaged potential is axisymmetric relative to the orbital
plane of the outer binary and so the angular momentum of the
inner binary along this symmetry axis is conserved;
(ii) the Hamiltonian describing the evolution of the
system is integrable (it has one degree of freedom);
(iii) the eccentricity of the outer binary $e_{\rm out }$ remains
constant.

These results break down when higher order terms
are included in the potential. 
In particular, when expanding up to the octupole approximation  
and considering an eccentric perturber ($e_{\rm out }>0$) the potential
is no longer axisymmetric and the corresponding Hamiltonian
has two degrees of freedom.
Recent work by \citet{katz11}, \citet{LN11}, and
\citet{naoz13a} show that under such 
conditions the KL oscillations are modulated on timescales
that are longer than $\tau_{\rm KL}$ in Equation (\ref{eq:tau_KL})
(by a factor of $\sim \epsilon_{\rm oct}^{-1}$ with 
$\epsilon_{\rm{oct}}=\frac{25}{16}\frac{a_{\rm{in}}}{a_{\rm{out}}}\frac{e_{\rm{out}}}{(1-e_{\rm{out}}^2)}
\frac{m_1-m_2}{m_1+m_2}$
in Eq. [A3]).
This longer-timescale modulation can give rise to a more
dramatic orbital evolution of the inner orbit, which includes
episodes of extremely high eccentricities and orbit flipping 
between retrograde and prograde relative
to the angular momentum vector of the outer binary.
Moreover, provided that the angular momentum of the outer orbit is not
much larger than the angular momentum of the inner orbit,
the eccentricity of the perturber does not necessarily 
remain constant (e.g., \citealt{naoz11}).

The KL cycles are driven by the interplay between the weak
tidal torque from the outer orbit and the shape of the inner
orbit. 
Thus, such cycles can be suppressed by extra forces that lead to
pericenter precession (see the precession rates  $Z_1$, $Z_2$, and 
$Z_{\rm GR}$ in Appendix A) in a timescale shorter than (or comparable to) 
 $\tau_{\rm KL}$ in Equation (\ref{eq:tau_KL}) 
(e.g., \citealt{WM03,FT07}). 

In the context of KL migration general relativistic (GR) precession 
is generally able to suppress the oscillations at the largest distances 
because the GR precession rate $Z_{\rm GR}$ in Eq. [A17] has the weakest
dependence on $a_{\rm in}$ compared to the other precession forces.
If the planetary orbit starts from $e\sim0$ then GR precession
suppresses  the KL cycles when $\tau_{\rm KL}Z_{\rm GR}\sim1$
\citep{FT07,DKS14}, which happens at a semi-major axis
\ba
a\sim 2.5 \mbox{ AU} 
  \left(\frac{m_1}{M_\odot}\right)^{1/2}
    \left(\frac{m_3}{M_\odot}\right)^{-1/4}
  \left(\frac{a_{\rm{out}}\sqrt{1-e_{\rm{out}}^2}}{1000\mbox{ AU}}\right)^{3/4}.
\label{eq:a_GR}
\ea

If the planetary orbits reaches a maximum eccentricity 
$e_{\rm max}$ that allows for migration down to a final 
semi-major axis $a_{\rm F}=a(1-e_{\rm max}^2)$, the KL 
oscillations are quenched once the planet migrates to a 
semi-major axis \citep{SKDT12}:
 \ba
a_{\rm  Q}&\sim& 2 \mbox{ AU} 
   \left(\frac{a_{\rm F}}{0.05\mbox{ AU}}\right)^{-1/7}	
  \left(\frac{m_1}{M_\odot}\right)^{4/7}
    \left(\frac{m_3}{M_\odot}\right)^{-2/7}\nonumber \\
   && 
  \left(\frac{a_{\rm{out}}\sqrt{1-e_{\rm{out}}^2}}{1000\mbox{ AU}}\right)^{6/7},
\label{eq:a_Q}
\ea
thereafter the eccentricity and semi-major axis decay
 at constant angular momentum.

\subsection{Previous work on KL migration}
\label{sec:KL_mig}

Based on the formalism developed by \citet{1998EKH}
and \citet{1998KEM} for KL cycles with tidal friction,  
\citet{WM03} carried out the first calculation of planet migration 
as a plausible explanation of history of the very 
eccentric ($e=0.93$)  planet HD 80606b.
Such migration ends with the formation of a Hot Jupiter:
a gas giant planet with semi-major axis $<0.1$ AU.

Subsequent work by \citet{FT07} and \citet{WMR07}
study the orbital distributions of Hot Jupiters that arise
from KL migration by building a large number of
star-planet-star systems.
In particular, \citet{FT07} show that HJs formed by KL migration
have orbits that are commonly misaligned with respect to the 
spin axes of their host stars, which has been observed for many HJ
systems (e.g., \citealt{al12}). 
A major limitation of this work is that the 
authors consider a population of stellar perturbers with a 
fixed semi-major
axis and zero eccentricity (similar to our simulation 
SMA500e0 in Table 1), which
is not representative of the observed binary population.
In contrast, \citet{WMR07} do consider a population of
perturbers based on the observed binary distributions, which allows
them to estimate that KL migration might account 
for $\sim10\%$ of the observed HJs.

The studies by \citet{FT07} and \citet{WMR07} have modeled 
the gravitational interactions by using the secular approximation, 
expanding the potential
up to quadrupole approximation. As discussed in \S\ref{sec:KL},
such approximation might be inaccurate for eccentric binaries for which
$a_{\rm in }/a_{\rm out }$ is not $\ll1$.
Indeed, \citet{naoz12} show that by considering the octupole-level
gravitational interactions the efficiency to produce HJs increases 
considerably (a factor of $\sim4-6$) relative to the Monte Carlo
simulations by \citet{WMR07}, while the obliquity distribution of HJs
broadens relative to  that predicted by \citet{FT07}.

All these studies of KL migration focus on the final states of the
planetary systems:
the simulations after several Gyrs or when the planet
has either migrated or been tidally disrupted.
Our approach is somewhat different because we study the steady-state 
distribution of the  planetary orbital elements due to KL migration.
This allows us to study the formation of not only HJs, but also
migrating planets at wider separations.

\subsection{Constraints on tidal dissipation}

The orbital period of the Jupiter-Io system is shorter than Jupiter's
spin period (super-synchronous rotation).
Assuming that the outward migration of Io to its
current location occurs within $4.5$ Gyr 
constrains the maximum amount of dissipation in Jupiter,
which in terms of the quality 
factor\footnote{The quality factor is defined as the
ratio between the energy dissipated during the tidal forcing 
cycle and the average tidal interaction energy.} 
$Q_{\rm J-I}>5.9\times10^4$ \citep{GS66,YP81,leconte10},
close to the measurement of
$Q_{\rm J-I}=(3.56\pm0.66)\times10^5$ using 
astrometric observations of the Galilean moons
\citep{laine09}.
This upper limit to the amount of dissipation 
of $Q_{\rm J-I}>5.9\times10^4$
can be translated into a 
limit on time-lag\footnote{The quality factor is related
to the viscous time by $1/Q=2|\Omega-n|\tau$ where
$\Omega$ is the planet's spin frequency and $n$ is the mean motion.
Similarly, the planets' viscous time $t_{V}$
and time-lag $\tau$ are
 related by $t_{V}=3(1+k_L)R^3/(Gm\tau)$ 
 where $k_L$ is the planets's Love number
 \citep{1998EKH}} 
 for the Jupiter-Io interaction $\tau_{\rm J-I}<0.062$ s or 
on the viscous time of $t_{V,{\rm J-I}}>15$ yr \citep{SKD12}.

 Note that the constraints on the amount of dissipation in 
Jupiter (e.g., viscous time) derived from the Jupiter-Io 
interaction depend only on its internal structure 
and can, therefore, be used as a benchmark for the tidal
interaction between a Jupiter-like planet orbiting its host star.

Recent attempts to calibrate the amount of tidal 
dissipation using the observed sample of giant planets  
have been carried out by \citet{hansen10,hansen12}. 
\citet{hansen10} evolves the planetary orbits from an initially 
prescribed eccentricity-period distribution
using the equilibrium tide theory \citep{hut} 
(similar to our treatment in Appendix A but ignoring the 
gravitational interactions due to a perturber) and constraints 
the amount of dissipation
required to fit the data after evolving the systems for 
3 Gyr.
By fitting the envelope of the observed period-eccentricity distribution
the author finds that for Jupiter-like planets the 
viscous time\footnote{The author uses another parametrization
in terms of an internal dissipation constant $\sigma$,
which is related to the viscous time as
$t_V=2(1+k_l)/(MR^2\sigma)$.} 
of the exoplanets should be $t_{V}\simeq230$ yr.
Similarly, by studying individual short-period systems with moderate 
eccentricities \citet{hansen10} finds similar values,
roughly $t_{V}\sim100-1000$ yr, which are consistent
similar estimates by \citet{quinn14}.

Similarly, \citet{SKD12} give a lower limit to the amount 
of dissipation (an upper limit to the viscous time
of the planet $t_{V}$) required to circularize an initially
highly eccentric planetary orbit at $a\sim5$ AU to a final  
circular orbit at $a\simeq0.06$ AU within 10 Gyr. 
Such limit  is $t_{V}\lesssim1.5$ yr for a Jupiter-like planet, 
which is smaller by an order of magnitude than the lower limit 
$t_{V}>15$ yr inferred from the Jupiter-Io interaction.

The discrepancy found by \citet{SKD12} might not pose a severe 
problem for high-eccentricity migration given the uncertainties in
the models of tidal dissipation and the physical 
properties of the exoplanets.
For instance, this apparent contradiction can be resolved by
the recently proposed model by \citet{SL13} in which 
the dissipation happens in the core of giant planets. 
In this model, the energy dissipated is a function of the
tidal forcing frequency and, depending on the
size of the core, the amount of dissipation
can match that inferred from Jupiter-Io, while at the same time
being orders of magnitude more efficient at longer periods,
allowing high-eccentricity migration to operate.

Our approach in this work is to parametrize the amount of
tidal dissipation in order to study how it changes the 
evolution of KL migration and thereby to asses
what amount, if any,  can best explain the observations.

\subsection{Plan of this paper}
\label{sec:plan}

In \S\ref{sec:slow_KL} and \S\ref{sec:fast_kozai} we describe the evolution
of planetary orbits in two migration regimes: fast and slow KL migration.
These sections serve as a basis to understand the results from our Monte 
Carlo simulations which we present in \S\ref{sec:simulations}.
In \S\ref{sec:SMA} and \S\ref{sec:psi} we show the results
for the semi-major axis and stellar obliquity distribution of Hot Jupiters
 from our simulations, which we compare with the observations.
In \S\ref{sec:companions}, \S\ref{sec:migrating}, and \S\ref{sec:non_mig}  
we characterize the production
of Hot Jupiters, intermediate period planets, and non-migrating planets.
Finally, in \S\ref{sec:discussion} we discuss the implications and limitations
from our results and we summarize our main findings in
\S\ref{sec:summary}.

\begin{figure}[h!]
   \centering
  \includegraphics[width=8.4cm]{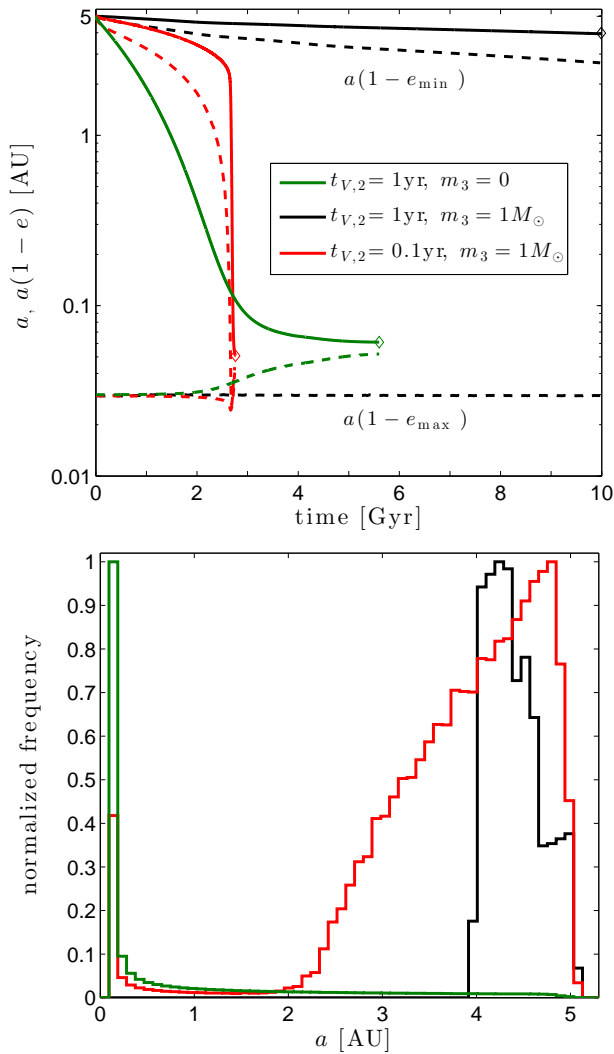}
  \caption{Evolution of migrating planets
  for different planetary viscous times $t_{V,2}$ and perturber
  masses $m_3$ as labeled.
{\it  Upper panel:}
  semi-major axis (solid lines) and pericenter 
  distance (dashed lines).
{\it  Lower panel:}  time spent in each semi-major axis bin
  normalized by the tallest bin.
  The green lines indicate a case in which 
there is no perturber ($m_3=0$) and the planet is initially
placed in a highly eccentric orbit  with $a=5$ AU, $e=0.994$.
 The black and red lines indicate the evolution when the perturber 
 has $m_3=1M_\odot$, $a_{\rm out}=500$ AU, $e_{\rm out}=0$, 
 and an initial inclination of $i_{\rm tot}=85^\circ.3$ (angle between 
 inner and outer angular orbital momenta), while the planet initially  
 has $a=5$ AU, $e=0.01,\omega_0=0,\Omega_0=0$, and a spin period
  of 20 days (the initial spin period of the host star is set to 10 days). 
Since these simulations show many KL cycles within a Gyr
(initially $\tau_{\mbox{\tiny{KL}}}\simeq 2.4\times10^6$ yr 
in Eq. \ref{eq:tau_KL})
we only display the minimum ($a[1-e_{\rm max}]$) and 
maximum  ($a[1-e_{\rm min}]$) 
pericenter distances  reached within each KL cycle.
All simulations start with zero obliquities (the angles between 
${\bf h}_{\rm{in}}$ and ${\bf \Omega}_{1}$ or ${\bf \Omega}_{2}$)
and they  reach the same minimum pericenter
  distance $a(1-e_{\rm max})=0.03$ AU during
  migration (the minimum and maximum pericenter 
  distances are the same for the green line).
The simulations are stopped once $a<0.1$ AU and 
$e<0.1$ or 10 Gyr have passed (the semi-major axis
at which the simulation is stopped is indicated with a 
diamond).
  }
\label{fig:tv_example}
\end{figure}   

\section{Slow Kozai-Lidov migration}
\label{sec:slow_KL}
In this section, we study a regime of KL migration 
that involves multiple KL cycles
with the semi-major axis shrinking by a small fractional 
amount during the high-eccentricity episodes of such 
cycles. 
We refer to this migration regime as ``slow" KL migration.

The model for orbit evolution used in this section is described in 
Appendices A and B.
In Appendix A, we explicitly show the equations of motion describing the
secular evolution of star-planet-star triple system including 
the effects from general relativistic precession, stellar and planetary 
spin evolution, and tides.
In Appendix B, we provide an analytical expression 
for the time-averaged eccentricity over
a KL cycle in Equations  (B6)-(B8).

For the sake of brevity, in some expressions we drop 
the sub-index ``in" when referring to the inner orbit.
Thus,  $e\equiv e_{\rm in}$
and $a\equiv a_{\rm in}$.

\subsection{Semi-major axis evolution
and migration rate}
\label{sec:a_KL}

We study the semi-major axis evolution of a slowly migrating 
planet and  calculate its migration rate in
the presence of KL cycles.

In the upper panel of Figure \ref{fig:tv_example}, we show the 
semi-major axis and pericenter evolution for migration tracks 
with different planetary viscous 
times $t_{V,2}=0.1$ yr (red lines) and $t_{V,2}=1$ yr (black lines).
In both cases, the KL mechanism forces the orbit to 
a maximum eccentricity of 0.994 and a
small pericenter distance of $ a(1-e)\simeq 0.03$AU.
Tides gradually extract orbital energy
during the eccentricity maximum of each cycle,
 so the orbit decays, reaching a final semi-major
axis of $a_{\rm F}\simeq 0.05$ 
AU\footnote{The minimum pericenter distance decreases from 
$ a(1-e)\simeq 0.03$AU to 0.025AU  due to the transfer of 
orbital angular momentum to the planet's spin 
that occurs after general relativistic precession
quenches the oscillations and migration speeds up.}.
For the simulation with  $t_{V,2}=0.1$ yr the planet reaches $a_{\rm F}$ after
$\sim3$ Gyr, while it takes more than $10$ Gyr for the simulation
with $t_{V,2}=1$ yr (the complete evolution is not shown in the figure). 

For comparison we include an evolutionary track when there is no perturber
and $t_{V,2}=1$ yr (green line in Figure \ref{fig:tv_example}).
Migration proceeds roughly at constant orbital angular momentum because
in this example the transfer of angular momentum between the 
spins and the orbit has little effect on the orbital evolution.
The evolution starts from a highly eccentric orbit with $e=0.994$, reaching
a pericenter distance of $ a(1-e)= 0.03$AU, identical (by construction)
to the minimum pericenter distance of the simulations undergoing 
KL cycles described above.
In this case, the migration to the final semi-major axis $a_{\rm F}= 0.06$ AU
 takes $\sim6$ Gyr.
By comparing this track to the one undergoing KL oscillations
with $t_{V,2}=1$ yr (black solid line), we observe that the former migrates
from $a=5$ AU to $\simeq4$ AU in $\sim 0.25$ Gyr, 
while the latter simulation  does so in 10 Gyr.
Thus, in our example migration proceeds initially $\sim40$ times more 
slowly when the KL oscillations are taken into account.

In the lower panel of Figure \ref{fig:tv_example}, we show a histogram 
of the time spent in each semi-major axis bin (or the time-averaged 
semi-major axis distribution) during the migration tracks
depicted in the upper panel. 
Note that we stopped the simulations shown by red and green lines once 
$a<0.1$ AU and  $e<0.1$ 
so they do not spend an arbitrary 
amount of  time as Hot Jupiters (red and green lines).
For the evolution at almost constant angular momentum (green line), 
the time-averaged semi-major axis distribution
follows a power-law $\propto a^{-1/2}$ 
during the evolution at high eccentricity, which is
expected from the orbit-averaged energy loss rate at fixed periastron 
distance, which results in $\dot{a}\propto a^{1/2}$ \citep{SKDT12}.

On the contrary, in the simulation with $t_{V,2}=0.1$ yr that undergoes KL
oscillations (red line) the time-averaged semi-major axis distribution
 peaks at $a\sim 4.5$ AU and
decays almost linearly from $a\sim 4.5$ AU to
$a\sim 2$ AU; for $a<1.6$ AU the KL oscillations are 
damped by general relativistic precession. 
According to the approximate expression in Equation (\ref{eq:a_Q}),
the oscillations should be completely damped for $a<1.1$ AU.
Once the oscillations are damped
migration proceeds at almost constant angular momentum and the
distribution follows the power-law $\propto a^{-1/2}$,
as expected. 
In contrast, by increasing the viscous time to 
$t_{V,2}=1$ yr (black line), the time-averaged 
semi-major axis distribution is restricted to $a>4$ AU
since the migration timescale to become a Hot Jupiter
is longer than 10 Gyr.

\begin{figure}[t!]
   \centering
  \includegraphics[width=8.7cm]{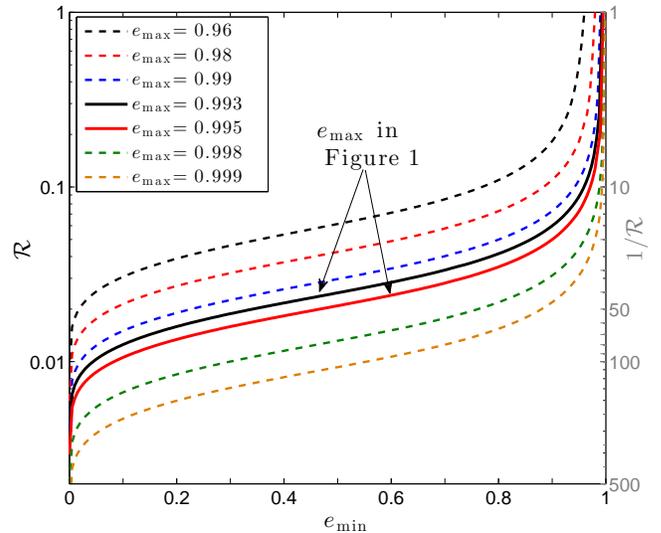}
  \caption{Ratio between the migration rate 
  $|\dot{a}/a|$
  averaged over a KL cycle 
and the migration rate at constant angular momentum 
($\mathcal{R}$ from Eq. \ref{eq:ratio})
as a function of the minimum eccentricity reached in a
KL cycle $e_{\rm min}$, for different values of the
maximum eccentricity $e_{\rm max}$ as labeled. 
The right $y-$axis in gray shows the values of the reciprocal 
$1/\mathcal{R}$
and the solids curves (red and black) indicate the tracks 
(range of $e_{\rm max}$) of our example Figure \ref{fig:tv_example}.
The migration rate at constant angular momentum is 
computed at fixed eccentricity $e_{\rm max}$.
We assume that all the dissipation occurs inside the planet
and that the planet and the star are non-rotating.
  }
\label{fig:ratio}
\end{figure}   

We can analytically quantify by how much is  migration slowed 
down by KL oscillations relative to migration 
at a constant angular momentum (with
same minimum pericenter distance 
or maximum eccentricity $e_{\rm max}$). 

In Appendix B we derive an exact expression for the time-averaged 
eccentricity distribution $n_e(e|e_{\rm min},e_{\rm max})$ 
during a KL cycle in the quadrupole 
approximation (Equation B8), 
which depends only on the minimum and maximum eccentricities
reached in a cycle $e_{\rm min}$ and $e_{\rm max}$.
By assuming that all the dissipation occurs in the planet (i.e., in $m_2$), 
from Equations (A4)-(A5), (A11)-(A12), and (A16) we can compute the 
migration rate (inverse of the migration timescale $\tau_a$)
 as:
\ba
\tau_a^{-1}&=&\left |\frac{1}{a}\frac{da}{dt}\right|=
\left | 2V_2\frac{e^2}{1-e^2} +2W_2\right|\\
 &\equiv& \mathcal{F}(e)/t_{F2}(a),
 \label{eq:a_dot}
\ea
where $\mathcal{F}(e)\propto \left(1-e\right)^{-15/2}$
as $e\to1$, which reflects the strong dependence of the 
tidal dissipation rate on the pericenter distance. 

Thus, the ratio $\mathcal{R}$
between the migration rate averaged over a KL cycle 
and the migration rate at constant angular momentum 
($e=e_{\rm  max}$) is
\ba
\mathcal{R}(e_{\rm min},e_{\rm  max})=\frac{1}{\mathcal{F}(e_{\rm{max}})}
\int_{e_{\rm min} }^{e_{\rm  max}} 
de\mathcal{F}(e)
n_e(e|e_{\rm min},e_{\rm  max});\nonumber\\
\label{eq:ratio}
\ea
note that the dependence on the semi-major axis disappears.
 
In Figure \ref{fig:ratio}, we show the ratio $\mathcal{R}$ from 
Equation (\ref{eq:ratio}) as a function of the minimum eccentricity reached 
in a KL cycle $e_{\rm  min}\in[0,e_{\rm max}]$ for different values of the
maximum eccentricity $e_{\rm  max}$.
We observe that $\mathcal{R}\to0$ as $e_{\rm min}\to0$ because in this limit
the  eccentricity remains small for an arbitrarily  large fraction of the KL 
cycle. Also, $\mathcal{R}\to1$ when the oscillations are quenched
(i.e., $e_{\rm  min}\to e_{\rm  max}$), as expected.

From Figure \ref{fig:ratio}, we observe that the ratio
$\mathcal{R}$ decreases as $e_{\rm  min}$ ($e_{\rm max}$)
decreases (increases).
This is because when the eccentricity oscillates with
larger amplitude (i.e.,  smaller $e_{\rm  min}$ or larger
$e_{\rm  max}$) the planet spends a larger fraction of the KL cycle at 
pericenter distances that are too large for tidal dissipation to occur.
Additionally, as we increase $e_{\rm max}$ to more extreme values, 
significant tidal dissipation is constrained to a smaller vicinity
around $e_{\rm max}$, so the planetary orbit spends
a shorter fraction of the KL cycle dissipating energy.

We can compare our estimates of the migration rate reduction
(i.e., $\mathcal{R}$ Eq. \ref{eq:ratio})
with our simulations in Figure \ref{fig:tv_example} 
(black and green lines in the upper panel).
As discussed above, the simulations show that migration from 
$a=5$ AU to 4 AU happens $\sim 40$ times more slowly 
when KL oscillations are present. 
Also, the eccentricity reaches a minimum of 
$e_{\rm min}\simeq0.1-0.3$ during the KL cycles at 
$a\simeq4-5$ AU.
We set $e_{\rm  max}=0.993-0.995$, such that
for $a=5$ AU the minimum pericenter becomes 
$a(1-e_{\rm  max})=0.025-0.035$ AU, bracketing the value 
0.03 AU from the simulation.
Thus, from the black and red solid lines in Figure \ref{fig:ratio} 
we get $\mathcal{R}\simeq 0.01-0.02$, meaning 
that migration with KL oscillations happens $\sim50-100$ 
more slowly than that with no oscillations, roughly consistent
with the numerical simulation, which results in factor of 
$\sim 40$.

From our numerical example with $t_{V,2}=0.1$ yr in Figure 
\ref{fig:tv_example}  (red line), we observe that during 
migration from $a=5$ AU to 2 AU the KL oscillations are 
gradually damped, while the minimum pericenter 
distance remains constant.
This means that the KL oscillations evolve towards larger values of
$e_{\rm min}$ and smaller values of $e_{\rm max}$
(upward and rightward evolution of $\mathcal{R}$
in Figure \ref{fig:ratio}).
Such evolution increases the value $\mathcal{R}$ and, therefore, 
speeds up the migration relative to that at constant angular momentum, 
which explains why the time-averaged semi-major axis distribution 
in the lower panel of Figure \ref{fig:tv_example} decreases as the 
system migrates from $a=5$ AU to 2 AU.
For instance, the time-averaged semi-major axis distribution peaks at 
$a\simeq4.5$ AU (red line in lower panel of Figure \ref{fig:tv_example}) 
where the KL oscillations have typically
$e_{\rm  min}\simeq0.15$, while it is reduced a by factor of two 
when   $a\simeq3.2$ AU where $e_{\rm min}\simeq0.55$.
This is consistent with the results for $\mathcal{R}$ in Figure
\ref{fig:ratio}, where we observe that an evolution from
$e_{\rm min}=0.15$ to $e_{\rm  min}=0.55$ increases $\mathcal{R}$
by a factor of $\simeq2$ when $e_{\rm  max}=0.993-0.995$
(black and red solid lines).

In summary, our numerical example and analytical calculations
show that KL oscillations typically slow down the migration by $\sim2$
orders of magnitude relative to migration at constant angular momentum.
This implies that a migrating planet spends most of its life either
at a few AU undergoing KL oscillations or as 
a Hot Jupiter (i.e., $a<0.1$ AU). 
Thus, migration at a few AU becomes the ``bottleneck'' 
that limits the ability of slow migration to produce Hot Jupiters.
Once the KL cycles are quenched, 
migration occurs much faster.

\subsection{Condition for slow migration}
\label{sec:slow_condition}

Slow migration occurs only if the secular torque 
from the companion is strong enough that it can change
periastron distance before
tidal dissipation is able to shrink  
the semi-major axis significantly.
 
From Equations (A4)-(A5),  the 
characteristic timescale in which the secular torque 
changes the periastron distance $r_{\rm p}=a(1-e)$
in the limit $1-e\ll1$ is
\ba
\tau_{\rm p}\equiv  \left|\frac{1}{r_{\rm p}}\frac{dr_{\rm p}}{dt}\right|^{-1}\simeq
\frac{(1-e^2) \tau_{\mbox{\tiny{KL}}}}
{5 |({\bf \hat{q}}_{\rm{in}}\cdot {\bf \hat{h}}_{\rm{out}})
({\bf \hat{e}}_{\rm{in}}\cdot {\bf \hat{h}}_{\rm{out}})|},
\label{eq:tau_rp}
\ea
where the denominator is of order
unity during an eccentricity maximum of the KL 
cycle\footnote{At maximum eccentricity the inclination
is minimum and it roughly corresponds to the critical inclination
for KL oscillations: 
$|\cos(i_{\rm tot})|={\bf \hat{h}}_{\rm{in}}\cdot
{\bf \hat{h}}_{\rm{out}}\simeq\sqrt{3/5}$, implying that
$({\bf \hat{q}}_{\rm{in}}\cdot {\bf \hat{h}}_{\rm{out}})^2+
({\bf \hat{e}}_{\rm{in}}\cdot {\bf \hat{h}}_{\rm{out}})^2\simeq2/5$.}
 and $\tau_{\mbox{\tiny{KL}}}$ is the KL timescale given in
Equation (\ref{eq:tau_KL}).

There is a critical pericenter distance 
$r_{\rm p,c}$ at which the migration
timescale $\tau_{a}$ in Equation (\ref{eq:a_dot}) equals 
$\tau_{\rm p}$ in Equation (\ref{eq:tau_rp}). 
By assuming that all the dissipation happens inside the planet 
and that the planet is non-rotating ($\Omega_2=0$), we get
\ba
r_{\rm p,c}&\simeq& 0.005 \mbox{ AU} \left[ \left(\frac{m_1}{M_\odot}\right)^{2.5}
\left(\frac{M_J}{m_2}\right)^2\left(\frac{0.1\mbox{ yr}}{t_{V,2}}\right) 
 \left(\frac{R_2}{R_J}\right)^{8}   \right.\nonumber\\
&&\left.
\left(\frac{5\mbox{ AU}}{a}\right)^3	
  \left(\frac{a_{\rm{out}}\sqrt{1-e_{\rm{out}}^2}}{100\mbox{ AU}}\right)^3	
  \left(\frac{M_\odot}{m_3}\right)
\right]^{1/6.5}.
\label{eq:r_p}
\ea

Clearly, slow migration requires that $\tau_{a}\gg\tau_{\rm p}$, which
implies that during the KL cycles the planetary orbit 
has to avoid pericenters distances that are too
close to $ r_{\rm p,c}$.
Given the strong dependence of tidal dissipation on 
the pericenter distance, keeping the pericenter distance
$a(1-e)>2 r_{\rm p,c}$ might be enough to
meet the condition for slow migration. 

From our example in Figure \ref{fig:tv_example},
we observe that the minimum pericenter in the black and
red lines is $a(1-e)= 0.03$ AU, while  from Equation (\ref{eq:r_p})
we get $r_{\rm p,c}=0.0074$ AU and 
$r_{\rm p,c}=0.0105$ AU, respectively.
Thus, our example is well outside the critical pericenter 
and is in the slow migration regime, as expected
from the evolution depicted in the figure.

\subsection{Constraints on tidal dissipation}
\label{sec:tides}

In this subsection, we constrain the minimum level of dissipation 
required to form a Hot Jupiter within the lifetime of the planetary 
system.
This approach is similar in principle to the calculation in 
\citet{SKD12}, but considering the presence of KL cycles. 

We have shown that the ``bottleneck'' that limits the production of
Hot Jupiters during migration is the phase in which the
planetary orbit is undergoing KL oscillations.
This allows us to give an approximate condition for KL migration
to occur: the migration timescale $\tau_a=|a/\dot{a}|$
of a planet at a few AU undergoing KL oscillations 
has to be less than $\sim10$ Gyr. 
From Equations (\ref{eq:a_dot}) and  (\ref{eq:ratio}), 
taking the limit $1-e_{\rm max}\ll 1$, we can write the time-averaged 
migration timescale over a KL cycle as
\ba
\langle \tau_a \rangle_{\mbox{\tiny{KL}}}  \simeq \frac{2^{25/2}}{3861}
\frac{t_{F,2}(a) \left(1-e_{\rm max}\right)^{15/2}}
{\mathcal{R}(e_{\rm min},e_{\rm max})}.
\ea

If migration proceeds slowly and at constant minimum
angular momentum ($\propto [a(1-e_{\rm max}^2)]^{1/2}$)
down to the final semi-major axis 
$a_{\rm F}=a(1-e_{\rm max}^2)$,
we get the following condition for slow KL migration:
\ba
\frac{\langle \tau_a \rangle_{\mbox{\tiny{KL}}}}{10\mbox{ Gyr}}
&\simeq& 
\left(\frac{t_{V,2}}{0.4\mbox{ yr}}\right) 
\left(\frac{0.01}{\mathcal{R}}\right) 
\left(\frac{a_{\rm{F}}}{0.06\mbox{ AU}}\right)^{15/2}
\left(\frac{a}{5\mbox{ AU}}\right)^{1/2}	\nonumber \\
&\times&
\left(\frac{m_2}{M_J}\right)^2\left(\frac{M_\odot}{m_1}\right)^2
\left(\frac{R_J}{R_2}\right)^8<1,
\label{eq:kl_condition}
\ea
where $\mathcal{R}$ in Equation (\ref{eq:ratio}) 
contains the details of the KL oscillations
(see Figure  \ref{fig:ratio} for reference).

For the fiducial parameters in 
Equation (\ref{eq:kl_condition}), we constrain the amount of tidal 
dissipation to  $t_{V,2}<0.4$ yr, while for similar parameters 
\citet{SKD12} find that migration 
from 5 AU to $\simeq0.06$ AU at constant angular momentum
requires $t_{V,2}<1.5$ yr.
Thus, our constraint is $\sim4$ times more stringent than that in \citet{SKD12}.
Note that the bottleneck in \citet{SKD12} is the migration at $a\lesssim1$ AU
since the migration at constant angular momentum
is much slower for $a\sim0.06-1$ AU than for $a\sim1-5$ AU  
(see green line in Figure \ref{fig:tv_example}).
In contrast, the migration timescale with KL cycles is determined
by that at $a\gtrsim1$ AU, where the planet is undergoing KL oscillations.
Therefore, we do not expect to recover the condition by
\citet{SKD12} by just suppressing the 
KL cycles (i.e., setting $\mathcal{R}=1$) in Equation
(\ref{eq:kl_condition}).


As an example, we can apply the constraint in Equation
(\ref{eq:kl_condition}) to the 
hypothetical evolutionary track of HD 80606b depicted by 
\citet{FT07} (Figure 1 therein) in which the authors use
stellar and planetary masses of 
$m_1=1.1M_\odot$ and $m_2=7.8M_J$.
The planet migrates from $a=5$ AU to a final 
semi-major axis of $a_{\rm{F}}=0.071$ AU. 
Thus, Equation (\ref{eq:kl_condition})
yields $t_{V,2}<0.0023$ yr, consistent with the value
used by the authors of $t_{V,2}=0.001$ yr.
   
We have limited our analysis to the case in which the 
secular gravitational interaction is approximated up to the 
quadrupole level in $a_{\rm in}/a_{\rm out}$
($\epsilon_{\rm oct}=0$ in Eq. A3), which is valid when 
$a_{\rm in}\ll (1-e_{\rm out})a_{\rm out}$.
Stellar binaries have a wide eccentricity
and, therefore, our approximation might break down in some cases.
We still expect that our constraint on the amount of dissipation
remains valid because the eccentricity modulation from the octupole 
happens on timescales that are longer than the KL timescale
\citep{katz11,naoz12}. 
Thus, this longer-timescale eccentricity
modulation would make the planet spend a larger fraction of a 
KL cycle at pericenter distances that are too large for tidal dissipation to 
occur, slowing down migration even longer compared to
the quadrupole-level estimates.
We study the effect of the octupole and the
other relevant forces using Monte Carlo simulations
in \S\ref{sec:simulations}.
 
 \section{Fast Kozai-Lidov migration}
\label{sec:fast_kozai}

In this section, we study the regime of KL migration 
in which tidal dissipation is strong enough that
the semi-major axis shrinks significantly before the
secular torque from the companion has time to
change the periastron distance.
We refer to this migration regime as 
``fast" KL migration.

We expect that fast KL migration occurs roughly when 
the migration timescale ($\tau_{a}$ in Eq. [\ref{eq:a_dot}])
equals the timescale for the secular torque to change the periastron 
distance significantly ($\tau_{\rm p}$ in Eq. [\ref{eq:tau_rp}]).
We showed in \S\ref{sec:slow_condition} that 
both timescales are equal at a critical pericenter 
distance $r_{\rm p,c}$ given by 
Equation (\ref{eq:r_p}).
 Once this critical pericenter is reached, migration
occurs at  roughly constant angular momentum, so 
the final semi-major axis becomes 
$a_{\rm F}\simeq 2r_{\rm p,c}$.
Thus, by assuming that all the dissipation happens inside the planet 
and that the planet is non-rotating ($\Omega_2=0$), we get
\ba
a_{\rm  F}&\simeq& 0.012 \mbox{ AU} \left[ \left(\frac{m_1}{M_\odot}\right)^{2.5}
\left(\frac{M_J}{m_2}\right)^2\left(\frac{0.1\mbox{ yr}}{t_{V,2}}\right) 
 \left(\frac{R_2}{R_J}\right)^{8}   \right.\nonumber\\
&&\left.
\left(\frac{5\mbox{ AU}}{a}\right)^3	
  \left(\frac{a_{\rm{out}}\sqrt{1-e_{\rm{out}}^2}}{100\mbox{ AU}}\right)^3	
  \left(\frac{M_\odot}{m_3}\right)
\right]^{1/6.5},
\label{eq:a_f_2}
\ea
where we have slightly changed the pre-factor from 
0.01 AU that results from $a_{\rm F}\simeq 2r_{\rm p,c}$
 (Eq. [\ref{eq:r_p}])
to 0.012 AU to better reproduce the results 
from a set of numerical simulations with fast-migrating
planets.
 
A similar expression for $a_{\rm F}$ in
Equation (\ref{eq:a_f_2}) is found by  \citet{WMR07} 
by equating the eccentricity forcing to the
eccentricity damping due to tides.

Fast  KL migration is expected to be the source of
the shortest-period  Hot Jupiters and these are 
particularly susceptible to collisions with the star or 
tidal disruption.	
Note that for the fiducial parameters in  
Equation (\ref{eq:r_p}) we have a critical
pericenter distance $r_{\rm p,c}\simeq R_\odot$,
meaning that the planet can collide before tides 
become strong enough to shrink the semi-major axis
significantly.
Moreover, planets that approach the star too closely 
can be tidally disrupted at even larger separations of
$\sim2-3R_\odot$  \citep{GRL11,Liu12}.
 Thus, one might expect that an important fraction of
the fast migrating planets are likely to be destroyed.
We show in  \S\ref{sec:simulations} by means of Monte Carlo
simulations  that this is indeed the case.

We note that $a_F$ in Equation (\ref{eq:a_f_2}) 
depends strongly on the radius of the planet, as
$a_F\propto R_2^{8/6.5}$. 
Interestingly, gaseous giant planets are expected to form 
with an initially larger radii, which is then shrunk as the planet
cools down. 
This suggests that radius shrinkage of fast-migrating planets
can shift $a_{\rm F}$  to larger values \citep{WMR07}. 
Note, however, that by increasing the planetary radii the 
distance at which the planet is tidally disrupted also increases
and the effect of time-varying radius on the migration track
 is far from clear.
In \S\ref{sec:shrink}, we test the effect of radius shrinkage 
on our simulations.


\begin{table*}
\begin{center}
\caption{Summary of simulated systems and outcomes}
\begin{tabular}{c|ccccc|c|ccccc}
\hline
\hline
Name  & $a_{\rm{out}}$ [AU]&$m_3$ [$M_\odot$]  & $f(e_{\rm{out}})$ &$R_p$ $[R_J]$ &$t_{V,2} $[yr]&$N_{syst}$ & HJ (\%)& Mig. (\%)& TD (\%) & Non-Mig. (\%)  \\
\hline
MC-tv0.1 &$100-1500$ & 1&$U(e_{\rm{out}}^2;0,0.81)$ & 1& 0.1& 6009&3.4& 0.3 &25.0 &71.3  \\
MC-tv0.1-nt$^{a}$ &$100-1500$ & 1&$U(e_{\rm{out}}^2;0,0.81)$ & 1& 0.1& 6009&26.1& 0.3 &~2.3&71.3  \\
MC-tv1     &$100-1500$ & 1&$U(e_{\rm{out}}^2;0,0.81)$ &1& 1& 4052 &1.5& 0.1 &26.1 &72.3   \\
MC-tv0.01 &$100-1500$ & 1&$U(e_{\rm{out}}^2;0,0.81)$ & 1& 0.01& 4095&7.4& 0.3 &21.7 &70.6   \\
Ecc-tv0.1 &$100-1500$ & 1&$U(e_{\rm{out}};0,0.95)$ & 1& 0.1& 4235&3.2 &0.4   &19.2  &77.2  \\
Mass-tv0.1& $100-1500$ & 0.1&$U(e_{\rm{out}}^2;0,0.81)$ & 1&0.1&3912& 5.0 &0.2 & 19.5 & 75.3   \\
Spin-tv0.1$^{b}$& $100-1500$ & 1&$U(e_{\rm{out}}^2;0,0.81)$ & 1&0.1&5824& 3.4 &0.3 & 28.1 & 68.2   \\
Rp-tv0.1 &$100-1500$ & 1&$U(e_{\rm{out}}^2;0,0.81)$& $1+e^{-t/\left(3\cdot10^7\mbox{yr}\right)}$ 
& 0.1&5212& 2.4 &0.9& 21.4 &75.3   \\
Rp-tv0.01 &$100-1500$ & 1&$U(e_{\rm{out}}^2;0,0.81)$& $1+e^{-t/\left(3\cdot10^7\mbox{yr}\right)}$ 
& 0.01&4213& 6.2 &1.1& 19.4 &73.3   \\
Rp-tv0.03$^{c}$ &$100-1500$ & 1&$U(e_{\rm{out}}^2;0,0.81)$& $1+\onehalf e^{-t/\left(3\cdot10^7\mbox{yr}\right)}$ 
& 0.03&5485& 2.3 &0.7& 25.7 &71.3   \\
SMA500e0 &$500$ & 1&$e_{\rm{out}}=0$& $1$ 
& 0.01&7070& 10.2 &1.5& 0 &88.3   \\
\hline
 \multicolumn{11}{l}{ Note: all simulations have $m_1=M_\odot$, $m_2=M_J$, and $t_{V,1} =50$ yr.
$ U(x; x_{\rm min}, x_{\rm max})$ is the uniform distribution with $x_{\rm min}<x<x_{\rm max}$.}\\
 \multicolumn{11}{l}{We classify the outcomes as: {\it Hot Jupiters} (HJ) ($a<0.1$ AU),
  {\it Migrating} (Mig.) (0.1 AU$<a<4.5$ AU),  {\it Tidally Disrupted} (TD) }\\
   \multicolumn{11}{l}{ (Mig.) (minimum pericenter $<R_t$ with $R_t$ 
given by Eq. [\ref{eq:RL}]  with $f_t=2.7$),
  {\it Non-Migrating} (Non-Mig.) ($a>4.5$ AU). }\\
     \multicolumn{11}{l}{ $^{(a)}$MC-tv0.1-nt is identical to MC-tv0.1, but assuming that planets are only 
     disrupted by collisions with the star, i.e., $R_t=R_\odot+R_J$}\\
          \multicolumn{11}{l}{in Equation (\ref{eq:RL}).}\\
 \multicolumn{11}{l}{ $^{(b)}$In Spin-tv0.1 identical to MC-tv0.1, but  we change the initial host star's spin period from 10 days to
 3 days.}\\
  \multicolumn{11}{l}{ $^{(c)}$In Rp-tv0.03, we set $f_t=3.2$ in Eq. [\ref{eq:RL}].}
 \end{tabular}
\end{center}
\label{table:all_sim}
\end{table*}

\section{Numerical experiments}
\label{sec:simulations}
We run numerical experiments for the evolution of triple systems 
consisting of a wide stellar binary  with masses $m_1=1 M_\odot$ and 
$m_3$ and a planet orbiting $m_1$ with mass $m_2=1 M_J$.
The radii are taken to be solar for the host star and in most 
simulations the planet has a Jupiter radius (see Table 1).
The equations of motion are fully described in Appendix A.

\subsection{Initial conditions}

The relative initial inclination of the orbits is drawn from an isotropic 
distribution, i.e.,  
$\cos(i_{\rm tot})={\bf \hat{h}}_{\rm{in}} \cdot {\bf \hat{h}}_{\rm{out}}$ uniformly 
distributed in $[-1,1]$.
The longitude of the argument of pericenter and 
longitude of the ascending node are chosen
randomly  for the inner and outer orbits. 
The planets start at a fixed semi-major axes of 5 AU with an 
eccentricity that follows a Rayleigh distribution:
\ba
dp=\frac{e\,de}{\sigma_e^2} \exp\left(-\frac{1}{2} e^2/\sigma_e^2\right),
\label{eq:sigma_e}
\ea
where $\sigma_e$ is an input parameter chosen to be 0.01 
in most simulations.

The semi-major axes of the stellar binaries is chosen to be uniform in
$\log(a_{\rm out})$, motivated by the observed 
binary period distributions at semi-major axis $>$100 AU 
(e.g., \citealt{DM91,ragha11}). 
We consider a semi-major axis in the range
of 100-1500 AU.
The binary eccentricity is taken to be uniformly distributed in 
$e_{\rm{out}}^2$
(constant phase-space density at fixed orbital energy). 
We impose a maximum binary eccentricity of 0.9 
in most simulations.
We also run one simulation (Ecc-tv01 in Table 1)
with a uniform distribution in $e_{\rm{out}}$,  
motivated by the distribution observed
in solar-type stars by \citet{ragha11}.
We discard systems that do not satisfy the stability condition
\citep{MA01}:
\ba
\frac{a_{\rm{out}}}{a_{\rm{in}}}&>& 
2.8(1+\mu)^{2/5}\frac{(1+e_{\rm out})^{2/5}}{(1-e_{\rm out})^{6/5}}  
\left(1-0.3\frac{i_{\rm tot}}{180^\circ}\right)
\ea
where $\mu=m_3/(m_1+m_2)$.

The host star and the planet start spinning with 
periods of 10 days  (expect for Spin-tv0.1) and 10 hours, respectively,
both along the ${\bf \hat{h}}_{\rm{in},0}$ axis, 
implying that the
initial obliquities (the angles between ${\bf \hat{h}}_{\rm{in}}$
and ${\bf \Omega}_{1}$ or ${\bf \Omega}_{2}$),
are zero.

\begin{figure*}
   \centering
  \includegraphics[width=18cm]{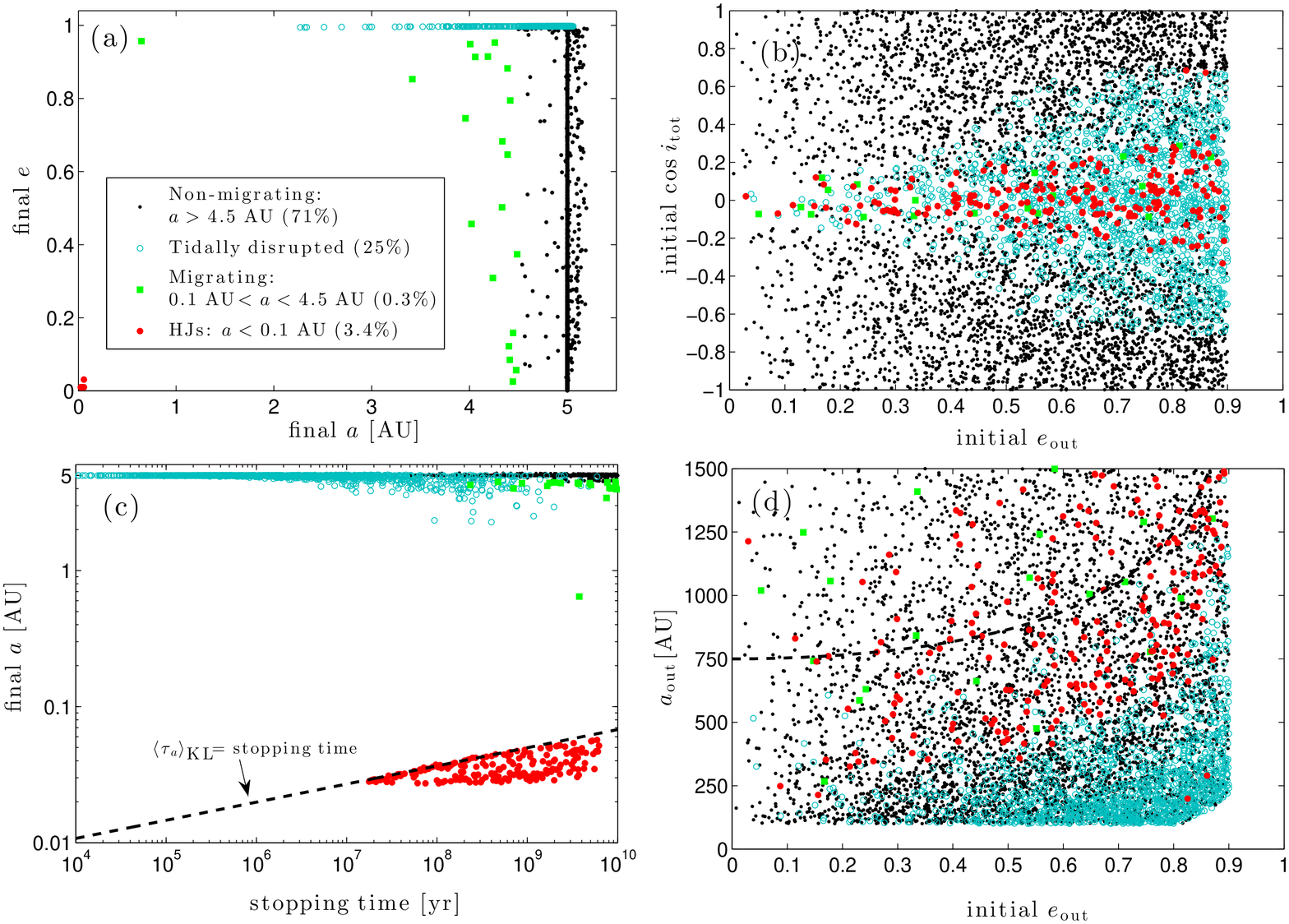}
  \caption{Outcomes for the fiducial Monte Carlo simulation MC-tv0.1
  (see Table 1), as labeled in panel (a).
  The stellar binary has a uniform distribution in $\log(a_{\rm out})$ in 
  [100,1500] AU and an eccentricity distribution that  is uniform in 
  $e_{\rm{out}}^2$ in $[0,0.81]$.
  We set the stellar and planetary viscous times to $t_{V,1}=50$ yr 
  and $t_{V,2}=0.1$ yr, respectively. 
  The planetary orbits are initialized with $a=5$ AU and their
  eccentricities are drawn from the distribution in Equation (\ref{eq:sigma_e})
  with $\sigma_e=0.01$, corresponding to nearly circular orbits.
  The planets are labeled as disrupted (blue circles) if 
  the pericenter becomes smaller than $0.0127$ AU 
  ($R_t$ with $f_t=2.7$ in Eq. [\ref{eq:RL}]).
  Panel (a): final semi-major axis versus final eccentricity of the planetary orbit.
  Panel (b): initial eccentricity of the stellar binary $e_{\rm out}$
 versus  initial cosine of inclination of the inner and outer orbits (angle between
  the orbital angular momentum vectors).
   Panel (c):  time at which the simulation is stopped 
   versus the final semi-major axis. The dashed line indicates the boundary
   at which $\langle \tau_a \rangle_{\mbox{\tiny{KL}}}$ in Equation 
   (\ref{eq:kl_condition}) equals the stopping time.
Panel (d):  initial eccentricity of the stellar binary $e_{\rm out}$ versus
the semi-major axis of the stellar binary. The dashed line indicates
the boundary $a_{\rm{out}}(1-e_{\rm{out}}^2)^{1/2}=750$ AU 
(Eq. \ref{eq:a_pert}) for tidal disruptions. 
  }
\label{fig:MC_001}
\end{figure*}   

\subsection{Stopping conditions}

We stop our simulations when at least 
one of the following conditions happens:
\begin{enumerate}
\item A maximum time has passed.
We choose this maximum time
uniformly distributed between 0 and 10 Gyr,
which is meant to provide a final sample 
of planetary systems as it 
would be observed if the rate of star formation is 
constant.

\item A Hot Jupiter in a circular orbit is formed:
the planetary orbit reaches $a<0.1$ AU and $e<0.01$.
 Following these systems 
for a long times is computationally expensive and a significant evolution 
is mostly seen in the eccentricity, which decreases to even lower values
(see, however, the discussion in \S\ref{sec:tide_star} about tidal
dissipation in the host star). 

\item The planet is tidally disrupted.
Disruption can happen  when the planet reaches
small pericenter distances. 
Hydrodynamic simulations for 
Jupiter-like planets 
 \citep{faber05, GRL11,Liu12} 
show that the pericenter distance at which a 
planet in a highly eccentric orbit gets 
disrupted\footnote{Depending on the
orbital eccentricity, the planet can be either
ejected after substantial mass loss or completely disrupted
\citep{GRL11}. We refer to both outcomes as tidal disruptions.
} is
\ba
R_{t}=f_tR_{p}\left(\frac{m_1}{m_1+m_2}\right)^{-1/3},
\label{eq:RL}
\ea
where the dimensionless coefficient $f_t$ is of
order unity.
The simulations by \citet{GRL11} and \citet{Liu12} result in $f_t=2.7$,
and, as argued by these authors, this value
might be regarded as a 
lower limit since  disruptions at wider separations
might still happen
over timescales longer than in their simulations.
Similar simulations by \citet{faber05} report
a disruption distance that is slightly smaller: $f_t=2.16$. 
However, \citet{naoz12} have carried Monte Carlo simulations 
similar to the ones we present here, but have used 
a different and much less restrictive disruption distance:
$f_t=1.66$.

In order to study how the results depend on the disruption 
criterion we decided to stop most simulations only when
the planet collides with star (i.e., $a(1-e)<R_1+R_2$)
and record the time whenever a new
minimum pericenter is reached.
For the simulations Rp-tv0.1 and Rp-tv0.01 
where the radius of the planet varies in time 
we use the disruption criterion as $f_t=2.7$ in  
Equation (\ref{eq:RL}), while in Rp-tv0.03 we use 
$f_t=3.2$.

Note that all simulations (except for MC-tv0.1-td and Rp-tv0.03) 
use $f_t=2.7$ in  Equation (\ref{eq:RL}) to classify a system as 
tidally disrupted.

\end{enumerate}

\subsection{Results}
\label{sec:sim_results}

We classify the outcomes from our simulations into four categories;
\begin{enumerate}
\item systems that form a {\it Hot Jupiter} (HJ), defined as  
those that migrate to $a<0.1$ AU; 
\item systems with {\it migrating} (Mig.) planets, defined as 
those with 0.1 AU$ <a<4.5$ AU at the end of the simulation; 
\item systems with {\it tidally disrupted} (TD) planets, defined 
as those that reach pericenters smaller than 
$R_t$ with $f_t=2.7$ in Equation (\ref{eq:RL});
\item systems with {\it non-migrating} (Non-Mig.) planets, 
defined as those with $a >4.5$  AU 
at the end of the simulation.
\end{enumerate}

In Table 1, we show the parameters of our 
simulations and the branching ratios into each category
of outcome.
We choose MC-tv0.1 as our fiducial simulation and discuss 
this first, then compare it with the rest of the simulations.

From Table 1 and Figure \ref{fig:MC_001}, we observe 
that the most common outcome in MC-tv0.1 is systems 
with non-migrating planets (black dots, $\simeq 71\%$). 
From panel (b) in Figure \ref{fig:MC_001},  this outcome is 
preferentially found in systems in which
the outer companion has relatively 
low inclination and eccentricity, so 
the KL mechanism\footnote{The critical mutual inclination for 
KL oscillations to occur is given by 	
$|\cos(i_{\rm tot})|\leq\sqrt{3/5}\simeq0.78$ or 
$i_{\rm tot}\in[39.2^\circ,140.8^\circ]$.} 
does not produce high enough eccentricities for efficient tidal 
dissipation to occur.
Also, non-migrating systems are slightly more common when the  
stellar perturbers are more widely spaced: the median of the 
outer semi-major axis is $\simeq515$ AU compared to 
$\simeq410$ AU for all the systems. 
This might be because the KL timescale increases
as $\propto a_{\rm out}^3$ (Eq. [\ref{eq:tau_KL}]), 
which allows for the extra forces that cause apsidal precession
(e.g., general relativity) to limit the
eccentricity growth more easily for more distant companions.

From panel (a) in Figure \ref{fig:MC_001}, we observe that some 
systems have migrated outwards by 
$\lesssim 10\%$ (black dots with final $a>$5 AU) as 
a result of the angular momentum transferred from the planet's spin 
to a highly eccentric  planetary orbit during the
simulation.

The second most common outcome is tidal disruptions 
(blue circles, $\simeq 25\%$), in systems where the planetary 
orbit becomes extremely eccentric reaching $a(1-e)<0.0127$ AU 
(i.e., $R_t$ with $f_t=2.7$ in Eq. [\ref{eq:RL}]).
In \S\ref{sec:fast_kozai} we argued that disruptions happen
only if the timescale at which the KL mechanism changes the 
pericenter distance is shorter than the migration timescale.
From Equation  (\ref{eq:r_p}) the condition 
above translates into $r_{\rm p,c}<a(1-e)$, where
 $r_{\rm p,c}$ is the critical pericenter at which 
both timescales are equal.
Thus, a necessary condition for disruptions is
$r_{\rm p,c}<R_t$, which for the parameters in our 
fiducial simulation constrains the perturber to
\ba
  a_{\rm{out}}\sqrt{1-e_{\rm{out}}^2}<750 \mbox{ AU},
  \label{eq:a_pert}
\ea
which is satisfied by all the systems with disrupted planets 
(see dashed line and blue points in panel d
 of  Figure \ref{fig:MC_001}).
Thus, the disrupted systems are preferentially found in 
more eccentric and tighter binaries (panel  d):
the median $e_{\rm out}$ ($a_{\rm out}$) in disrupted 
systems is $\simeq0.72$ ($\simeq225$ AU)
compared to $\simeq 0.62$ ($\simeq 410$ AU)
for the whole sample.
Note also that for more eccentric and tighter perturbers
the octupole-level gravitational perturbations
($\epsilon_{\rm oct}$ in Eq. [A3]) become stronger, allowing 
for more phase-space volume 
at which the planets can reach very high eccentricities
\citep{naoz11,katz11}. 
For instance, from panel (d) we observe that  
the distribution of the initial mutual inclinations in disrupted 
systems widens as we increase $e_{\rm out}$, reaching 
$i_{\rm tot}\sim 50^\circ-130^\circ$
for $e_{\rm out}>0.6$. 

From panel (c) in Figure \ref{fig:MC_001}, 
we observe that most planets get tidally disrupted early on in 
the simulation and some of them migrated inwards 
by as much as 3 AU before crossing the tidal disruption radius. 
From Equation (\ref{eq:kl_condition}) we can estimate the shortest
migration timescale by setting $a_{\rm F}=2R_t=0.0254$ AU, which
results in $\langle \tau_a \rangle_{\mbox{\tiny{KL}}}\sim6\cdot10^6$ yr.
This is consistent with the simulation where  
all the planets that migrate to $a<4$ AU before being disrupted
have evolved for $>10^7$ yr (panel c).

The third most common outcome is the systems that form a 
Hot Jupiter (red dots, $\simeq 3.4\%$).
Compared to the systems that have disruptions, almost all ($\simeq99\%$)
of the HJs are formed in systems with an initially narrow range of 
mutual inclinations: $|i_{\rm tot}-90^\circ|<20^\circ$ (panel b),
 and larger semi-major axis of the perturber: 
 mean and median $a_{\rm out}$ of 890 AU 
 and 845 AU (panel d).
 These observations might be linked because
 for more distant perturbers the contribution from the 
 octupole-level gravitational interactions becomes weaker 
 and more easily quenched by general relativistic precession
\citep{naoz13b}, implying that the initial mutual inclinations 
need to be closer to $90^\circ$ for the
quadrupole-level interactions can force the orbit to very 
high eccentricities.
Additionally, planets reaching very high eccentricities
in systems with weaker perturbers 
(i.e., larger values of $a_{\rm{out}}(1-e_{\rm{out}}^2)^{1/2}$) 
are less likely to be disrupted (Eq. [\ref{eq:kl_condition}]),
and can instead form HJs. 
 
All of the HJs are formed after $\sim10^7$ yr 
as required for the minimum migration timescale of
$\sim6\cdot10^6$ yr,
discussed above.
As seen in panel (c) of Figure \ref{fig:MC_001}, HJs with larger
final semi-major axis are formed later in an average sense, 
as expected. 
We show the boundary at which 
$\langle \tau_a \rangle_{\mbox{\tiny{KL}}}$
 in Equation (\ref{eq:kl_condition}) equals the stopping time  
 (i.e., time it takes to form a HJ).
Recall that $\langle \tau_a \rangle_{\mbox{\tiny{KL}}}$ is the minimum 
timescale for a planet undergoing slow KL migration
 to reach a final semi-major
axis $a_{\rm F}$ and, therefore, at fixed $a_{\rm F}$ we expect 
HJs to be formed in longer timescales
(i.e., below this boundary), which is consistent 
with the simulations.

The least common outcome is the systems with migrating planets 
($\simeq0.3\%$).
These systems require that their migration timescale 
is slightly longer than their actual ages (or stopping
time in our simulations).
From panel (c), we see that there is only one planet 
in the migration track $a(1-e^2)\sim 0.05$ AU 
(system with $a=0.65$ AU and $e=0.96$), while
the rest have $a(1-e^2)> 0.4$ AU and are, therefore, 
off the migration track undergoing KL oscillations.
For reference,  the typical migrating planet has a companion
with $a_{\rm out}(1-e_{\rm out})^{1/2}\sim 500$ AU for which 
$\tau_{\mbox{\tiny{KL}}}$ in Equation (\ref{eq:tau_KL}) 
becomes equal to the GR precession timescale at 
$a \sim1.5$ AU and the oscillations would be quenched
completely for smaller $a$.
Recall from \S\ref{sec:a_KL} that once the KL oscillations 
are quenched, migration proceeds much faster (a factor of $\sim50-100$)
and, therefore, almost no migrating planets are expected at $a<1.5$ AU, 
consistent with the simulations.
The small number of migrating planets is striking
and will be further discussed  in \S\ref{sec:migrating}.

\begin{figure}
   \centering
  \includegraphics[width=8.5cm]{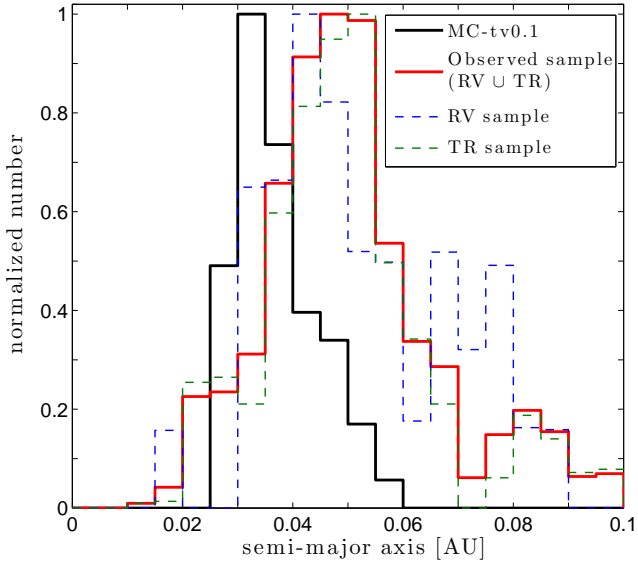}
  \caption{Semi-major axis distribution for the Hot Jupiters formed 
  in the fiducial simulation MC-tv0.1 (solid black line) and the 
  observed sample of planets
  with $M\sin(i)>0.1M_J$ and $a<0.1$ AU
  detected in radial velocity (dashed blue line) and
  transit (green dashed line) surveys, and 
  the combination of both observed samples (solid red line).
  We correct the transit sample by the geometric selection 
  bias.
  All histograms are normalized by the tallest bin.
The bin width is $0.005$ AU.
}
 \label{fig:a_tv01}
\end{figure}  

\section{Semi-major axis distribution of Hot Jupiters}
\label{sec:SMA}

As of January 2013, the observed sample of Hot 
Jupiters\footnote{Taken from The Exoplanet Orbit Database 
\citep{wright11}.} (planets with $M\sin(i)>0.1M_J$ and $a<0.1$ AU)
contains 196 planets, 159 planets detected in 
 transit (TR) and 37 detected by radial velocity (RV) surveys.
The former sample has a mean (median) semi-major axis
of 0.045 AU (0.045 AU), while the latter sample has
a mean semi-major axis of 0.052 AU (0.048 AU).
After accounting for the geometric selection 
bias\footnote{Extra selection biases might 
affect the derived semi-major distribution for the transit
planets. 
For instance, by accounting for the detection probability for a 
given S/N, \citet{gaudi05} find a detection efficiency
$\propto a^{-5/2}$, which yields
a mean (median) semi-major axis of
0.057 AU (0.052 AU) in our sample of planets detected in
transit surveys. 
We ignore this extra correction to the semi-major
axis distribution in this section, but use it in 
\S\ref{sec:sma_rate}. Note that a semi-major
axis distribution skewed to larger values would
strengthen the main result of this section: KL migration 
generally produces semi-major axes that are smaller 
than those observed.  }
 for the planets detected in transit 
surveys, the mean (median) semi-major
axis is 0.050 AU (0.049).
In contrast, RV surveys generally have uniform 
sensitivity in the semi-major axis range of our sample
and selection effects should not affect the
semi-major distribution significantly 
\citep{butler06,cumming08}.

Strictly speaking, our results should be compared with
the HJ systems with detected binary companions. 
However, such sample is too small (16 systems) to make 
a statistical comparison with our results 
and the sample of detected binary companions is 
likely to be fairly incomplete.
Thus, we ignore whether a HJ in our sample has a detected 
binary companion or not.
We note, however, that our conclusions might not differ 
significantly by considering the sample of HJ systems with detected 
stellar companions since their
mean and median semi-major axes is $\sim0.048$ AU, roughly 
consistent with our much larger sample of all HJ
systems.

 A Kolmogorov-Smirnov (KS) test between the RV and TR semi-major 
axis distributions shows that these distributions are consistent
($p-$value $\simeq0.4$).
Based on these findings, we combine the sample from TR
(corrected by the geometric selection bias) and 
RV surveys, which leaves
a sample of 196 planets with mean (median) semi-major axis
of 0.050 AU (0.049 AU). 
We use this distribution for our subsequent analysis.

A useful measure for subsequent analysis is the ratio
between the number of Very Hot Jupiters (VHJs, $a<0.04$ AU or
orbital periods $<3$ days) and
the total number of Hot Jupiters ($a<0.1$ AU):
\ba
\mathcal{F}_{\rm VHJ}=\frac{\#\{a<0.04\mbox{ AU}\}}{\#\{a<0.1\mbox{ AU}\}}.
\label{eq:f_vhj}
\ea
The RV (TR) sample has $\mathcal{F}_{\rm VHJ}\simeq0.25$
($\mathcal{F}_{\rm VHJ}\simeq0.24$), while the combined sample
has $\mathcal{F}_{\rm VHJ}\simeq0.24$.
Similarly, by combining RV and TR samples \citet{gaudi05}
find that the ratio between the number of HJs  with 
$a\simeq0.02-0.04$ AU and the number of HJs with 
$a\simeq0.04-0.085$ AU is $\sim0.1-0.2$, which
roughly translates into $\mathcal{F}_{\rm VHJ}\sim 0.09-0.17$. 
Similarly, \citet{gould06} find that the 
ratio between the number of HJs  with 
$a\simeq0.02-0.04$ AU and the number of HJs with 
$a\simeq0.04-0.06$ AU is $\simeq 0.45$ (i.e.,
$\mathcal{F}_{\rm VHJ}\lesssim0.3$).

In Figure \ref{fig:a_tv01} we show the semi-major axis distribution 
for the HJs formed in MC-tv0.1 (black line) and that from the observed 
sample (red dashed line).
The former distribution has a semi-major axis range  
$\simeq0.025-0.06$ AU
and is highly skewed towards low values: median of 0.034 AU
and $\mathcal{F}_{\rm VHJ}\simeq0.75$ (compared to $\simeq0.24$
in the observed sample).
The peak at $\sim 0.03$ AU is set by the disruption boundary, 
which implies that HJs are constrained to have a minimum
semi-major axis $>2R_t\simeq0.025$ AU (Eq. \ref{eq:RL}).
This suggests that by having a less restrictive disruption distance,
the peak should move to even lower values. We discuss this in the
subsequent section.

The observed semi-major axis distribution is wider with a range
 $\simeq0.015-0.1$ AU.
The overall shape of the observed distribution differs from that of the 
simulation, peaking towards larger values 
$\simeq0.04-0.05$ AU and with a more symmetrical
shape around this peak. 

We conclude that the semi-major axis distribution 
the of Hot Jupiters formed in our fiducial simulation 
disagrees with the observations because it produces
too many planets with $a<0.04$ AU relative to the
number of planets with $a>0.04$ AU compared to the 
observed sample.

In what follows we study how the different parameters in our
simulation (Table 1) can affect the semi-major axis distribution.
We ignore Spin-tv0.1 in this analysis
 because the semi-major axis 
distribution of the HJs in this simulation is essentially the same as 
that in our fiducial simulation.

\begin{figure}
   \centering
  \includegraphics[width=8.5cm]{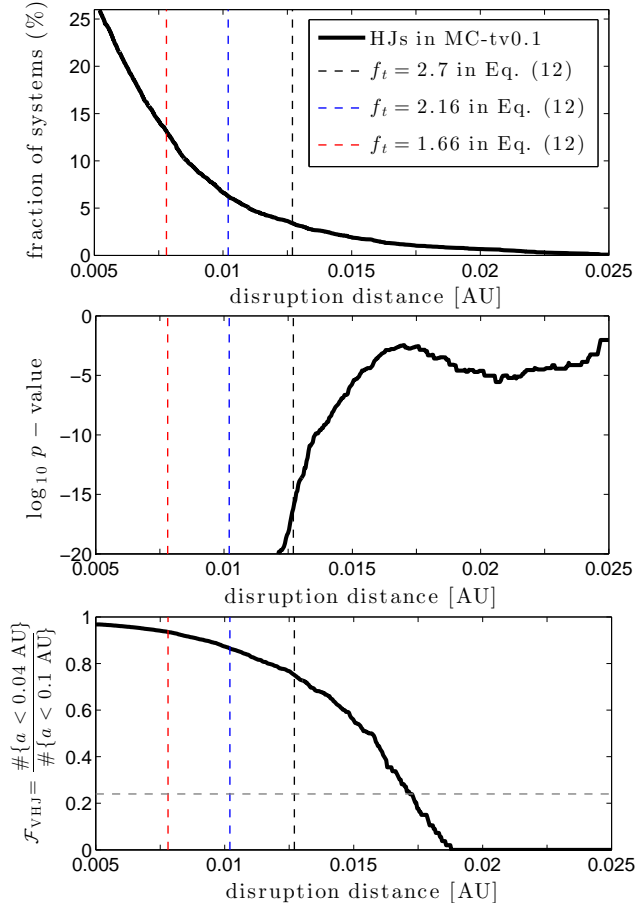}
  \caption{Hot Jupiters  formed in MC-tv0.1 as a function
  of the  disruption distance (or different values of $R_t$ by
  changing $f_t$ in Eq. [\ref{eq:RL}]).
  {\it Upper panel:} fraction of systems forming Hot Jupiters.
   {\it Middle panel:} $p-$value from a K-S test between the
   observed semi-major distribution and that from the simulation.
   {\it Lower panel:}  ratio between the number of Very Hot Jupiters 
   ($a<0.04$ AU) and the number of Hot Jupiters ($a<0.1$ AU)  
   ($\mathcal{F}_{\rm VHJ}$ in Eq. [\ref{eq:f_vhj}]), where the 
   horizontal dashed line shows this ratio for the observed
   sample.
  The fraction of systems that
  would form a HJ if the disruption boundary is 
  $\simeq0.005$ AU (star-planet collision) is   
  $\simeq 26\%$ (see MC-tv0.1-td in Table 1). 
  The vertical black, blue, and red dashed lines indicate 
  $R_t$ from Equation (\ref{eq:RL}) with $f_t=2.7$, 
  $f_t=2.16$, and $f_t=1.66$, respectively.}
 \label{fig:HJ_Rt}
\end{figure}

\subsection{Dependence on the disruption distance}
\label{sec:hj_dis}

Recall that we evolve the planets in the simulation MC-tv0.1
even after they crossed the disruption distance
$R_t$ with $f_t=2.7$ in Equation (\ref{eq:RL}) 
and stopped the simulation only when the planet collides with star
(i.e., $a(1-e)<R_\odot+R_J$).

The limiting scenario in which tidal disruption happens
only when the planet collides with the star is given by MC-tv0.1-td
where we set the tidal disruption distance 
to $R_t=R_\odot+R_J$ (Eq. [\ref{eq:RL}]).
Collisions in MC-tv0.1-nt
happen in $\simeq2\%$ of the systems (Table 1)
when the perturber is strong enough: median
$a_{\rm out}(1-e_{\rm out})^{1/2}\simeq 78$ AU, 
consistent with the expectation that
collisions happen when $r_{\rm p,c}<R_\odot+R_J\simeq0.005$ AU
(Eq. [\ref{eq:r_p}]). 

We record the time whenever a new minimum 
pericenter is reached, which allows us to study the effect 
of the disruption distance on the production of Hot Jupiters.

In the upper panel of Figure \ref{fig:HJ_Rt}, we show the fraction 
of HJs formed for a given disruption distance (i.e., different
values of $R_t$ or of $f_t$ in Eq. [\ref{eq:RL}]).
The maximum fraction of HJs that can be formed is $\simeq 26\%$,
by assuming that planets are only disrupted by collisions with
the star.
We observe that the fraction decreases dramatically as
$R_p$ increases. 
For $f_t=2.7$ in Equation (\ref{eq:RL}), we have 
the fraction of $3.4\%$ HJs reported in Table 1, while using 
$f_t=2.12$ this fraction increases to $6.3\%$.
Moreover, for the disruption distance of 0.0078 AU (i.e.,
$f_t=1.66$ in Eq. [\ref{eq:RL}]) 
used in \citet{naoz12} we get a fraction of $\simeq 13\%$
consistent with their similar simulation 
SMARan\footnote{These authors used a
different planetary viscous time of $t_{V,2}=1.5$ yr and
minimum semi-major axis of the perturber of
$a_{\rm out}=51$ AU.} which produces HJs in $13\%$ 
of the systems.

In the middle panel of Figure \ref{fig:HJ_Rt}, we show the
$p-$value from a K-S test comparing the
observed semi-major axis distribution and that from the simulation
for a given disruption distance.
Restricted to sample sizes of at least 10 HJs in the simulation,
the maximum $p-$value is $\simeq0.0035$ for $R_t=0.017$ AU
(i.e., $f_t=3.6$ in Eq. [\ref{eq:RL}]),
which would result in a population of HJs with final semi-major axis
$>0.034$ AU that represents a fraction of $\simeq1\%$ of
all the systems.

In the lower panel of Figure \ref{fig:HJ_Rt}, we show the
ratio between the number of Very Hot Jupiters 
($a<0.04$ AU) and the total number of Hot Jupiters ($a<0.1$ AU)  
($\mathcal{F}_{\rm VHJ}$ in Eq. [\ref{eq:f_vhj}]). 
By decreasing disruption distance one allows for the formation
of more HJs at smaller semi-major axis and, therefore,
$\mathcal{F}_{\rm VHJ}$ increases.
Recall that the observed sample has 
$\mathcal{F}_{\rm VHJ}\simeq0.24$ (horizontal dashed line)
and this ratio is reached in fiducial simulation by setting 
the disruption distance to $\simeq0.017$ AU 
(i.e., the same value for maximum $p-$value above).
This ratio increases from $\simeq0.24$ at $R_t\simeq0.017$ AU 
to $\simeq0.75$ ($\simeq0.94$) when setting 
$f_t=2.7$ ($f_t=1.66$) in Equation (\ref{eq:RL}).

As we decrease $R_t$ from $0.017$ AU to 0.005 AU, 
the $p-$value drops dramatically, while $\mathcal{F}_{\rm VHJ}$
departs from the observed value of $\simeq0.24$.
This is because the planets that reach small pericenters
tend to migrate fast and their final semi-major axis is 
approximately twice the minimum pericenter distance reached 
during the simulation (see \S\ref{sec:fast_kozai}). 
This behavior implies that there is a pile-up of HJs
at $a\simeq2R_t$, like the one observed in 
Figure \ref{fig:a_tv01}.  
Thus, by decreasing $R_t$ this pile-up is shifted to
lower values deviating from the observed peak at
$\simeq0.04-0.05$ AU.

In summary, as we decrease the tidal disruption distance 
relative to our fiducial value the efficiency to produce 
HJs increases rapidly, while the semi-major axis distribution
of the HJs formed starts rapidly deviating from that in
the observed sample because too many planets are
formed at very small semi-major axis.
The observed semi-major axis distribution
differs to that in the simulation for any disruption
distance, while a coefficient of $f_t\simeq3.6$ in
Equation (\ref{eq:RL}) gives the least bad fit to
the data.

\begin{figure}
   \centering
  \includegraphics[width=8.5cm]{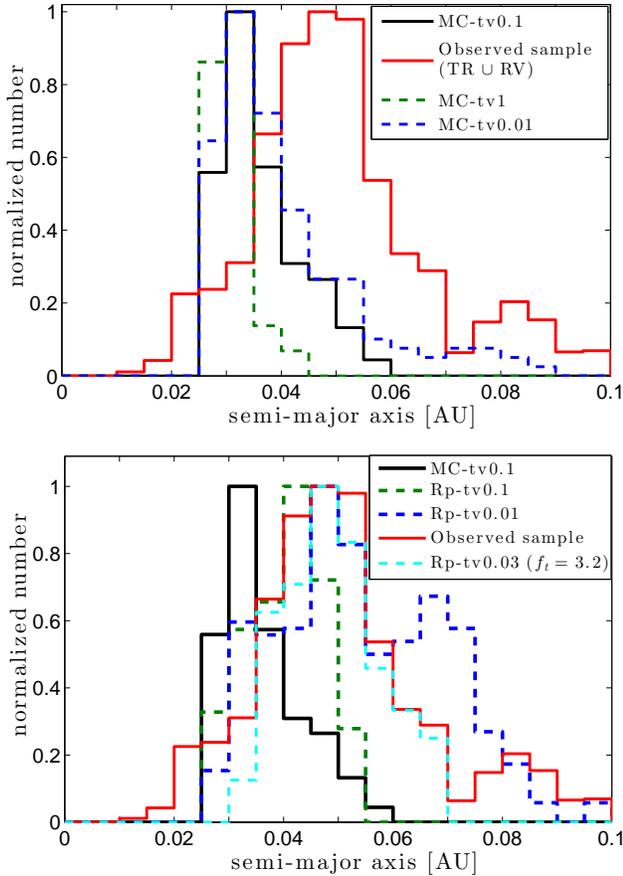}
  \caption{Semi-major axis distribution for the Hot Jupiters formed 
  in the simulations MC-tv0.1 (black solid lines, upper and lower 
  panels) and the observed sample (red solid lines, upper and 
  lower panels).
  {\it Upper panel:} semi-major axis distribution in 
  MC-tv1 (green dashed line) and MC-tv0.01 (blue dashed line).
  {\it Lower panel:} semi-major axis distribution in
Rp-tv0.1 (green dashed line), Rp-tv0.01 (blue dashed line),
and Rp-tv0.3 (light blue dashed line).
  All histograms are normalized by the tallest bin.
The bin width is $0.005$ AU.}
 \label{fig:HJ_tv}
\end{figure}  

\subsection{Dependence on tides}
\label{sec:hj_tides}

In the upper panel of Figure \ref{fig:HJ_tv}, we plot the semi-major axis 
distribution of Hot Jupiters for MC-tv1, MC-tv0.1, and MC-tv0.01 
(see Table 1).

We observe that the fiducial simulation MC-tv1 is only able to form
HJs with $a<0.045$ AU, which is consistent with our constraint 
in Equation (\ref{eq:kl_condition}) where by setting  $t_{V,2}=$1 yr 
we find that the maximum semi-major axis that a planet can reach
is $\simeq0.053$ AU.
Since the semi-major axis distribution is narrower than
in the fiducial simulation, the agreement with the observations
is even worse. For instance, 
$\mathcal{F}_{\rm VHJ}$ in Equation (\ref{eq:f_vhj})
increases from $\simeq0.75$ in our fiducial simulation
to $\simeq0.97$ (compared
to 0.24 in the observed sample).

On the contrary, the simulation MC-tv0.01 gives rise to HJs
up to $\simeq0.09$ AU, which is consistent with  the prediction
of Equation (\ref{eq:kl_condition}) that the maximum
final semi-major axis is $\simeq0.098$ AU for 
 $t_{V,2}=$0.01 yr. 
 The semi-major axis distribution widens relative to
 the fiducial, which results in a slight decrease in
$\mathcal{F}_{\rm VHJ}$ from $\simeq0.75$ in the fiducial simulation 
to $\simeq0.63$ (compared to 0.24 in the observed sample).
However, the disagreement with the data is still present 
and is mostly due to the strong pile-up
of planets at $\sim0.03$ AU. 

Similar to the previous subsection, we vary the disruption
distance in MC-tv0.01 and determine which value 
gives the best agreement with data, based on a KS test
and the ratio $\mathcal{F}_{\rm VHJ}$ in Equation (\ref{eq:f_vhj}).
We find that the maximum $p-$value is $\simeq0.027$ 
for $R_t=0.016$ AU (i.e., $f_t=3.4$ in Eq. [\ref{eq:RL}]),
which would result in a population of HJs with final semi-major axis
$>0.032$ AU that represents a fraction of $\simeq3.6\%$ of
all the systems with $\mathcal{F}_{\rm VHJ}=0.24$ (similar to
to that in observed sample).
This maximum $p-$value is higher than that from our 
fiducial simulation ($\simeq0.0035$) because for more 
efficient tidal dissipation HJs can be formed at larger 
semi-major axis, giving a better fit to the data.

In summary, we observe that the main effect of increasing the 
efficiency from tidal dissipation is to widen the semi-major axis 
distribution, which improves only slightly the agreement with the 
observations.
However, the dominant feature of the 
semi-major axis distribution in our simulations is the strong
pile-up of systems at the disruption boundary, which 
is off-set relative to the observed peak for our fiducial
disruption distance.
A model with a  tidal disruption distance of 
0.016 AU ($f_t\simeq3.4$ in
Eq. [\ref{eq:RL}]) and a planetary
viscous time of 0.01 yr  results in the least bad agreement with
the observations.

\subsection{Dependence on radius shrinkage}
\label{sec:shrink}

Gaseous giant planets are expected to form with large radii,
which then shrink to the current observed values of 
$\simeq 1R_J$ as the planet cools down.

In Rp-tv0.1 and in Rp-tv0.01 (Table 1) 
we prescribe the evolution of the planetary radius 
following \citet{WMR07} as
\ba
R_p(t)=R_J\left[1+\exp\left(-t/3\cdot10^7\mbox{yr}\right)\right],
\label{eq:r_pl}
\ea
which is an arbitrary functional form, but roughly 
describes the radius evolution from cooling models of 
Jupiter-like planets which predict shrinking timescales 
of $\sim 10^{7}$ yr (e.g., \citealt{burrows97}).

In the lower panel of Figure \ref{fig:HJ_tv},  we show the semi-major 
axis  distribution of the HJs formed in Rp-tv0.1 ($t_{V,2}=$0.1 yr) and 
Rp-tv0.01 ($t_{V,2}=$0.01 yr). We compare this distribution
with that from our fiducial simulation MC-tv0.1 and the 
observed sample.

We observe that the semi-major axis distribution 
in Rp-tv0.1 peaks at $a\sim 0.04$ AU and has a
symmetrical shape in the range $a\simeq 0.025-0.55$ AU:
the mean (median) semi-major axis is 0.042 AU (0.043 AU).
This is in sharp contrast with the fiducial simulation where
we observe a strong pile-up of systems at $a\simeq 0.025$ AU,
which is set by the disruption distance (see \S \ref{sec:hj_dis}).
Since the disruption distance is proportional to the
planetary radius (Eq. [\ref{eq:RL}]), such
distance in Rp-tv0.1 is initially larger by a factor of
two compared to that from the fiducial simulation 
($t=0$ in Eq. [\ref{eq:r_pl}]).
Therefore, if a planet migrates within
$10^7$ yr its semi-major axis is constrained to
$\gtrsim 0.043$ AU 
(set $t<10^7$ yr in Eqs. [\ref{eq:a_f_2}],[\ref{eq:RL}],
and [\ref{eq:r_pl}]), 
which explains the deficit of planets at $\simeq 0.025-0.03$ AU relative to 
the fiducial simulation. 
Note that an important fraction of planets are indeed
able to migrate fast up to a final semi-major axis 
$\gtrsim 0.043$ AU since tidal dissipation is greatly increased 
by the larger planetary radii: by setting $R_2=2R_J$ in Equation
(\ref{eq:a_f_2}) we get $a_{\rm F}>0.043$ AU
for a perturber with 
$a_{\rm{out}}(1-e_{\rm{out}}^2)^{1/2}>250$ AU, which
corresponds to $\sim 60\%$ of the systems.

The semi-major axis distribution of HJs in Rp-tv0.1
is able to reproduce the position of the peak
in the observed sample: the median semi-major axis in
 Rp-tv0.1 is $\simeq0.043$ AU, while the observed HJs have
 a median of $\simeq0.045 $ AU.
 Also the fraction of Very Hot Jupiters decreases from 
 $\mathcal{F}_{\rm VHJ}\simeq 0.75$ to  $\mathcal{F}_{\rm VHJ}\simeq 0.42$
 (compared to $\simeq0.24$ in the observations).
However, there are no planets at $a>0.06$ in Rp-tv0.1 because 
tides are not strong enough to allow planets to migrate to these
separations, similar to what happens in
the fiducial simulation (from Eq. [\ref{eq:a_f_2}]
the migration timescale to form a HJ at 
$\simeq 0.07$ AU is $\sim 10$ Gyr whether or not
shrinkage according to Eq. [\ref{eq:r_pl}] is included).

In Rp-tv0.01 we have increased the efficiency of tides,
so HJs at $a>0.06$ AU are able to form.
From the lower panel of Figure \ref{fig:HJ_Rt}
we observe that the semi-major axis 
distribution widens significantly compared to Rp-tv0.1: 
the mean (median) semi-major axis in Rp-tv0.01 is 0.062 AU (0.062 AU),
 while $\mathcal{F}_{\rm VHJ}\simeq 0.19$.
Thus, the semi-major axis distribution has shifted to values
that are larger than the observed distribution
and a KS test between the observed sample and Rp-tv0.01
yields $p-$value of $\sim3\cdot 10^{-4}$.

The fact that the mean and median of observed semi-major axis is
bracketed by the results from Rp-tv0.1 and Rp-tv0.01, suggests that
there should be a value of $t_{V,2}=0.01-0.1$ and/or functional form
of $R_p(t)$ that fits the observations better.
We have not carried a parameter survey, but we tried an extra 
simulation $Rp-tv0.03$ in which $t_{V,2}=0.03$ yr, 
$R_p(t)=R_J\left(1+0.5\cdot\exp\left(-t/3\cdot10^7\mbox{yr}\right)\right)$,
and $f_t=3.2$ in Equation (\ref{eq:RL}) that
fits the semi-major axis distribution much better
($p-$value is $\sim0.1$).

In summary, the planetary radius shrinkage reduces the
production of HJs with small semi-major axes, while
shifting the peak of the distribution to larger values.
These effects result in a better fit to the observations compared
to a model in which the planet's radius remains constant in time.

\begin{figure}
   \centering
  \includegraphics[width=8.5cm]{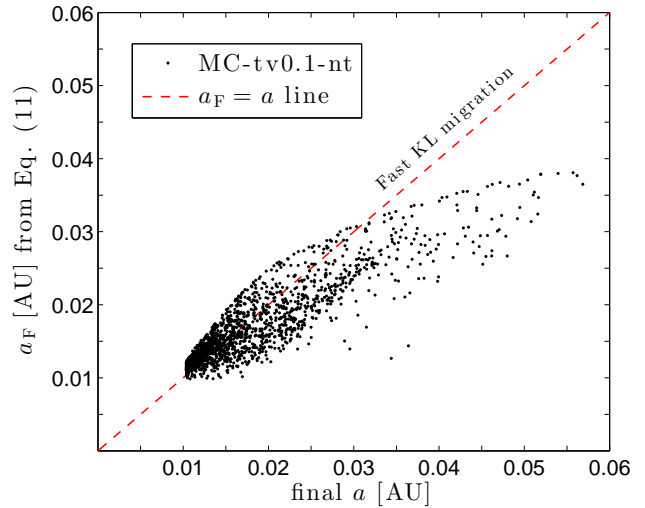}
  \caption{Final semi-major axis predicted from fast 
  KL  migration ($a_{\rm F}$ from
  Equation [\ref{eq:a_f_2}]) as a function 
  of the final semi-major axis obtained 
  for the Hot Jupiters in MC-tv0.1-nt (see Table 1).
   We calculate $a_{\rm F}$ using the parameters in MC-tv0.1-nt 
   and the properties of the perturber (i.e.,  $a_{\rm{out}}(1-e_{\rm{out}}^2)^{1/2}$) 
   corresponding to each system.
 The red dashed line is the straight line $a_{\rm F}=a$
 for visual comparison.}
 \label{fig:af_vs_a}
\end{figure}   

\subsection{Dependence on the perturber }
\label{sec:perturber}

In \S\ref{sec:fast_kozai} we argued that planets undergoing
fast KL migration would reach a final semi-major axis $a_{\rm F}$ 
given by Equation (\ref{eq:a_f_2}). 
This expression for $a_{\rm F}$ is only approximate and ignores 
the planetary and stellar spins, which can change the angular 
momentum of the planetary orbit during migration.
 
In Figure \ref{fig:af_vs_a}, we show the final semi-major 
axis $a_{\rm F}$ from Equation (\ref{eq:a_f_2}) as a function 
of the final semi-major axis reached in MC-tv0.1-nt
(planets are only destroyed by collisions).
For each HJ formed in MC-tv0.1-nt we calculate $a_{\rm F}$ 
using the same input parameters that go into  Equation 
(\ref{eq:a_f_2}) and use the corresponding
semi-major axis and eccentricity of the perturber, so:
$a_{\rm F}=0.012 \mbox{ AU }(
a_{\rm{out}}(1-e_{\rm{out}}^2)^{1/2}/100\mbox{AU})^{3/6.5}$. 
 
We observe that the analytical expression gives a fair 
description of the results in MC-tv0.1-nt at small $a$: 
relative differences between $a_{\rm F}$ and final $a$ 
are within $\sim30\%$ ($\sim50\%$) for $a<0.02$ AU 
($a<0.03$ AU).
This indicates that most of these planets have undergone 
fast KL migration and since $\simeq90\%$ of the HJs
in MC-tv0.1-nt have $a<0.03$ AU, we conclude that fast 
(as opposed to slow) KL migration is the dominant 
migration regime.

In contrast, the agreement with the analytical expression 
gets worse as we increase $a$ 
(see deviations from the red dashed line).
In particular, we observe that HJs at $a\gtrsim 0.03$ AU have
migrated to larger separations relative to what is predicted 
by $a_{\rm F}$. This is because most of these planets 
have undergone slow KL migration
and, therefore, reach larger semi-major axis by 
dissipating orbital energy through a sequence of many
high-eccentricity cycles.

Since fast KL migration is the dominant migration channel,
our results suggest that by having a weaker 
perturber (lower mass and angular momentum)  
we can shift the final semi-major distribution to larger 
values (see $a_{\rm F}$ from Eq. [\ref{eq:a_f_2}]). 
We test this hypothesis by running the simulations 
Mass-tv0.1 and Ecc-tv0.1 (Table 1).

In Mass-tv0.1 we decrease the mass of the perturber from
$m_3=1M_\odot$ in the fiducial simulation to $m_3=0.1M_\odot$,
which implies an increase of $\sim40\%$ in $a_{\rm F}$ 
($\propto m_3^{-1/6.5}$) from 
Equation (\ref{eq:a_f_2}) relative to the fiducial simulation.
The semi-major axis distribution (not shown) is almost identical to the 
fiducial simulation (see Figure \ref{fig:a_tv01}) because even by
shifting the semi-major axis of the planets at small separations
to larger distances these still pile-up at the disruption distance.
In particular, the fraction of VHJs decreases only slightly from
 $\mathcal{F}_{\rm VHJ}\simeq 0.75$ in the fiducial simulation to  
 $\mathcal{F}_{\rm VHJ}\simeq 0.70$ in Mass-tv0.1.
The maximum $p-$value is $\simeq0.001$ for $R_t=0.017$ AU
(i.e., $f_t=3.6$ in Eq. [\ref{eq:RL}]), which is similar to the 
value found in the fiducial simulation of $\simeq0.003$.
However, the fraction of HJs formed using $R_t=0.017$ AU 
increases from $\simeq1\%$ in the fiducial simulation
to $\simeq2\%$.
Similarly, from Table 1 we observe that the number of HJs 
formed  ($f_t=2.7$ in Eq. [\ref{eq:RL}])
also increases from $\simeq 3.5\%$ in the fiducial
simulation to $\simeq 5\%$. 
This difference is at the expense of having less tidal disruptions
per HJ in systems with less massive perturbers: for weaker 
perturbers planets undergoing fast KL migration 
reach larger final semi-major axis (Eq. [\ref{eq:a_f_2}]) and can 
more easily avoid disruption.

In Ecc-tv0.1 we change the eccentricity distribution of the
binary from thermal with maximum eccentricity 
of 0.9 ($U(e_{\rm{out}}^2;0,0.81)$) in the fiducial 
simulation to a uniform distribution with maximum 
eccentricity of 0.95 ($U(e_{\rm{out}};0,0.95)$).
The semi-major axis distribution (not shown) is almost identical
to that in the fiducial simulation (see Figure \ref{fig:a_tv01}).
The fraction of VHJs decreases only slightly from
 $\mathcal{F}_{\rm VHJ}\simeq 0.75$ in the fiducial simulation to  
 $\mathcal{F}_{\rm VHJ}\simeq 0.67$ in Mass-tv0.1.
The maximum $p-$value is $\simeq0.007$ for $R_t=0.018$ AU
(i.e., $f_t=3.8$ in Eq. [\ref{eq:RL}], $\simeq 1.2\%$ of HJs), which 
is slightly  higher than the value found in the fiducial simulation 
of $\simeq0.003$.
Since a uniform distribution of binary eccentricities produces 
in average weaker perturbers compared to a thermal distribution, 
the number of disruptions per HJ decreases slightly relative to the
fiducial simulation.

In summary, most HJs have undergone fast, instead of slow,
KL migration and  the approximate expression
for the final semi-major axis (Eq. [\ref{eq:a_f_2}]) works well for HJs with
 $a\lesssim0.03$ AU.
According to this expression, weaker perturbers (i.e.,
less massive, less eccentric, and/or more distant) produce HJs 
at larger semi-major axes, as we observe in the simulations, 
and are more likely to match the observations better.
However, decreasing the mass by a factor of 10  and 
having a uniform eccentricity distribution of the perturber 
did not change the results significantly because the planets 
still pile-up at the disruption distance, a feature that dominates
the semi-major axis distribution.

\begin{figure*}
   \centering
  \includegraphics[width=8.5cm]{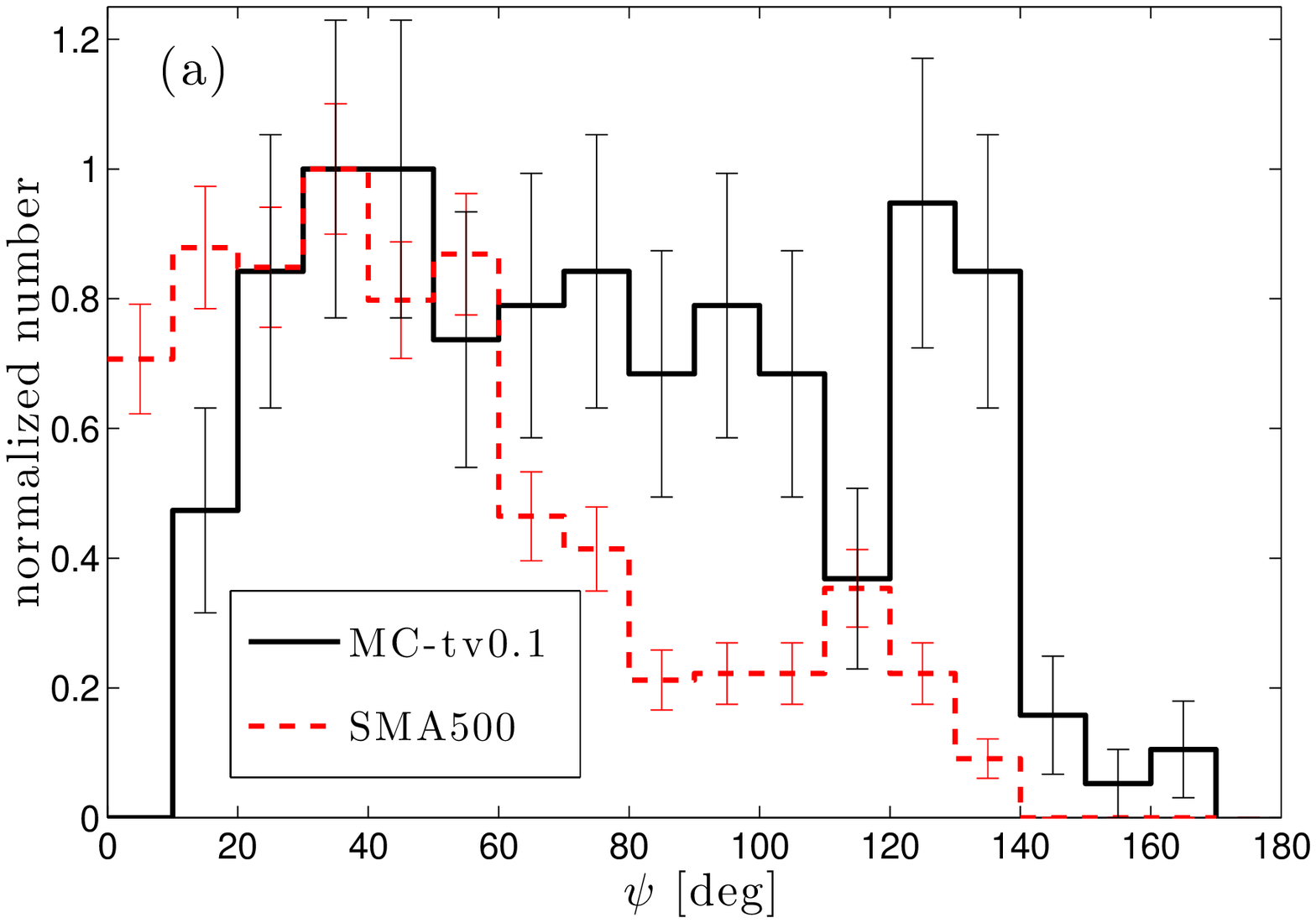}
  \includegraphics[width=8.5cm]{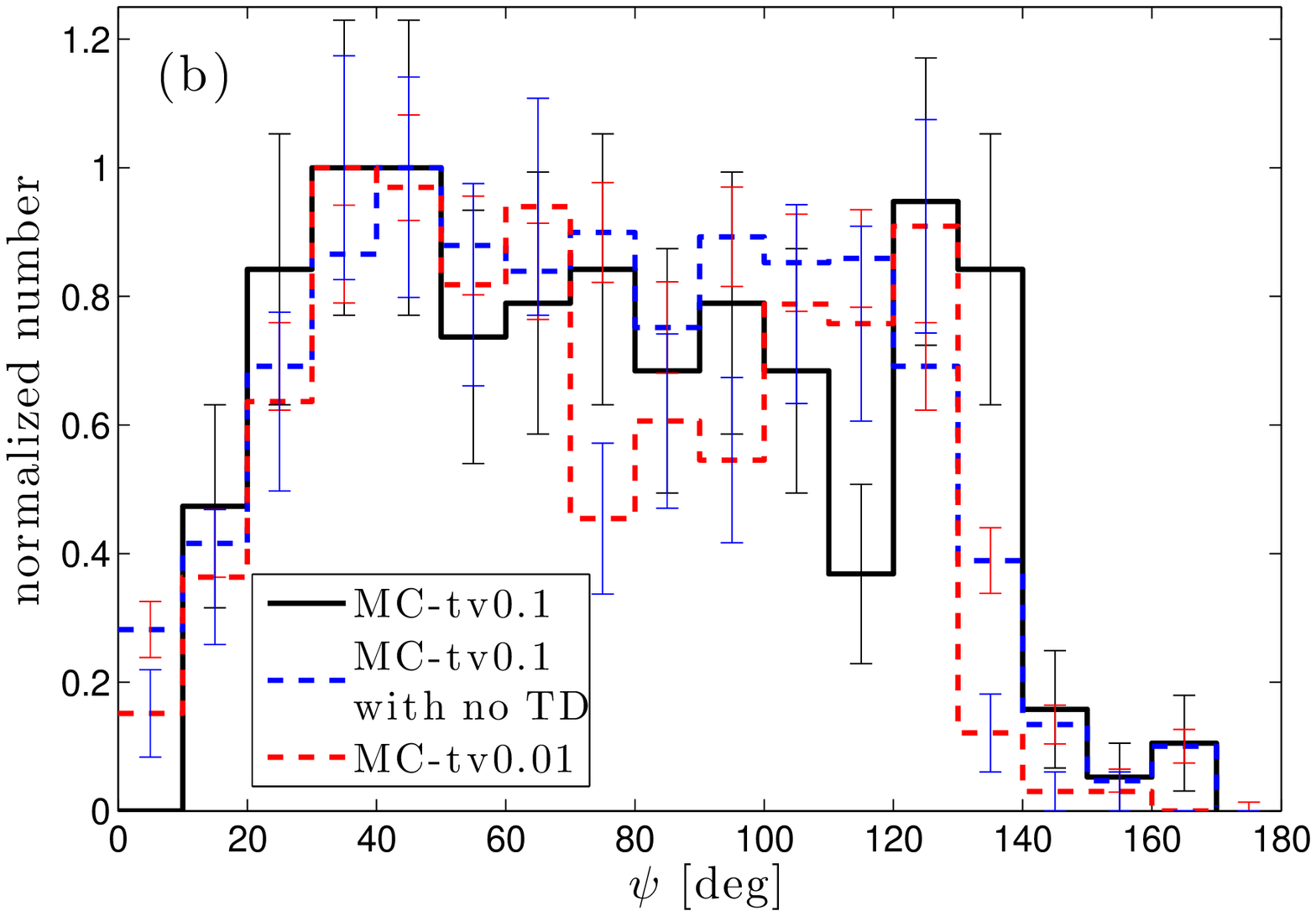}\\
    \includegraphics[width=8.5cm]{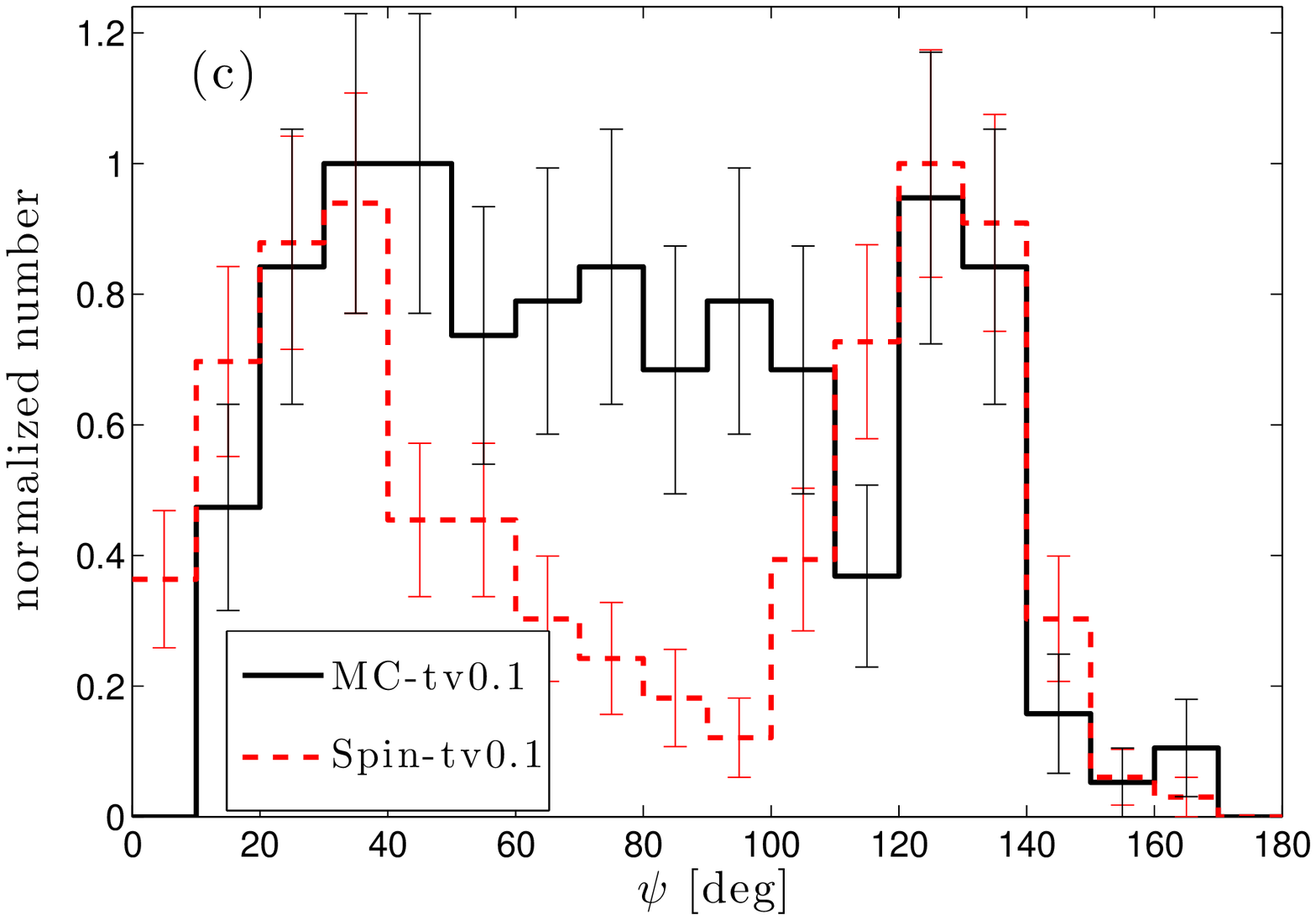}
      \includegraphics[width=8.5cm]{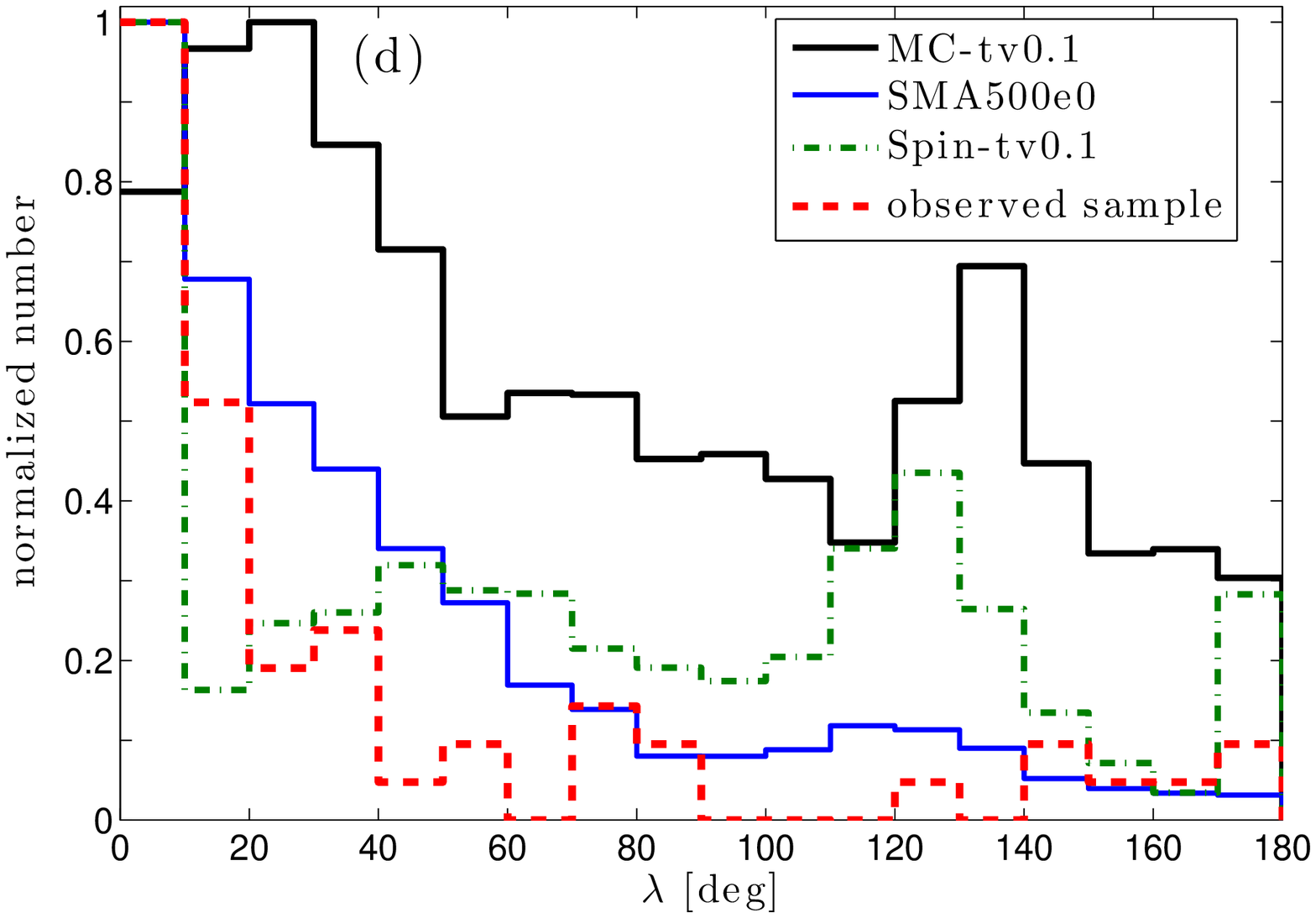}
  \caption{Final distribution of the angle between the spin
  of the host star  and the normal of the planetary orbit $\psi$
  (often called stellar obliquity or
  misalignment angle) and its sky-projected 
  value $\lambda$
  for the HJs formed in the simulations 
  and observations as labeled.
  {\it Panel a:} the results for MC-tv0.1 (solid
  black line) and SMA500e0 (red dashed line).
    {\it Panel b:} the results for MC-tv0.1
  (solid black line), MC-tv0.01 (blue dashed line),
  and MC-tv0.1-nt (red dashed line).
      {\it Panel c:} the results for MC-tv0.1
  (solid black line)
  and Spin-tv0.1 (red dashed line).
        {\it Panel d:} the results for MC-tv0.1
  (solid black line), SMA500e0 (solid blue line),
  Spin-tv0.1 (green dot-dashed line),
  and the observed sample of HJs (red dashed line).
  The error bars indicate the $1\sigma$ confidence limits from
  the Poisson counting errors for each 
  bin. The bin width is $10^\circ$.
  }
 \label{fig:HJ_psi}
\end{figure*}  

\section{Stellar obliquity distribution of Hot Jupiters}
\label{sec:psi}

In Figure  \ref{fig:HJ_psi} we show the final distribution 
of the angle between the spin axis of the host
star and the normal of the planetary orbit $\psi$
(often called the stellar obliquity 
angle or misalignment angle)
for the Hot Jupiters formed in MC-tv0.1 (solid black line
in upper and lower panels),
SMA500e0 (red dashed line in upper panel),
 MC-tv0.01 (blue dashed line in middle panel), 
MC-tv0.1-nt (red dashed line in middle panel),
and Spin-tv0.1 (red dashed line in lower panel).

We observe that the distribution of the stellar obliquity in 
our fiducial simulation MC-tv0.1 
ranges from $\psi\simeq10^\circ-170^\circ$ and 
is roughly flat
 for $\psi\simeq10^\circ-140^\circ$, with a mean (median) of 
 $\simeq78^\circ$ ($\simeq74^\circ$).
The fraction of retrograde systems ($\psi>90^\circ$) 
in MC-tv0.1 is $\simeq 38\%$, which is consistent
within Poisson the errors with the fraction of $\simeq 44\%$ 
found by \citet{naoz12} in similar simulations.

\subsection{Quadrupole- vs octupole-level 
gravitational interactions}

In the panel a of Figure  \ref{fig:HJ_psi}
we compare the results of our fiducial simulation
MC-tv0.1 with SMA500e0. 
The latter simulation
is identical to that from \citet{FT07} (see figure 10 therein), 
which has a perturber with fixed semi-major axis $a_{\rm out}=500$ AU
and eccentricity $e_{\rm out}=0$
so the octupole-level contribution
to the gravitational interaction vanishes.
Thus,  comparing SMA500e0 to our fiducial  simulation 
illustrates the role of these higher-order 
contributions to the potential in shaping the obliquity 
distribution.
 
We observe that the distribution of the obliquity in SMA500e0 
ranges from $\psi\simeq0^\circ-140^\circ$ 
and is skewed  towards low values relative
to MC-tv0.1: the mean (median) in SMA500e0 is $\simeq49^\circ$
($\simeq43^\circ$), compared to $\simeq78^\circ$
($\simeq74^\circ$) in MC-tv0.1.
 SMA500e0 has a significant 
deficit of systems relative
to MC-tv0.1 at all obliquities $\psi\gtrsim60^\circ$.

The fraction of retrograde systems ($\psi>90^\circ$) 
in MC-tv0.1 is $\simeq 38\%$, while it decreases
to $\simeq 15\%$ in SMA500e0.

We conclude that the octupole-level gravitational interactions can
have a significant effect on broadening the obliquity distribution.
This result is consistent with  \citet{naoz12}, who
argue that by adding the octupole order term in the
potential  the planetary orbit can reach more
extreme mutual inclinations, which produce 
a flatter distribution of $\psi$. 

\subsection{Effect of tidal dissipation
and disruption distance}

In the panel b of Figure  \ref{fig:HJ_psi} we compare the 
obliquity distribution of our fiducial simulation
MC-tv0.1 with  MC-tv0.01 and MC-tv0.1-nt 
to study the effect of the strength of tidal dissipation
and the disruption distance, respectively.
 
The obliquity distribution has a similar overall shape in all 
three simulations and is more or less flat for
$\psi\simeq20^\circ-130^\circ$.
However, both MC-tv0.01 and MC-tv0.1-nt
show a marginally significant 
excess (deficit) of systems with
$\psi=0-10^\circ$ and $\psi=110^\circ-120^\circ$ 
($\psi=130^\circ-140^\circ$) relative to the fiducial
simulation.
 The mean (median) in MC-tv0.01 and MC-tv0.1-nt
 is $\simeq72^\circ$ ($\simeq67^\circ$) and 
 $\simeq75^\circ$ ($\simeq74^\circ$), compared to $\simeq78^\circ$
($\simeq74^\circ$) in MC-tv0.1.
 The fraction of retrograde systems ($\psi>90^\circ$) 
in MC-tv0.01 is $\simeq 35\%$ and $\simeq 38\%$ in
MC-tv0.1-nt, similar to the fiducial
simulation with $\simeq 38\%$.
 
A KS test comparing the obliquity distribution
obtained from the fiducial simulation MC-tv0.1
and MC-tv0.01 (MC-tv0.1-nt)
results in a $p-$value of $\simeq0.1$ ($\simeq0.13$).
Thus, at  the $90\%$ confidence level
the obliquity distributions in MC-tv0.1, MC-tv0.01,
and MC-tv0.1-nt are all consistent with one another.
 
In conclusion, changing the amount
of tidal dissipation and ignoring planetary tidal disruptions
have no significant effect on the distribution of
stellar obliquities.
 
 \subsection{Effect of host star's spin period}

In the panel c of Figure  \ref{fig:HJ_psi} we compare the 
obliquity distribution of our fiducial simulation
MC-tv0.1 with  Spin-tv0.1
to study the effect of decreasing 
the initial host star's spin period from our fiducial 
value of 10 days to 3 days.
We note that host stars with giant planets are observed
to have rotation periods of $\sim2-20$ days 
(e.g., \citealt{wb13}).

We observe that the obliquity distribution 
of HJs  in Spin-tv0.1 differs significantly relative 
to our fiducial simulation and becomes bimodal
peaking at $\psi\sim20^\circ-40^\circ$ 
and $\psi\sim110^\circ-140^\circ$. 
By decreasing the period of the host star the distribution
shows a significant 
deficit of HJs with $\psi\sim40^\circ-110^\circ$
relative to MC-tv0.1.  

This bimodality in the obliquity distribution
has been recently discovered by \citet{storch14}
and it has been attributed to the chaotic behavior
of the host star's spin axis during KL migration. 
As \citet{storch14} describe, this behavior happens 
when the precession frequency due to planetary orbit
and the rotation-induced stellar quadrupole (proportional
to the host star's spin frequency)
roughly matches the precession frequency of the planet's
orbital angular momentum vector due to the binary
companion.
In our fiducial simulation, 
the former frequency is generally 
less than the latter during the formation of the HJs
and, therefore, the spin axis of the host star remains 
roughly unaffected by the torque due to the planet.
On the contrary, by increasing host star's spin frequency
in Spin-tv0.1 these precession frequencies 
can be comparable in most cases during the formation of the HJ 
leading to a different host star's spin axis behavior
and modifying the obliquity 
distribution.

We conclude that by decreasing the host star's spin period
the distribution of obliquities becomes bimodal in our
simulations.
This result is consistent with \citet{storch14}
and, based on their work, a similar behavior 
is expected when increasing the mass of the planets
in our simulations.

 
 \subsection{Comparison with observations}

As of January 2013, the observed sample of 
Hot Jupiters\footnote{Taken from The Exoplanet Orbit Database 
\citep{wright11} } (planets with $M \sin(i) > 0.1M_J$ and 
$a < 0.1$ AU) contains 56 planets
with projected stellar obliquity measurements
$\lambda$.
 
In panel d Figure \ref{fig:HJ_psi} we show the
projected obliquity distribution from the observed sample and that
from our fiducial simulation MC-tv0.1 and SMA500e0. 
The projection of
$\psi$ (see Figure \ref{fig:HJ_psi}) is calculated by taking $10^5$
random orbital configurations relative to a fixed observer
for each simulated system  (see e.g., \citealt{FW09}).
We observe that the fiducial simulation produces too 
many systems with large projected obliquities relative to
the observed sample.
For instance, the observations show that only $\simeq40\%$ 
($\simeq35\%$) 
of planets have $\lambda >20^\circ$ ($\lambda >30^\circ$), 
while $\simeq83\%$ ($\sim73\%$ ) of the HJs in MC-tv0.1
are in the same range of projected obliquities.
If we assume that all the misaligned ($\lambda >20^\circ$) 
HJs are due to KL migration, one  gets a crude upper limit to the
fraction of systems that can be explained by the theory as:
$40\%$ from the misaligned sample
plus a fraction of $0.4/0.83$ of the $\simeq17\%$
of aligned planets in the theoretical distribution.  
This results in an upper limit of $\simeq 48\%$ and a similar
upper limit of $\simeq 47\%$ results by considering the planets
with $\lambda >30^\circ$ as the misaligned sample.

Our crude upper limit is consistent with  \citet{naoz12} 
who found that KL migration can account for at most $\sim60\%$
(figure 4 therein) of the obliquity distribution, while it most likely 
accounts for $\sim30\%$ of the systems. 
These authors follow \citealt{morton11} by taking into account the errors 
in the observed projected obliquities and they represent the theoretical 
obliquity distribution as the sum of three migration mechanisms:
 KL migration, planet-planet scattering, and disk migration.

In contrast, the simulation SMA500e0 which considers
a non-eccentric stellar binary (i.e., no octupole-level gravitational 
interactions) produces more systems with small obliquities
relative to the fiducial simulation. 
Therefore, the obliquity distribution of SMA500e0
compares more favorably with the observations than
our fiducial simulation.
A KS test comparing the obliquity distribution
of SMA500e0 and the observed sample results in a $p-$value 
of $\simeq0.008$.

Similarly, in our simulation with shorter initial host star's spin periods
Spin-tv0.1 we get a better agreement with the observations
compared with our fiducial simulation because the
distribution of $\lambda$ is skewed towards lower values which
peaks at $\lambda<10^\circ$. 
However, the HJs
in Spin-tv0.1 tend to still be more misaligned than in the observations:
$\simeq40\%$  ($\simeq35\%$) 
of planets have $\lambda >20^\circ$ ($\lambda >30^\circ$)
in the observed sample, 
while $\simeq70\%$ ($\simeq63\%$ ) of the HJs in Spin-tv0.1
are in the same range of projected obliquities.

In summary, KL migration is unable to reproduce the obliquity
distribution of HJs and by assuming that all the observed 
misaligned planets are due to KL migration, this mechanism
can produce at most $\sim50\%$ of the planets. 
Simulations with only quadrupole-level gravitational interactions
and  shorter initial host star's spin periods
show a better agreement with the data.

\begin{table*}
\begin{center}
\caption{Hot Jupiter systems: maximum contribution to the 
observed semi-major axis distribution from 
$\mathcal{F}_\lambda$ in Eq. (\ref{eq:fraction_model}), production 
rate, and semi-major axis of stellar binary}
\begin{tabular}{c|cc|cc}
\hline
\hline
Name & $\mathcal{F}_\lambda$ in Eq. (\ref{eq:fraction_model}) &HJ production rate: & $a_{\rm out}$ [AU] \\
           & RV - TR - TR2$^{a}$ ($\%$)
           &$f_{<0.1}$ from Eq. [\ref{eq:f_hj}] ($\%$) & mean - median \\
\hline
MC-tv0.1& 44 - 37 -  19  &0.092&890 - 845 \\
MC-tv0.1-nt & ~6 -~ 2 - 0.2 &0.73~~&373 - 270 \\
MC-tv1& 12 -  31 - 11  &0.042& 832 - 801 \\
MC-tv0.01& 58 - 43 - 23  & 0.21~~& 656 - 598\\
Ecc-tv0.1& 49 - 40 - 21 & 0.089 &757 -  725 \\
Mass-tv0.1& 48 - 35 - 19 &0.14~~  &490 - 470 \\
Spin-tv0.1& 43 - 36 - 19 &0.092  &830 - 811 \\
Rp-tv0.1&  64 - 61 - 41   &0.067 & 701 - 678 \\
Rp-tv0.01& 81 - 67 - 61 &0.17~~ &608 - 537 \\
Rp-tv0.03& 82 - 83 - 73 &0.064 &834 - 793  \\
\hline
     \multicolumn{4}{l}{ $^{(a)}$ $\mathcal{F}_\lambda$ in Eq. (\ref{eq:fraction_model}) is
     calculated using $N=10$ bins for different observed samples: RV}\\
      \multicolumn{4}{l}{(radial velocity planets), TR (transit planets corrected by
      geometric bias), and TR2}\\
      \multicolumn{4}{l}{(transit planets corrected by assuming
      a detection probability $\propto a^{-5/2}$).} 
\end{tabular}
\end{center}
\label{table:three_pl}
\end{table*}

\begin{figure}[h]
   \centering
  \includegraphics[width=8.4cm]{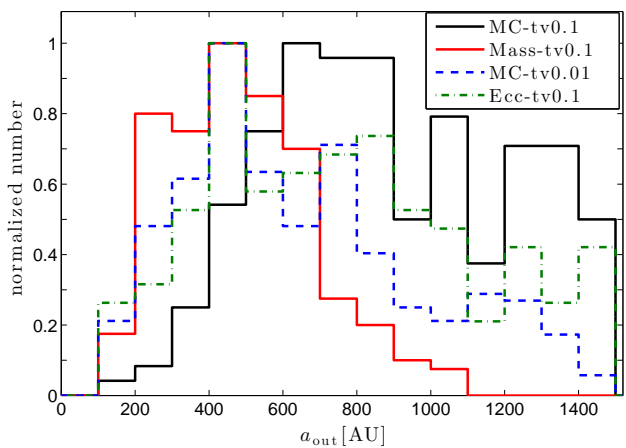}
  \caption{Distribution of the semi-major axis of the stellar 
  binary for the systems that formed a Hot Jupiter 
  in the simulations MC-tv0.1 (solid black line),
  Mass-tv0.1 (solid red line), MC-tv0.01 (blue dashed line),
  and Ecc-tv0.1 (green dotted-dashed line).
  See Table 1 for the description of each simulation.
  The bin size is 100 [AU].}
\label{fig:HJ_comp}
\end{figure}   

\section{Hot Jupiter systems: formation rate
 and binary companions}
\label{sec:companions} 

\subsection{Efficiency constraint from the semi-major
axis distribution}
\label{sec:sma_rate}

As discussed in \S\ref{sec:SMA}, the semi-major axis distribution
of HJs formed by KL migration does not match the observed 
distribution unless we change the disruption distance or start
the simulation with an inflated planet.
In this subsection we use this mismatch to place constraints 
on the maximum contribution
from KL migration to the observed population of HJs.

We build an approximate measure of the maximum 
fraction of systems $\mathcal{F}_\lambda$ that the theoretical distribution 
$Y_{\mbox{\small{t}}}$ can explain from the observed distribution 
$Y_{\mbox{\small{obs}}}$ as
\ba
\mathcal{F}_\lambda=1- \underset{{\lambda\in[0,1]}}\min
\sum_{i=1}^N
\left|Y_{\mbox{\small{obs}}}^i-
\lambda Y_{\mbox{\small{t}}}^i\right|,
\label{eq:fraction_model}
\ea
where $Y_{\mbox{\small{obs}}}^i$ and $Y_{\mbox{\small{t}}}^i$ 
are the fraction of systems with $a\in0.1\mbox{ AU}[(i-1)/N,i/N]$ 
with $i=1,..,N$ in the observed sample 
and the theoretical distribution.
We calculate $\mathcal{F}_\lambda$ by letting  
$Y_{\mbox{\small{obs}}}^i$ take values within its 
$1-\sigma$ Poisson error bar for each bin $i$, but
constrained to be normalized to one: 
$\sum_{i=1}^N Y_{\mbox{\small{obs}}}^i=
\sum_{i=1}^N Y_{\mbox{\small{t}}}^i=1$.

In Table 2, we show the values of $\mathcal{F}_\lambda$ 
for a discretization of the semi-major axis in $N=10$ bins
and using both the RV and transit planets (TR and TR2).
In TR we correct the observed distribution by
geometric bias ($\propto a^{-1}$) only (as in the rest of the paper), 
while in TR2 we correct the observed distribution 
with a steeper function of the semi-major axis
$\propto a^{-5/2}$, which takes into account
both the transit probability and the 
detection probability for a given S/N \citep{gaudi05}. 
Our results for $\mathcal{F}_\lambda$  change only slightly 
(by $\sim10\%$) for $N=5$ and  $N=20$ bins. 

The constraint from the RV sample is weaker (larger
$\mathcal{F}_\lambda$) than that 
from TR (and TR2) because the former has a smaller
 sample (37 planets) than the latter (159 planets).
Moreover, TR2 is more constraining than TR because
the former produces a distribution that favors planets
with larger semi-major axes. 

From Table 1, our fiducial simulation MC-tv0.1 can produce 
up to $44\%$ of the HJs based on the RV sample and
up to $37\%$ ($19\%$)  based on the TR (TR2) sample. 
Such fractions decrease to $<6\%$ in MC-tv0.1-nt 
(i.e., ignoring tidal disruption).
Thus, even though the number of HJs formed in 
MC-tv0.1-nt is higher than MC-tv0.1 by a factor of
$\simeq 8$ (Table 1), the production can be strongly
 constrained by the semi-major axis distribution.

 Based on the TR sample we observe from Table
 1 that the KL migration can explain up to $\sim40\%$
 of the HJs if we ignore the simulations where the
 planetary radius shrinks (Rp-tv0.1, Rp-tv0.01, and
 Rp-tv0.03).
This fraction goes down to  $\sim20\%$ if
we consider a steeper semi-major axis dependence
for the selection biases.

We conclude that the semi-major axis distribution of HJs
in the observed population places an upper limit
of  $\sim20-40\%$ to the overall production of HJs 
from KL migration due to binary companions. 
Such contribution can increase
up to $\sim 70\%$ if we start the simulations
with an inflated planet.

\subsection{Formation rate}
\label{sec:rate}

Following  \citet{WMR07} we estimate the fraction of stars with Hot
Jupiters ($a<0.1$ AU) that have migrated by the KL mechanism as:
\ba
f_{<0.1}=f_b\cdot f_p\cdot f_{\rm KL},
\label{eq:f_hj}
\ea
where $f_b$ is the fraction of stars in binaries,
 $f_p$ is the fraction of Solar-type
stars hosting a gas giant planet at a few AU, and $f_{\rm KL}$ is 
the fraction of Hot Jupiters formed by the KL mechanism in our 
simulations.

The fraction of solar-type stars in binary systems is $\sim 65\%$
\citep{DM91,ragha11}, while the fraction of binary systems with 
$a_{\rm out }>100$ AU estimated from the observed 
semi-major axis distribution is $\sim30\%$  
\citep{ragha11,egg11}.
Note that the efficiency from KL migration to form HJs
in tighter binaries ($a_{\rm out}<100$ AU) is  negligible since 
HJs are typically  formed in binaries with $a_{\rm out}>200-400$ AU
(see Figure \ref{fig:HJ_comp}  and the 
recent simulations by \citealt{MB14} for tighter 
binaries).
Thus, we approximate the fraction of stars in binaries by restricting 
ourselves to wide binaries only, which results in
 $f_b\sim65\%\times30\%\sim20\%$.
Note also that the fraction of stars hosting planets in single stars
is similar to that in wide binary stars
\citep{ragha11}.

The fraction of solar-type stars hosting a planet more massive 
than $\simeq 0.15M_J$ on a orbit with a period shorter than 
10 years is $\simeq 14\%$ \citep{mayor11}, and most of these 
are beyond $\sim1$ AU.
This fraction is consistent with the lower limit of $7\%$ previously found
by \citet{marcy05}.
Thus, the fraction of solar-type stars hosting a gas giant planet 
at a few AU is $f_p\simeq14\%$. 
This number might be regarded as an upper limit
since a fraction of these systems
have multiple planets in which case
the KL mechanism might not operate (e.g., \citealt{inn97,FT07}).

The fraction of Hot Jupiters $f_{\rm KL}$  formed by the KL mechanism in our 
simulations is given in Table 1 and 
spans the range $\simeq1.4-7.4\%$.
This range is consistent with the previous estimates by \citet{WMR07}
who find $f_{\rm KL}\simeq2.5\%$ using similar Monte Carlo 
simulations (although with a different prescription of 
tides\footnote{The authors assume that during migration $Q$ 
is constant as opposed to a constant viscous time (or time-lag) that we
assume in our simulations.} and
ignoring the effect from octupole-level gravitational interactions). 

As discussed in \S\ref{sec:hj_dis}, the fraction of HJs formed 
in our simulations increases rapidly for smaller values of the tidal disruption 
distance (see upper panel of Figure \ref{fig:HJ_Rt}) and by considering 
a distance of 0.0078 AU (i.e.,
$f_t=1.66$ in Eq. [\ref{eq:RL}]) as in \citet{naoz12} we get a fraction 
of $f_{\rm KL}\simeq 13\%$ in our fiducial simulation, consistent
with the results from these authors. 
However, as we decrease the tidal disruption distance the
semi-major axis distribution from our model deviates more 
and more from the observed distribution (see middle and 
lower panel of Figure \ref{fig:HJ_Rt}). 

In Table 2, we show the estimated fraction of stars with Hot
Jupiters produced by KL migration $f_{<0.1}$ in Equation
(\ref{eq:f_hj}) by taking $f_b=0.2$, $f_p=0.14$, and 
$f_{\rm KL}$ from Table 1 (i.e., with $f_t=2.7$ in Eq. [\ref{eq:RL}]).
We observe that such fraction ranges in $f_{<0.1}=0.042-0.21\%$.

In contrast, the estimated overall occurrence rate of Hot Jupiters in 
FGK dwarfs is  $\simeq0.9-1.5\%$ from RV surveys
 \citep{marcy05,mayor11,wright12} and
 $\simeq0.3-0.5\%$ from TR surveys \citep{gould06,howard12}. 
The difference in the estimated rates from RV and TR surveys might
 be due to a difference in metallicity in both samples
\citep{wright12,Dawson13}.

Based on the rates of HJs produced by KL migration and the
observed rates in RV (TR) surveys, 
we infer that the overall contribution from KL
migration to the Hot Jupiter population
is $\sim3-23\%$ ($\sim8-70\%$).
If tidal disruptions are ignored in our simulations 
(MC-tv0.1-nt), KL migration can contribute up to $\sim80\%$
($\sim100\%$) based on the RV (TR) surveys, but such production
rate is limited to $<6\%$ based on the semi-major axis distribution
(see \S\ref{sec:sma_rate}).
Note that the overall fraction $f_p$ of stars with giant planets
is estimated from RV surveys and, therefore, 
it might more appropriate to compare our model with the HJ rate
observed in the RV surveys rather than in the TR surveys.

In summary, based on the RV surveys only, KL
migration due to stellar companions
forms $\sim3-23\%$ of the Hot Jupiters.
Such rate can increase up to $\sim80\%$ 
by decreasing the disruption distance, but is strongly 
limited by the observed semi-major axis distribution.

\subsection{Semi-major axis of the stellar binaries}
\label{sec:sma_bin}

In Figure \ref{fig:HJ_comp} we show the distribution of the
semi-major axis of the stellar binary  $a_{\rm out}$ for 
systems that formed a Hot Jupiter
  in  MC-tv0.1 (solid black line),
  Mass-tv0.1 (solid red line), MC-tv0.01 (blue dashed line),
 and Ecc-tv0.1 (green dotted-dashed line).
In Table 2 we show the mean and median $a_{\rm out}$
for these systems.
 
As discussed in \S\ref{sec:sim_results} and shown 
in Figure \ref{fig:HJ_comp}, the fiducial simulation
MC-tv0.1 forms HJs in systems with wide binaries 
(typically $a_{\rm out}=500-1500$ AU), which is consistent
with the simulations by \citet{naoz12}.

From Figure \ref{fig:HJ_comp} we observe that as we decrease 
the mass of the perturber from 1$M_\odot$ in
the fiducial simulation to 0.1$M_\odot$ in Mass-tv0.1 the distribution 
of the semi-major axis of the binary in Hot Jupiter systems
narrows and shifts to lower 
values: the median $a_{\rm out}$ decreases from 
845 AU in the fiducial simulation to 
470 AU in  Mass-tv0.1 (Table 2).
In contrast, most HJs in Mass-tv0.1 are found in binaries
with $a_{\rm out}=200-700$ AU, which is roughly the range 
found in our fiducial simulation of  $a_{\rm out}=500-1500$ AU 
divided by $\sim2$. This is expected because the mass of
the perturber only affects the KL timescale as
$\tau_{\rm KL}\propto a_{\rm out}^3/m_3 $ (Eq. [\ref{eq:tau_KL}]), so 
at fixed KL timescale $a_{\rm out}\propto m_3^{1/3}$, which
corresponds to a factor of $\sim2$ decrease in 
 $a_{\rm out}$ when the mass of the binary companion 
 is reduced by 10.

By changing the binary eccentricity distribution
 from thermal in the fiducial simulation to uniform in Ecc-tv0.1
 we observe that the distribution of the 
semi-major axis of the binary narrows, but only slightly 
(Figure \ref{fig:HJ_comp}).

From Figure \ref{fig:HJ_comp} we also observe that by increasing the
amount of tidal dissipation from $t_{V,2}=0.1$ yr in the fiducial
simulation to  $t_{V,2}=0.01$ yr in MC-tv0.1 the distribution 
of $a_{\rm out}$ shifts to lower values: 
the median $a_{\rm out}$ decreases from 
845 AU in the fiducial simulation to 
598 AU in  MC-tv0.01 (Table 2). 
Also, the formation of HJs in MC-tv0.01 is more 
efficient by $\simeq2.2$  (Table 1)
and  by $\sim10$ in  $a_{\rm out}=100-400$ 
than in the fiducial simulation.
  
We conclude that stellar companions in systems with Hot 
Jupiters are expected to be preferentially found at wide separations:
$a_{\rm out}\simeq400-1500\left(m_3/1M_\odot\right)^{1/3}$ AU.
The formation of HJs in binaries with relatively small separations 
($a_{\rm out}\lesssim400$ AU) 
is more likely for planets with more efficient tidal
dissipation.
 
\begin{figure}
   \centering
  \includegraphics[width=8.5cm]{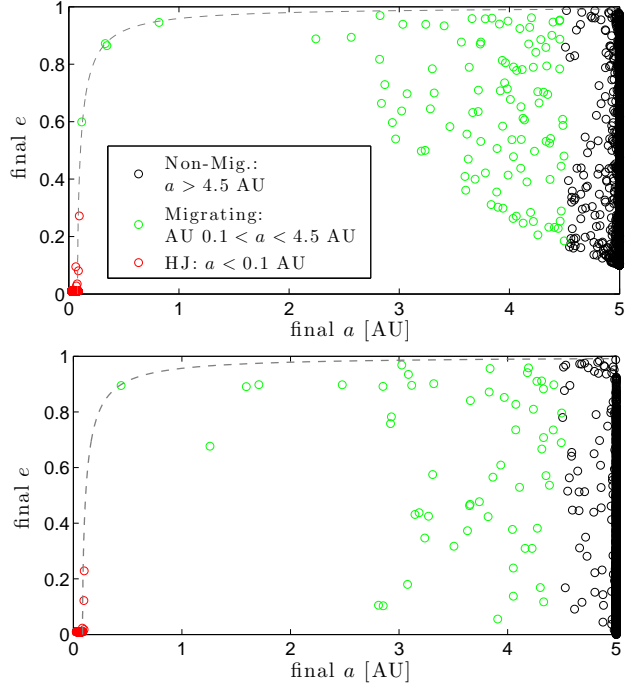}
  \caption{Final semi-major axis and eccentricity of the planetary
  orbit for simulations SMA500e0 (upper panel) and 
  Rp-tv0.01 (lower panel), similar to the panel (a) in Figure
 (\ref{fig:MC_001}).
 The dashed lines indicated the constant angular momentum 
 tracks: $a(1-e^2)=0.08$ AU.}
\label{fig:ss}
\end{figure}  

\section{Formation of intermediate period planets}
\label{sec:migrating}

From Table 1 we observe that the fraction of migrating planets 
($0.1\mbox{ AU }<a<4.5$ AU) is less than $\sim1\%$ in 
all simulations (see also panel (a) of Figure \ref{fig:MC_001}). 
Thus, KL migration is very inefficient at producing planets 
with intermediate periods ($\sim10 \mbox{ d}-3\mbox{ yr}$, or 
$a\sim0.1-2$ AU).

This lack of intermediate period planets is due to two 
different effects:
\begin{itemize}
\item many planets undergo fast KL migration (according to the
definition in \S\ref{sec:fast_kozai})
in which case the time spent at intermediate
periods is negligible compared to the planet's lifetime;
\item if the planets undergo slow KL migration
(see HJs close to the dashed line in panel (c) of Figure
\ref{fig:MC_001}), they spend most their lifetimes either undergoing 
KL oscillations at long periods ($>3$ yr) or as HJs
(see discussion in \S\ref{sec:slow_KL} and red line in the lower
panel of Figure \ref{fig:tv_example}).
\end{itemize}

In Figure \ref{fig:ss} we show the final semi-major axis and
eccentricity for the planets in the two simulations that
have the highest fraction of migrating planets (see Table 1):
SMA500e0 (upper panel) and Rp-tv0.01 (lower panel).

We observe that both SMA500e0 and 
Rp-tv0.01 have 4 planets with $a=0.1-2$ AU.
The former simulation has all four planets following
a constant angular momentum  track 
$a(1-e^2)\simeq0.08$ AU because the perturber has 
$a_{\rm out}=500$ AU and the oscillations are quenched by 
GR at $a\sim1.5$ AU (see  Figure \ref{fig:tv_example}).
The latter simulation has only one planet following the
track   $a(1-e^2)\simeq0.08$ AU, while the other
three are found off the migration track undergoing
KL oscillations. 
These three planets have a perturber with $a_{\rm out}\sim200$ AU 
so KL oscillations are quenched at $a\lesssim1$ AU and
given that the planetary radius shrinks, the migration occurs
much faster during the first $\sim10^7$ yr and then slows
down making it more likely to form intermediate period
planets.  

\subsection{Steady flow of migrating planets}

Following \citet{SKD12}, the expected
number density of planets 
with a given eccentricity in a constant
angular momentum track $J$ can be estimated as
(Eqs. 7 and 10 therein)
\ba
\frac{d\mathcal{N}_J(e)}{de}\simeq C_J\frac{1}
{e(1-e^2)^{3/2}(2.33+6.12e^3)},
\label{eq:dN_de}
\ea
where $C_J$ is a constant proportional to the current
of migrating planets with angular momentum in $(J,J+dJ)$.

By integrating Equation (\ref{eq:dN_de})
 one can estimate the ratio between
the number of moderately eccentric ($e=0.2-0.6$) planets 
 and the number of highly eccentric planets,
 which we define as 
 $e=0.85-0.99$ (at the migration track
 $a(1-e^2)\simeq0.08$ AU, the upper limit 
 of 0.99 implies $a<4$ AU).
By considering the planets that lie in
the angular momentum bin $a(1-e^2)=0.06-0.1$ AU,
this ratio is 1.5 and 1 in SMA500e0 and 
Rp-tv0.01, which is consistent with the $\simeq1.4$ 
that results from integrating Equation (\ref{eq:dN_de})
(although there are too few planets in the simulations
to make a more thorough comparison with 
\citealt{SKD12}).

In the RV sample there are 37 giant planets 
($M\sin i>0.1 M_J$)  with $a<0.1$ AU, 4 giant planets
with moderate eccentricities in $a(1-e^2)=0.06-0.1$ AU
and 2 giant planets in highly eccentric orbits:
HD 80606b \citep{naef01} and 
HD 20782b \citep{otoole09}, both of which 
are in binary stellar systems.
By limiting to orbital periods $<2$ yr and defining
a highly eccentric planet at $e>0.9$ \citet{SKD12}
argue that given the four moderately eccentric planets
one would expect $2-3$ planets with high eccentricities,
which is consistent with the observations.
Consistently, the simulation SMA500e0 has 2 moderately 
eccentric planets and one  highly eccentric planet.
Again, the simulations and the RV sample have too few
planets as to make a more thorough comparison and
the  RV surveys are biased against detecting
highly eccentric planets (e.g., \citealt{cumming04}).
In any case, the simulation SMA500e0 forms $\sim300$ HJs  per 
moderately eccentric planet which is very large compared 
with the $\sim10$ HJs in the RV sample. 

In order to avoid selection biases
against finding highly eccentric planets,
 \citet{SKD12} proposed to
study the flow of migrating planets using 
the {\it Kepler} sample.
 By constraining the eccentricity from transit light curves
in {\it Kepler} (see \citealt{DJ12}), \citet{Dawson12} report
a null detection of  highly
eccentric planets ($e>0.9$) in sample of 41
candidates with $a\simeq0.2-1.6$ AU 
(orbital periods of $\simeq 36 \mbox{ d}-2\mbox{ yr}$).
In order to compare this observation with the expected number
of highly eccentric planets, these authors consider a sample of 
$31$ (19) Hot Jupiters with $a\simeq0.04-0.06$ AU 
($a\simeq0.06-0.1$ AU) and use the RV sample to
estimate a number of moderately eccentric planets.
Their estimates indicate that KL migration can only 
produce $0.15_{-0.11}^{+0.29}$ of the HJs in {\it Kepler}.

Our simulations show that the lack of highly eccentric
planets observed by \citet{Dawson12}
is expected from KL migration.
For instance, the fiducial simulation MC-tv0.1 is able
to form 51 HJs with $a\simeq0.04-0.06$ AU 
(all with $e\simeq0$)  and only one
highly eccentric planet with $a=0.2-1.6$ AU  
(see panel a of Figure \ref{fig:MC_001}).
Similarly, Rp-tv0.01 forms 151 (122) HJs
with $a=0.04-0.06$ AU ($a=0.06-0.01$ AU)
and only one moderately eccentric HJ
and one highly eccentric planet 
(see lower panel of Figure \ref{fig:ss}). 
Note, however, that the total number of HJs relative to 
the number of moderately eccentric HJs is much
larger in the simulations than in the RV sample.
Since the eccentricities of the four moderately eccentric
planets in the observed sample are $e\sim0.3-0.5$ 
over-estimates due to bias in
RV measurements are unlikely to account
for the difference between
our simulation and the data \citep{zakamska11}.

In summary, despite the low number of planets
in a migration track at constant angular momentum
found in our simulations,  our results are
 consistent with the steady-state flow 
 of migrating planets given by \citet{SKD12}.
Since KL migration happens either fast compared 
to the planet lifetime or slow but spending a small
fraction of the time in the migration track with
$a<2$ AU, our simulations form $\sim3$ highly eccentric planet 
and $1$ moderately eccentric HJ for every  
$\sim300$ HJs in circular orbits. 
This result is consistent with lack of highly eccentric
planets at intermediate periods seen in  {\it Kepler}
\citep{Dawson12}, but is inconsistent with the number of 
moderately eccentric HJs in the RV sample and,
therefore, with the expected number of highly eccentric 
planets in {\it Kepler} predicted by \citet{SKD12}.

\begin{figure}
   \centering
  \includegraphics[width=8.5cm]{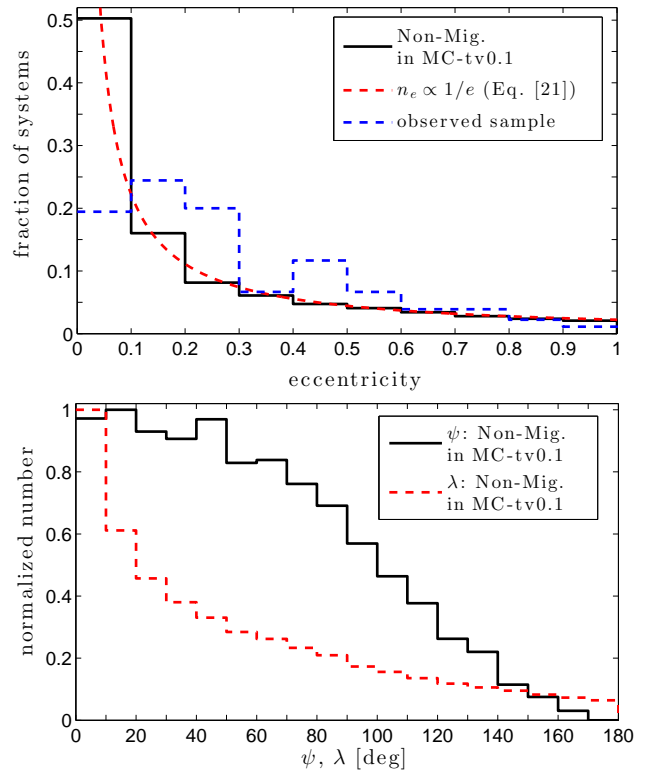}
  \caption{Final eccentricity distribution (upper panel) 
  and final obliquity distribution (lower panel) for the 
  non-migrating planets in MC-tv0.1.  
  {\it Upper panel:} the analytic time-averaged distribution (red
  dashed line)  from Equation (\ref{eq:n_e_iso}), normalized so 
  it coincides with the center of the first bin in the simulation, and
 the observed sample (blue dashed line) of 180 RV-detected 
  planets with $M\sin i>0.1M_J$, $a>1$ AU, and no detected 
  planetary companions.
   {\it Lower panel:} stellar obliquity $\psi$ and projected
   stellar obliquity $\lambda$ (normalized by the tallest bin).
  }
\label{fig:non_mig}
\end{figure}  

\section{Non-migrating planets}
\label{sec:non_mig}

From Table 1 we observe that the fraction of non-migrating 
planets ($a>4.5$ AU) in all the simulations is 
$\sim70-90\%$ or $\sim8-50$ non-migrating planets for every HJ
formed, which is roughly consistent with the ratio of 
$\sim 10-16$ between the number of giant planets at $a>1$ AU 
and the number of HJs derived from RV surveys (see \S\ref{sec:rate}).
Here, we show the steady-state eccentricity and obliquity  
distributions of these planets, which can potentially
be compared with the observed giant planets at a few AU.

In Equation (\ref{eq:n_e_i}), we calculate the time-averaged 
eccentricity distribution over a KL cycle $n_e(e_{\rm{in}}|i_0)$ 
for a planet that starts with $e^2\ll1$, where the perturber has an 
inclination $i_0$. 
In this limit, an isotropic distribution of perturbers produces a 
population of planets with a time-averaged eccentricity 
distribution given by 
\ba
n_e(e)\propto \int
\frac{d\cos i_0}{e
\left[\frac{3}{5}(1-e^2)-\cos^2 i_0\right]^{1/2}}\propto\frac{1}{e},
\label{eq:n_e_iso}
\ea
where the integral is taken over
$\cos^2 i_0<\frac{3}{5}(1-e^2)$ and the eccentricity has lower bound,
say $e>0.01$, otherwise $n_e(e)$ diverges (the eccentricity 
remains small for an arbitrarily large fraction of the KL
cycle as $e\to0$).

In the upper panel of Figure \ref{fig:non_mig}, we show the
eccentricity distribution of the non-migrating planets in our
fiducial simulation MC-tv0.1 and the time-averaged distribution
from Equation (\ref{eq:n_e_iso}).
We observe that the distributions coincide, which 
indicates that the modulation from the octupole-level
gravitational interactions (ignored in
the analytic derivation) do not play a significant role at shaping 
the steady-state eccentricity distribution.
Recall from \S\ref{sec:sim_results} that the octupole-level interactions
do play a major role at producing extremely high eccentricities
and, therefore, HJs and tidal disruptions.

In Figure \ref{fig:non_mig}, we compare this eccentricity distribution 
with that from a sample
of planets detected in RV surveys with no detected planetary 
companions. 
We observe that the KL mechanism overproduces
systems with low eccentricities 
(mean  and median of $\simeq0.2$ and $\simeq0.1$), while is 
produces too few systems with intermediate eccentricities 
($e\sim0.1-0.3$ and $e\sim0.4-0.6$ )
relative to the observations.
This is in agreement with the previous numerical work by
\citet{TR05} who find that the KL mechanism leaves
roughly $\sim50\%$ of the systems with $e\lesssim0.1$
and produces a deficit of planets with $e\simeq0.1-0.6$
relative to the observations.
Contrary to \citet{TR05}, we do not find that
the KL  mechanism produces an excess of planets 
with $e>0.6$ (except for the $0.9-1$ bin in which the difference 
is not statistically significant).
This difference might be due to the different observed
sample considered by these authors (72 RV
planets with $a>0.1$ AU excluding multiple-planet
systems) and the fact that in our simulations the planets
can migrate and get disrupted, which would tend to remove
planets with the highest eccentricities.

In the lower panel of Figure \ref{fig:non_mig}, we show the
the obliquity $\psi$ and the projected obliquity $\lambda$
distribution of the non-migrating planets.
Note that even in the absence of KL oscillations the 
the obliquity evolves due to the planetary nodal precession 
caused by an inclined companion.
The mean and median values of $\psi$ ($\lambda$)
are $\simeq58^\circ$ ($\simeq52^\circ$) and $\simeq53^\circ$
($\simeq38^\circ$), respectively.

We conclude that the eccentricity distribution of the
non-migrating planets follows the simple expression
derived in Appendix B and differs from the observed
distribution of planets at a few AU because the KL
mechanism produces too many (few) planets
with $e\lesssim0.1$ ($e\sim0.1-0.6$) relative to the
observations.
Also, these planets have a wide distribution of
stellar obliquities.

\section{Discussion}
\label{sec:discussion}

The main result from our study is that Kozai-Lidov 
migration in stellar binaries produces Hot Jupiters with a 
semi-major axis distribution that is shifted towards low values 
compared to the observations. 

We show (Figure \ref{fig:af_vs_a}) that the dominant migration 
channel is ``fast" KL migration (as opposed to ``slow" KL migration; 
see description of these regimes in \S\ref{sec:slow_KL}-\ref{sec:fast_kozai}).
This result implies that planets migrate preferentially to
small ($a<0.03$ AU) semi-major axis (see Eq. [\ref{eq:a_f_2}]), 
where the observed sample contains only  
$\sim3\%$ and $\simeq 9\%$  of  HJs detected in RV 
and transit surveys, respectively.
The tidal disruption distance sets the minimum 
semi-major axis attainable by a fast-migrating HJ.
As a result, the final semi-major axis distribution of the HJs formed 
in our Monte Carlo simulations (see Figure \ref{fig:a_tv01})
shows a significant pile-up of planets at roughly twice
the tidal disruption distance ($R_t$ in Eq. [\ref{eq:RL}]). 

In contrast, the subdominant slow KL migration 
channel is responsible for the HJs with large 
($0.05 \mbox{ AU}\lesssim a\lesssim0.1$ AU) semi-major axes,
where the observed sample 
contains $\simeq46\%$ and $\simeq45\%$ of the 
observed HJ population
detected in RV and transit surveys, respectively.
In \S\ref{sec:a_KL} we show that the ``bottleneck'' that limits the
production of such HJs is the phase in  which the planet is
undergoing KL oscillations at a few AU, which constrains
the minimum amount of tidal dissipation required to
form these planets (Eq. [\ref{eq:kl_condition}]).
From our simulations (upper panel of  Figure \ref{fig:HJ_tv})
and analytical estimates we get
a lower limit to the amount of tidal dissipation in the planet
of $t_{V,2}<0.1$ yr (or a time lag
$\tau_2>9$ s) for a Jupiter-like planet.

A way to bring the results into better agreement with the observed
semi-major axis distribution is to limit the production of HJs with 
the shortest periods.
One can achieve this in our model by either
\begin{itemize}
\item setting the tidal disruption distance at large 
enough values, $f_t\simeq3.2-3.6$ in 
$R_t$ from Equation (\ref{eq:RL}) is preferred
by the data, or 
\item considering that Jupiter-like planet starts inflated
with a radius of $\sim1.5-2R_J$ and then shrink to $R_J$,
as we show in \S\ref{sec:shrink}.
\end{itemize}

Note that the most recent calculations of the tidal 
disruption distance in Jupiter-like planets
by \citet{GRL11,Liu12} result in $f_t\simeq2.7$, 
which according to these authors might
be regarded as a lower limit since disruptions
can still happen at wider separations from repeated 
pericenter passages.

Planetary cooling models predict that 
Jupiter-like planets are born with larger radii and 
shrink in timescales of $\sim10^7$ yr 
(e.g., \citealt{burrows97}).
Gas giant planets can also be inflated due to energy 
dissipation during migration (e.g., \citealt{gu03}),
which can also limit the formation of the shortest-period
planets. We ignore this effect in our simulations and
discuss its possible consequences in \S\ref{sec:inflation}.

Evidently, limiting the production of the shortest-period planets
comes at the expense of significantly reducing the overall efficiency 
of KL migration to produce HJs (see Figure \ref{fig:HJ_Rt}).

The overall rate of HJ production derived from our simulations 
is $\sim2-7\%$ per binary systems harboring a giant planets
at a few AU (Table 1). 
As discussed in \S\ref{sec:rate}, the previous rate can account
for $\simeq3-23\%$ of the Hot Jupiter population based on the
planets discovered in RV surveys.

In \S\ref{sec:sma_rate},  we compare the semi-major axis distribution 
from our simulations and the observed sample, and find that  
KL migration accounts for at most $\sim 20\%$ ($\sim 40\%$) 
based on the planets discovered in transit surveys 
and corrected by a detection probability $\propto a^{-5/2}$
($\propto a^{-1}$, the geometric bias). 
This fraction ignores the simulations
in which the planetary radii varies in time (Rp-tv0.1 and Rp-tv0.01)
in which case the fraction rises to $\sim 40\%$ ($\sim 70\%$).

By simultaneously using the constraint from occurrence rate 
and the semi-major axis distribution of HJs (Table 2), we find 
that the simulation that can produce the highest fraction of
the observed HJs is MC-tv0.01 with a contribution of at most
$\simeq23\%$. 

Other results from our Monte Carlo simulations include:
\begin{enumerate}
\item The stellar obliquity angle distribution 
is significantly flatter in $\psi\simeq0^\circ-140^\circ$ 
than that obtained from simulations with
quadrupole-level gravitational interactions only
($\epsilon_{\rm oct}=0$ in Eq. [\ref{eq:phi}]) and
the distribution is fairly insensitive to the amount of tidal
 dissipation and the tidal disruption distance.
 The theoretical distribution fails to describe the observations
 because it produces too many misaligned planets, while
 a large fraction of the observations have projected 
 obliquities $\lambda<30^\circ$. This result places a crude 
 upper limit of $\sim50\%$ to the 
 contribution from KL migration to the observed 
 obliquity distribution.
 By considering the simulation with $\epsilon_{\rm oct}=0$ or
 with shorter initial host starÕs spin periods
 the distribution compares better with the observations 
 (see panel d of Figure \ref{fig:HJ_psi}).
 
Note that the disagreement between our simulations and 
the observations does not pose a major problem for our model
since we find that KL migration might only account for 
up to $\sim20\%$ of HJs, as discussed above.
Also,  as shown by \citet{morton11} and \citet{naoz12} the obliquity 
distribution might be explained by a 
population of misaligned systems from KL migration and/or
 planet-planet scattering plus a population of aligned
 systems from disk-migration (or secular interactions
 of nearly coplanar planets proposed by
 \citealt{petro14}).
 Moreover, the observed distribution of misalignments
 might be sculpted by tides in the star and this 
 would difficult the comparison between our
 models and the observations \citep{dawson14}.
  
 \item KL migration produces too few intermediate period
 planets ($0.1 \mbox{ AU}<a<2\mbox{ AU}$): 
roughly $\sim1$ intermediate period
 planet for every $\sim50-200$ HJs. 
As discussed in \S\ref{sec:migrating}, this is due to the
prevalence of fast KL migration and also due to nature
of slow KL migration where migrating planets spend most
of their lifetimes undergoing KL oscillations or as a HJ in
a low-eccentricity orbit.
This result is consistent with the lack of intermediate 
period planets found in {\it Kepler} \citep{Dawson12},
however, the predictions might be sensitive to the details
of tidal dissipation in the planet at very high eccentricities
(see discussion in \S\ref{sec:dyn_tide}).
 \item KL migration produces a large population
 of non-migrating planets ($\sim70\%$ of the simulated systems)
 with an eccentricity distribution that can be well-described
 by the analytic distribution $n_e\propto 1/e$ (see Appendix B).
 Such distribution produces an excess (deficit) of planets 
 with $e<0.1$ ($e\sim0.1-0.6$) relative to the observations,
 as previously found by \citet{TR05}. 
Contrary to the results of these authors, we do not find that 
the KL  mechanism produces a significant  excess of planets in 
very eccentric orbits relative to the observations.
\end{enumerate}

\subsection{Relation to previous work on KL migration}

The first studies of planetary KL migration in ensembles of
stellar binary systems are those by \citet{WMR07} and \citet{FT07}.

\citet{FT07} describe the distribution of spin-orbit misalignment
angles from KL migration by considering an isotropic distribution
of stellar perturbers with fixed eccentricity $e_{\rm out}=0$ and 
semi-major axis $a_{\rm out}=500$ AU (identical to our simulation 
SMA500e0). 
Such distribution can roughly describe the observed 
obliquity distribution of Hot Jupiters
(panel d Figure \ref{fig:HJ_psi}). 
However, the eccentricity and semi-major axis distributions of stellar 
binaries are wide and as shown by \citet{naoz12}
the higher-order terms in the gravitational potential can significantly 
modify the obliquity distribution (see panel a of
Figure \ref{fig:HJ_psi}).

Similarly, \citet{WMR07} studied the 
KL migration based on an observationally-motivated distribution
of binary eccentricities and semi-major axes.
These authors predicted that KL migration can 
produce HJs in $\simeq 2.5\%$ of their simulations.
However, they ignored the effect from the octupole-level
gravitational interactions, which according to  \citet{naoz12}
increases the formation rate of HJs in the simulations 
up to $\sim15\%$.
As we show here, such high efficiency of forming HJs in 
\citet{naoz12} is due to the small tidal disruption distance
used in their simulations ($f_t=1.66$ in Eq. [\ref{eq:RL}]).
The formation rate is a steep function of the disruption
distance and by decreasing the disruption distance, the
formation rate can be as high as $\sim26\%$
(Figure \ref{fig:HJ_Rt}). 
Using our fiducial disruption distance ($f_t=2.7$ in Eq. [\ref{eq:RL}])
we find that our simulations produce HJs in only 
$\simeq1.5-7.4\%$ of our simulations.

Unlike \citet{naoz12}, we have studied the semi-major axis
distribution of HJs from KL migration. 
This has allowed us to place constraints
that are complementary to those provided by the formation rate because
the semi-major axis distribution, unlike the obliquity distribution,
is very sensitive to the disruption distance. 
 \citet{WMR07} did study the semi-major axis 
distribution and showed that KL migration in binaries can 
reproduce the observed semi-major axis distribution when the 
simulation starts with an inflated planet, which then shrinks
(similar to our simulation Rp-tv0.03). 
However, their study ignores the effect from the octupole in the 
gravitational interactions, which,  as we show here,
is important to describe the semi-major axis distribution
of HJs because it gives rise to the strong pile-up at twice the 
disruption distance. 
Recall that we observed that $\sim 20-30\%$ our simulations
have disruptions, while the simulations by \citet{WMR07} 
had none.

Different from all these previous works, we have studied 
the orbital distribution of planetary systems due to
KL migration, focusing not only on the formation of HJs
but also on the expected distributions of planets at wider 
separations: the intermediate period planets 
and non-migrating planets.

More recently, \citet{MB14} studied the formation of HJs 
due to KL migration in stellar binaries. 
The orbital elements of the binary systems
result from stellar scattering from a third star.
The authors claim that the 3-day pile-up is a natural
outcome from tidal trapping (roughly equivalent to fast 
migration in our formalism) from quasi-parabolic orbits.
We observe, however, that their simulations produce HJs
that have not yet circularized for $a\gtrsim0.03$ AU
and are still on the migration track
$a(1-e^2)\simeq 0.03$ AU.
Thus, the formation of HJs at $a\gtrsim0.04$ AU is 
at the expense of eccentric planets that have not finished 
migration.
However, based in either the RV or the transit sample, the observed 
semi-major axis distribution is centered around $\simeq 0.04-0.06$ 
AU (see \S\ref{sec:SMA}) and most ($\gtrsim 95\%$) of these 
planets have small eccentricities ($e<0.2$).

The major difference between the simulations by 
 \citet{MB14}  and this  work is given by the prescription 
 for tidal dissipation.
In our work, we have used the equilibrium tide 
model for the whole evolution, while \citet{MB14} 
implement a recipe to describe the effect of dynamical 
tides calculations by  \citet{IP11} at high eccentricities ($e>0.9$)
and the equilibrium tides at smaller eccentricities.
The main consequence of doing this is that migration slows 
down considerably when equilibrium tides take over
and, therefore, many planets appear to be still in the migration 
track compared to our simulations.

\subsection{Effects ignored in our simulations}

In what follows we describe extra effects that have been ignored
in our simulations and how these could change our results.

\subsubsection{Orbital evolution of Hot Jupiters due to tidal
dissipation in the host star}
\label{sec:tide_star}

Once a Hot Jupiter in a circular orbit
has formed in our simulations, the planet rotates
synchronously with the orbital frequency 
and there is essentially no subsequent orbital evolution 
due to tides raised in the planet.
However, tides raised in the host star can still lead to 
evolution of the Hot Jupiter.

The characteristic evolution timescale of the spin host star 
and the orbit  of the Hot Jupiter  is roughly given 
by the tidal friction timescale\footnote{Note that the
orbital decay timescale is shorter than the tidal friction 
timescale due to the strong 
dependence of the tidal friction timescale on $a$.}. 
From Equation (\ref{eq:t_f_1}), we get
\ba
t_{F,1}&\simeq&15 \mbox{ Gyr} \left(\frac{t_{V,1}}{50 \mbox{ yr}}\right)
\left(\frac{a}{0.03 \mbox{ AU}}\right)^8
\left(\frac{R_\odot}{R_1}\right)^8 \nonumber \\
&\times&\left(\frac{M_1}{M_\odot}\right)\left(\frac{M_J}{M_2}\right),
\label{eq:t_f_star}
\ea
where $t_{V,1}$ is the viscous time of the host star.
For the parameters in Equation (\ref{eq:t_f_star}), 
the tidal friction timescale is equivalent to that calculated 
using a quality factor of $Q\simeq3\times10^5$
for a Hot Jupiter at $a=0.03$ AU. 
Note that such value of $Q$ is typically low compared to the
theoretical values of $\sim10^6-10^7$
(see references in \citealt{lai12}) and the tidal friction
timescale is likely to be longer than that 
in Equation (\ref{eq:t_f_star}).

From Equation (\ref{eq:t_f_star}) we see that 
tidal dissipation in the host star might have little 
effect on the host star spin\footnote{Assuming a solar-type star with 
spin period of $\sim1-10$ days.}  and the orbit of the Hot 
Jupiter (e.g., $\psi$ and $a$) if these planets are formed at 
$a\gtrsim0.03$ AU (although see \citealt{lai12}).
On the contrary, for Hot Jupiters closer-in ($a\lesssim0.03$ AU)
tides can in principle change the values of $\psi$ and $a$
we derive in our simulations.
Recent calculations by \citet{VF14a,VF14b} show that  
the obliquity and semi-major axis of observed HJ systems 
can evolve significantly. 
In particular, systems that lie within twice the Roche limit
($a<2R_t$ with $f_t=2.16$ in Eq. [\ref{eq:RL}]) might have 
spiraled in from an initial semi-major axis that is larger
than twice the Roche limit \citep{VF14b}. 

The host star's spin period in Hot Jupiter systems is typically 
larger than the orbital period so the orbit would decay in 
most cases.
In principle, this effect can limit the production of planets
with small semi-major (e.g., \citealt{LVW09,jackson09})
and change the obliquity distribution.
Note that the obliquity of Hot Jupiter systems 
is expected to decrease for prograde orbits ($\psi<90^\circ$),
while it can either increase or decrease for retrograde
orbits ($\psi>90^\circ$) depending on the ratio of the
orbital and the spin angular momenta.
As shown by \citet{RL13}, the retrograde HJs would typically
evolve towards $180^\circ$ and, therefore, the subsequent 
tidal dissipation in the star is not expected to bring our results
for the obliquity distribution in better agreement with the 
observations.
Finally, this re-aligment process might be more 
efficient in HJs orbiting cooler stars than hot star and could, 
in principle, account
for the observed trend of obliquity with surface 
temperature (e.g., \citealt{al12}).

We conclude that the semi-major axis distribution of Hot 
Jupiters formed in our simulations is expected to change
only slightly by subsequent tidal dissipation with the
host star because such planets are formed
with large enough semi-major axis. 
However, if these planets do experience some orbital 
decay our predicted semi-major distribution would evolve 
towards lower values making our results stronger.
Also, the obliquity distribution might evolve but is not expected
to bring the results in better agreement with the observations
because we expect that most retrograde planets 
will evolve towards $180^\circ$.

\subsubsection{Planetary inflation}
\label{sec:inflation}

During KL migration the orbital energy dissipated is about 
$\sim10$ times ($\sim R_J/a_{\rm f} \cdot M_\odot/M_J$ with
$a_{\rm f}$ the final semi-major axis) 
planet's binding energy, while the tidal dissipation rate in the 
planet is expected to exceed that in the host star
(compare $t_{F,1}$ and $t_{F,2}$ in Eq. [A16]). 
Provided that this orbital energy is deposited deep enough in the 
planet (not immediately radiated away), the planet  might be subject to
inflation due to tidal heating \citep{bode01,bode03,gu03}.
 
 Given the strong dependence of the tidal dissipation rate on the
 planet ($1/t_{F,2}\propto R_2^8$ in Eq. [A16]), planetary inflation
 can substantially speed up the migration, while at the same time
 the inflated planet is more susceptible to being tidally disrupted. 
 This scenario is similar to our case in which the planets 
 initially have a larger radii and then shrink
 (see discussion in \S\ref{sec:shrink}), expect that for tidally 
 inflated planets  the change in radius depends on the orbital 
 evolution and details of tidal dissipation.

Based on simulations with a time-varying planetary 
radius (Rp-tv0.1 and Rp-tv0.01), we expect that planetary inflation 
would limit the production of HJs with small semi-major axes
($a<0.03$ AU): such planets would be more susceptible to
tidal disruptions because these HJs reach the smallest
pericenter distances during migration, 
which would inflate the planet more efficiently and increase the 
tidal disruption distance ($R_t$ in Eq. [\ref{eq:RL}]). 
Moreover, such short-period HJs might also be subject to 
mass loss episodes in which case the planet would migrate 
outwards as a result of total angular momentum conservation
\citep{gu03}. 

We conclude that considering planetary inflation in our models
would shift the semi-major axis distribution to larger values
in better agreement with the observations, similar to our models
in which the planet is initially inflated.

\subsubsection{Effect from dynamical tides}
\label{sec:dyn_tide}

The tidal distortion in planets at very 
high eccentricities is generated only at the pericenter and is 
negligible during the rest of the orbit so the planet
can no longer achieve equilibrium figures as described in
the equilibrium tide model \citep{hut}.
Instead, the energy deposited in normal modes (surface gravity
waves) during a  periastron passage can be dissipated
and lead to orbital decay, which is described 
in the dynamical tide model (e.g., \citealt{lai97,IP04}).
We have ignored the effect from dynamical tides
in our simulations.

Most importantly for our study,
the model of dynamical tides by  \citet{IP11}
predicts  migration timescales that depend more steeply
(exponentially)  on the pericenter
distance than the equilibrium tides from \citet{hut}, which
predict a migration timescale $\tau_a\propto [a(1-e)]^{15/2}$
(Eq. [\ref{eq:a_dot}]).

\citet{BN12} use a recipe for tidal dissipation 
that attempts to describe the effect of dynamical tides in the migration 
rates at high eccentricities using the calculations by  \citet{IP11}
and  the equilibrium tide
approximation we have used in this work.
To do so, the authors added an empirical correction factor to the 
migration timescale from the constant time-lag model of 
$10^{200e^2(a(1-e)-0.022 \tiny{\mbox{AU}})}$, which
has a significant effect only at high eccentricities.
This prescription enhances the migration rate when the 
pericenter distance is $<0.022$ AU, while it slows it
down for $>0.022$ AU 
(the specific change-over distance of $0.022$ AU
changes only slightly depending on the planetary viscous times
or time-lag).

Based on this description of tidal dissipation in the planet 
at high eccentricities,
we expect the following effects in our calculations:
\begin{itemize}
\item the formation of HJs with small semi-major axes ($a<0.03$ AU)
should increase because these planets result from tidal circularization
of highly eccentric orbits with pericenter distances
$a(1-e)<0.015$ AU, which are expected to migrate at least $\sim25$
times more rapidly than in our calculations
according to the prescription by  \citet{BN12}.
Additionally, this effect would allow many planets to migrate 
fast enough so that they are not tidally disrupted, implying
that the number of HJs increases at the expense of
having less tidally disrupted planets. 
\item the formation of HJs with large semi-major axis ($a>0.05$ AU)
should decrease relative to our simulations because 
these planets are formed by slow KL migration in which 
high-eccentricity orbits 
reach pericenter distances of $a(1-e)\gtrsim0.025$ AU
and, therefore, the migration timescale is expected to
be longer in the model by \citet{IP11}.
\item the number of intermediate period planets and 
eccentric HJs should increase because a large fraction
of fast-migrating planets would slow-down their migration
as their orbits circularize and enter the regime in which
equilibrium tides takes over.
\end{itemize}

From these speculative arguments we conclude that
that the model dynamical tides by \citet{IP11} would 
produce a semi-major axis distribution of the HJs
skewed towards lower values compared to our
simulations.
This result  would strengthen 
our main result that KL migration produces too many HJs
with small semi-major axis relative to the observations.

\subsubsection{The validity of the secular approximation}
\label{sec:secular}

We have approximated the dynamical evolution of the 
three-body system using double-orbit averaging.
Using direct $N$-body integrations, \citet{anto13} showed that 
this approximation might break down for 
$(a_{\rm{out}}/a_{\rm{in}})(m_1+m_2)/m_3\simeq20$ or lower
values (see also \citealt{anto14}).
In this regime, the authors observe that once the planet
reaches very high eccentricities in a KL cycle, extra eccentricity
oscillations occur with timescale given by the period of 
the outer body \citep{ivanov}.

The eccentricity reached due to these extra oscillations
seen in the N-body calculations 
is larger than that obtained from the double-averaging
calculations \citep{anto13}.  
Also, the timescale of such oscillations
is shorter than the KL timescale by factor 
$\sim P_{\rm{in}}/P_{\rm{out}}$ and this
short-period eccentricity forcing 
can more easily overcome the effect of tidal damping and 
precession of the pericenter from GR.

The previous observations have a few consequences for
our work.
First, the number of tidally disrupted planets is expected to 
increase and the HJ period distribution would shift to lower
values.
This would strengthen our result that KL migration 
produces a semi-major axis distribution skewed towards
low values relative to the observations.
Second, short-period changes to the eccentricity  
can promote migration of systems that do not reach
high enough eccentricities in the double-averaging
approximation.
This effect should not change our results because we observe 
that migration  happens for perturbers typically at 
$a_{\rm out}>250$ AU (Figure \ref{fig:MC_001}, panel d) and, 
therefore, $(m_1+m_2)/m_3(a_{\rm{out}}/a_{\rm{in}})>50$, where 
 double averaging is expected to be 
 a good approximation \citep{ivanov}.

\subsubsection{Fate of tidally disrupted planets}

We have shown that $\sim 20-25\%$ of the planets in
our Monte Carlo simulations are tidally 
disrupted (see Table 1). 
The overall rate can be estimated as in \S\ref{sec:rate}
and results in $\sim1\%$ of the solar-type stars having a 
tidal disruption event. 
This is a lower limit because we limit
our study to wide binaries ($a_{\rm out}>100$ AU), while both
the rate of disruptions in the simulations and the
binary fraction increase for tighter binaries.
Additionally, the rate of disruptions can be enhanced even further
if we were to start with planets in highly eccentric orbits and in
low inclination binaries \citep{li14}.

From panels (a) and (c) in Figure \ref{fig:MC_001}
we observe that planets get disrupted on 
very eccentric orbits ($e>0.99$)  at $a>2$ AU. 
Also, the distribution of stellar obliquities in our fiducial
simulation at the moment of disruption (not shown)
span the range $\psi =0^\circ-170^\circ$, while the median
is $ \simeq70^\circ$ and 
most systems ($\simeq 75\%$) are disrupted in prograde
orbits ($\psi <90^\circ$).

For a  Jupiter-like planet orbiting a solar mass host star
at  $a>2$ AU the orbital energy is
$>15$ times smaller than its self-binding energy.
According to \citet{GRL11}, a Jupiter-like planet crossing the disruption
distance ($R_t$ in Eq. [\ref{eq:RL}] with $f_t=2.7$)
at these separations loses slightly less than half of its initial mass 
which ends up being accreted by the host star, while the stripped
planet gets ejected from the system.
The gas falls into the host star through an accretion disk that
emits optical and UV radiation for $\sim$ days to a year with a
peak luminosity of $10^{36}-10^{37} \mbox{erg }\mbox{s}^{-1}$ 
\citep{metzger12}.
Also, the gas lost by the planet can have enough angular momentum
to significantly alter the spin rate of the host star
and even its axis of rotation.

We conclude that given the large number of planetary  
disruptions, KL migration might contribute 
to the population of free-floating planets by the ejection
of planetary cores after gas stripping, while the gas
that remains bounded to the host star might
provide a significant source of transient radiation.


\section{Summary}
\label{sec:summary}

We study the steady-state orbital distributions of
giant planets migrating through the combination of the  
Kozai-Lidov (KL) mechanism due to a 
stellar companion and friction due to tides raised on 
the planet by the host star

We find that KL migration cannot produce all Hot Jupiters
(HJs).
It can, however, produce a fraction constrained by the 
following observations:

\begin{enumerate}
\item The observed semi-major axis distribution is consistent
with KL migration (simulation Rp-tv0.03)
if the following are both true:
\begin{itemize}
\item a lower limit to the amount of tidal dissipation
in the planet, parametrized by 
the planetary viscous time, is: $t_{V,2}<0.1$ yr.
Otherwise, KL migration is unable to produce HJs
in the semi-major axis range 
$0.05\mbox{ AU} \lesssim a  \lesssim0.1 \mbox{ AU}$, which 
contains $\simeq46\%$ ($\simeq45\%$) of the 
observed HJ population
detected in RV (transit) surveys.
This lower limit to the amount of dissipation is larger by a factor
of $\simeq 150$  than 
the upper limit inferred from the Jupiter-Io interaction of 
$t_{V,2}>15$ yr.
\item the distance at which a Jupiter-like planet in a highly 
eccentric orbit gets tidally disrupted is $\gtrsim 0.015$ AU. 
Otherwise, too many planets migrate to $a<0.03$ AU, where
the observed sample contains only  
$\sim3\%$ ($\simeq 9\%$ ) of  HJs detected in RV (transit) surveys.
One can achieve this large disruption distance either by setting  
$f_t\gtrsim3.2$ in Equation (\ref{eq:RL}) for a Jupiter-like
planet  or by starting with an inflated planet, which then shrinks
as in \S\ref{sec:shrink}.
In principle, the tidal dissipation in the star can also prevent the 
formation of short-period planets. 
\end{itemize} 
 If the standard parameters are used ($f_t\simeq2.7$ and no
 radius shrinkage), KL migration can produce 
 at most $\sim20-40\%$ of the Hot Jupiters.

\item The observed occurrence rate of HJs is roughly consistent 
with KL migration if $f_t\lesssim1$ (almost no tidal
disruptions take place).
If our fiducial disruption distance ($f_t\simeq2.7$) is used,
KL migration produces HJs at a rate which is only 
$\sim3-20\%$ of the observed one.
\item The distribution of the stellar obliquity angles of HJs
is inconsistent with KL migration.
This distribution is fairly insensitive to the amount of 
tidal dissipation and disruption distances in our
simulations.
Based on the fraction of misaligned planets in
the observations, KL migration can produce at most 
$\sim50\%$ of the HJs, independent of 
$t_{V,2}$ and $f_t$. Better agreement with the data
is found when the host star's spin period
is initially shorter.
\end{enumerate}
By simultaneously considering the constraints from the
occurrence rate and the semi-major axis distribution above,
we find a maximum fraction of $\sim20\%$ 
of the Hot Jupiters can be formed by KL migration
due to binaries (simulation MC-tv0.01).

Additionally, KL migration in binaries
is unable to form intermediate-period 
planets ($0.1\mbox{ AU} \lesssim a  \lesssim2 \mbox{ AU}$) 
because migrating planets spend most of their lifetimes undergoing KL
oscillations at $a>2$ AU or as a Hot Jupiter at $a<0.1$ AU.

\acknowledgements 
 I acknowledge support from the 
CONICYT Bicentennial  Becas Chile fellowship. 
I am indebted to Scott Tremaine who has critically
and patiently read and commented on various versions 
of this paper.
I am also grateful to Rebekah Dawson, Subo Dong,
Chelsea Huang, Boaz Katz, Renu Malhotra, Tim Morton, and 
Smadar Naoz, for enlightening
discussions and comments.
I gratefully acknowledge an anonymous referee for 
constructive feedback.
All simulations were carried out using computers 
supported by the Princeton Institute of Computational 
Science and Engineering.

\appendix
\section{Equations of motion }

We explicitly show the secular equations of motion for a 
hierarchical triple system
considering gravitational interactions up to the 
octupole approximation (i.e., expanding the 
gravitational interactions between the inner and outer binaries
up to $a_{\rm{in}}^3/a_{\rm{out}}^4$).
We do not assume the total mass or angular momentum
is in two of the bodies.
We also include the  non-Keplerian effects and tidal dissipation 
relevant to our problem.

We define the inner orbit relative to the center of mass of bodies 1 and 2, while 
the outer orbit is defined relative to the center of mass of bodies 3 and the center of
mass of bodies 1 and 2.
We specify the orientation of the inner and outer orbits by their 
Laplace-Runge-Lenz vectors ${\bf e}_{\rm{in}}$ and ${\bf e}_{\rm{out}}$, 
whose magnitudes are $e_{\rm{in}}$ and $e_{\rm{out}}$. 
Similarly, we express the angular momentum vectors as
\ba
{\bf h}_{\rm{in}}&=&\frac{m_1m_2\sqrt{G(m_1+m_2)a_{\rm{in}}(1-e_{\rm{in}}^2)}}
{(m_1+m_2)}{\bf \hat{h}}_{\rm{in}}\\
{\bf h}_{\rm{out}}&=&\frac{(m_1+m_2)m_3\sqrt{G(m_1+m_2+m_3)a_{\rm{out}}(1-e_{\rm{out}}^2)}}
{(m_1+m_2+m_3)}{\bf \hat{h}}_{\rm{out}},
\ea
and by defining the vectors ${\bf \hat{q}}_{\rm{in}}={\bf \hat{h}}_{\rm{in}} \times {\bf \hat{e}}_{\rm{in}}$ and
${\bf \hat{q}}_{\rm{out}}={\bf \hat{h}}_{\rm{out}} \times {\bf \hat{e}}_{\rm{out}}$ we complete
the right-hand triad of unit vectors 
$({\bf \hat{q}}_{\rm{in}},{\bf \hat{h}}_{\rm{in}} , {\bf \hat{e}}_{\rm{in}})$ and 
$({\bf \hat{q}}_{\rm{out}},{\bf \hat{h}}_{\rm{out}} , {\bf \hat{e}}_{\rm{out}})$.
In our notation, the sub-index $1$ indicates the host star, $2$ the planet, 
and $3$ the outer perturber.

The double time averaging of the perturbing potential over 
the orbital periods of the inner and outer orbits yields
(e.g., \citealt{farago2010,correia11,TY14})
\ba
\phi_{\rm{oct}}&=&\frac{\phi_0}{(1-e_{\rm{out}}^2)^{3/2}} 
\left[ \frac{1}{2} (1-e_{\rm{in}}^2)({\bf \hat{h}}_{\rm{in}}\cdot {\bf \hat{h}}_{\rm{out}})^2  
+ \right(e_{\rm{in}}^2 -\frac{1}{6}\left)-
\frac{5}{2}  ({\bf e}_{\rm{in}}\cdot {\bf \hat{h}}_{\rm{out}})^2   \right]+\nonumber\\
&+&\frac{\epsilon_{\rm{oct}}\phi_0}{(1-e_{\rm{out}}^2)^{3/2}} 
\bigg\{ 
 ({\bf e}_{\rm{in}}\cdot {\bf \hat{e}}_{\rm{out}})\left[ \left(\frac{1}{5}-\frac{8}{5}e_{\rm{in}}^2\right)
 - (1-e_{\rm{in}}^2)({\bf \hat{h}}_{\rm{in}}\cdot {\bf \hat{h}}_{\rm{out}})^2+
 7  ({\bf e}_{\rm{in}}\cdot {\bf \hat{h}}_{\rm{out}})^2
\right]
	+ \nonumber\\
&-&	
2(1-e_{\rm{in}}^2)({\bf \hat{h}}_{\rm{in}}\cdot {\bf \hat{h}}_{\rm{out}})({\bf e}_{\rm{in}}\cdot {\bf \hat{h}}_{\rm{out}})
 ({\bf \hat{h}}_{\rm{in}}\cdot {\bf \hat{e}}_{\rm{out}})
\bigg\}, \nonumber \\
&&\mbox{where}~~~
\phi_0=\frac{3G}{4} \frac{a_{\rm{in}}^2}{a_{\rm{out}}^3} \frac{m_1 m_2 m_3}{m_1+m_2},\quad 
\epsilon_{\rm{oct}}=\frac{25}{16}\frac{a_{\rm{in}}}{a_{\rm{out}}}\frac{e_{\rm{out}}}{(1-e_{\rm{out}}^2)}
\frac{m_1-m_2}{m_1+m_2}.
\label{eq:phi}
\ea

This potential reduces to that given by \citet{katz11}
in the test particle limit ($m_2 \ll m_1, m_3$) when one fixes the outer 
perturber's orbit to 
${\bf h}_{\rm{out}}= h_{\rm{out}}\hat{z}$ and ${\bf e}_{\rm{out}}= e_{\rm{out}}\hat{x}$.
A similar expression can be found in \citet{ford2000} and \citet{naoz13a}, but
in terms of orbital elements rather than the eccentricity and
angular momentum vectors.

By following \citet{tremaine09} one can easily write the equations of 
motion for the eccentricity and angular momentum vectors by taking gradients
of the potential. These fully describe the secular evolution of the inner and 
outer orbits when only the gravitational interactions are taken into account.
The advantage of this procedure is that the equations of motion derived from
$\phi_{\rm{oct}}$ do not diverge as the inner orbit approaches 
radial ($e_{\rm{in}}\to1$) and circular orbits ($e_{\rm{in}}\to0$) as happens when 
using the equations of motion in terms of the orbital elements  \citep{ford2000,naoz13a}.

The full set of equations describing the secular gravitational interaction, 
precession from general relativity, stellar oblateness, tidal friction, and the rotation 
of the star and inner planet can be written in a similar way as in \citet{2001EK} as:  
\ba
\frac{d {\bf e}_{\rm{in}}}{dt} &=& (Z_1 + Z_2 + Z_{\rm{GR}})e_{\rm{in}} {\bf \hat{q}}_{\rm{in}} - 
(Y_1 + Y_2) e_{\rm{in}}{\bf \hat{h}}_{\rm{in}} - (V_1 + V_2) {\bf e}_{\rm{in}} \nonumber\\
  &+& \tau_{\rm{in}}^{-1}\left[
	(1-e_{\rm{in}}^2)^{1/2} {\bf \hat{h}}_{\rm{in}} \times \nabla_{e_{\rm{in}}} \tilde{\phi}_{\rm{oct}} + {\bf e}_{\rm{in}} \times \nabla_{h_{\rm{in}}}\tilde{\phi}_{\rm{oct}}   
	\right],  	\\
\frac{1}{ h_{\rm{in}}}\frac{d {\bf h}_{\rm{in}}}{dt} &=& (Y_1+Y_2){\bf \hat{e}}_{\rm{in}} - (X_1 + X_2) {\bf \hat{q}}_{\rm{in}} - (W_1 + W_2) {\bf \hat{h}}_{\rm{in}} \nonumber\\
&+& (1-e_{\rm{in}}^2)^{-1/2} \tau_{\rm{in}}^{-1}
\left[
	 {\bf e}_{\rm{in}} \times \nabla_{e_{\rm{in}}}\tilde{\phi}_{\rm{oct}}+
	 (1-e_{\rm{in}}^2)^{1/2}{\bf \hat{h}}_{\rm{in}}  \times \nabla_{h_{\rm{in}}}\tilde{\phi}_{\rm{oct}} 	\right], \\
 \frac{d {\bf e}_{\rm{out}}}{dt}&=& \tau_{\rm{out}}^{-1}\left[
	(1-e_{\rm{out}}^2)^{1/2}{\bf \hat{h}}_{\rm{out}}  \times \nabla_{e_{\rm{out}}} \tilde{\phi}_{\rm{oct}} +
	 {\bf e}_{\rm{out}} \times \nabla_{h_{\rm{out}}}\tilde{\phi}_{\rm{oct}}   
	\right],  \\
\frac{1}{h_{\rm{out}}}\frac{d {\bf h}_{\rm{out}}}{dt}&=&(1-e_{\rm{out}}^2)^{-1/2} \tau_{\rm{out}}^{-1}\left[
	 {\bf e}_{\rm{out}} \times \nabla_{e_{\rm{out}}}\tilde{\phi}_{\rm{oct}}+ 
	 (1-e_{\rm{out}}^2)^{1/2}{\bf \hat{h}}_{\rm{out}}  \times \nabla_{h_{\rm{out}}}\tilde{\phi}_{\rm{oct}} 	\right],~~~~~~~~~~~~~~~~
\ea
\ba
I_1 \frac{d {\bf \Omega}_1 }{dt} &=& h_{\rm{in}} (-Y_1{\bf \hat{e}}_{\rm{in}} + X_1 {\bf \hat{q}}_{\rm{in}} + W_1 {\bf \hat{h}}_{\rm{in}}), \\
I_2 \frac{d {\bf \Omega}_2 }{dt} &=& h_{\rm{in}} (-Y_2{\bf \hat{e}}_{\rm{in}} + X_2 {\bf \hat{q}}_{\rm{in}} + W_2 {\bf \hat{h}}_{\rm{in}}), \\
 \mbox{where}&& \tau_{\rm{in}} =\frac{m_1m_2\sqrt{G(m_1+m_2)a_{\rm{in}}}}{(m_1+m_2)\phi_0}, ~
	 \tau_{\rm{out}} =  \frac{(m_1+m_2)m_3\sqrt{G(m_1+m_2+m_3)a_{\rm{out}}}}{(m_1+m_2+m_3)\phi_0},
\ea
and the gradients are taken over the dimensionless
potential $\tilde{\phi}_{\rm{oct}}\equiv \phi_{\rm{oct}}/\phi_0$ (Eqs. [A18]-[A21]).

The terms due to tidal effects and rotation can be expressed for $m_1$ as
\ba
V_1 &=& \frac{9}{t_{F1}} \left[ \frac{1 + (15/4)e_{\rm{in}}^2 + (15/8)e_{\rm{in}}^4 + (5/64)e_{\rm{in}}^6}{(1-e_{\rm{in}}^2)^{13/2}} - \frac{11 \Omega_{1h}}{18 \dot{l}_{\rm{in}}} \frac{1 + (3/2) e_{\rm{in}}^2 + (1/8) e_{\rm{in}}^4}{(1-e_{\rm{in}}^2)^5} \right], \label{eq:V}\\
W_1 &=& \frac{1}{t_{F1}} \left[ \frac{1 + (15/2)e_{\rm{in}}^2 + (45/8)e_{\rm{in}}^4 + (5/16)e_{\rm{in}}^6}{(1-e_{\rm{in}}^2)^{13/2}} - \frac{\Omega_{1h}}{\dot{l}_{\rm{in}}} \frac{1 + 3 e_{\rm{in}}^2 + (3/8) e_{\rm{in}}^4}{(1-e_{\rm{in}}^2)^5} \right], \\
X_1 &=& -\frac{m_2 k_1 R_1^5}{\mu \dot{l}_{\rm{in}} a_{\rm{in}}^5} \frac{\Omega_{1h} \Omega_{1e}}{(1-e_{\rm{in}}^2)^2}  - \frac{ \Omega_{1q} }{ 2 \dot{l}_{\rm{in}} t_{F1}} \frac{1 + (9/2) e_{\rm{in}}^2 + (5/8) e_{\rm{in}}^4}{(1-e_{\rm{in}}^2)^5}, \label{Xeq} \\
Y_1 &=& -\frac{m_2 k_1 R_1^5}{\mu \dot{l}_{\rm{in}} a_{\rm{in}}^5} \frac{\Omega_{1h} \Omega_{1q}}{(1-e_{\rm{in}}^2)^2}  + \frac{ \Omega_{1e} }{ 2 \dot{l}_{\rm{in}} t_{F1}} \frac{1 + (3/2) e_{\rm{in}}^2 + (1/8) e_{\rm{in}}^4}{(1-e_{\rm{in}}^2)^5}, \label{Yeq} \\
Z_1 &=& \frac{m_2 k_1 R_1^5}{ \mu \dot{l}_{\rm{in}} a_{\rm{in}}^5} \left[ \frac{2 \Omega_{1h}^2 - \Omega_{1q}^2 - \Omega_{1e}^2 }{ 2 (1-e_{\rm{in}}^2)^2}  + \frac{ 15 G m_2}{ a_{\rm{in}}^3 } \frac{1 + (3/2) e_{\rm{in}}^2 + (1/8) e_{\rm{in}}^4}{(1-e_{\rm{in}}^2)^5}  \right],
\ea
where analogous equations are written for $m_2$ by swapping indices. 
Considering dissipation in $m_1$,
the tidal friction timescale is defined in terms of the viscous timescale $t_{V1}$  and
\begin{equation}
t_{F1} = \frac{t_{V1}}{9} \left( \frac{a_{\rm{in}}}{R_1} \right)^8 \frac{m_1^2}{(m_1+m_2) m_2} (1+2k_1)^{-2}.
\label{eq:t_f_1}
\end{equation}
Here, $k_1$ is the classical apsidal motion constant, a measure of quadrupolar 
deformability which is related to the Love number $k_L $ and the 
coefficient $Q_E$ given by \cite{2001EK}: $k_{1} = k_L/2=\onehalf Q_E/(1-Q_E)$.  
We use  $k_1 = 0.014$, valid for $n=3$ polytropes, to represent
 the host star and  $k_2 = 0.25$, valid for $n=1$ polytropes, to represent
  gas giant planets.  
  
Additionally, $Z_{\rm{GR}}$ is the GR precession rate given by
\ba
Z_{\rm{GR}}=\frac{3G^{3/2}(m_1+m_2)^{3/2}}{a_{\rm{in}}^{5/2}c^2(1-e_{\rm{in}}^2)}.
\ea 

The gradients of the dimensionless potential $\tilde{\phi}_{\rm{oct}}= \phi_{\rm{oct}}/\phi_0$ are
\ba
\nabla_{e_{\rm{in}}} \tilde{\phi}_{\rm{oct}}    &=&-\frac{1}{(1-e_{\rm{out}}^2)^{3/2}}\left \{
5({\bf e}_{\rm{in}}\cdot {\bf \hat{h}}_{\rm{out}})  
- 2 \epsilon_{\rm{oct}}\left[
7({\bf e}_{\rm{in}}\cdot {\bf \hat{h}}_{\rm{out}})  ({\bf e}_{\rm{in}}\cdot {\bf \hat{e}}_{\rm{out}}) -
(1-e_{\rm{in}}^2)({\bf \hat{h}}_{\rm{in}}\cdot {\bf \hat{h}}_{\rm{out}})  ({\bf \hat{h}}_{\rm{in}}\cdot {\bf \hat{e}}_{\rm{out}})  
\right]
\right\} {\bf \hat{h}}_{\rm{out}} \nonumber \\
&&+
\frac{\epsilon_{\rm{oct}}}{(1-e_{\rm{out}}^2)^{3/2}} 
\left[ \left(\frac{1}{5}-\frac{8}{5}e_{\rm{in}}^2\right) -
(1-e_{\rm{in}}^2)({\bf \hat{h}}_{\rm{in}}\cdot {\bf \hat{h}}_{\rm{out}})^2+
 7  ({\bf e}_{\rm{in}}\cdot {\bf \hat{h}}_{\rm{out}})^2
\right] {\bf \hat{e}}_{\rm{out}} \nonumber \\
&&+
 \frac{2 }{(1-e_{\rm{out}}^2)^{3/2}} \left[
 1- \frac{8}{5} \epsilon_{\rm{oct}}
 ({\bf e}_{\rm{in}}\cdot {\bf \hat{e}}_{\rm{out}}) 
 \right] {\bf e}_{\rm{in}} ,
\ea 
\ba
\nabla_{h_{\rm{in}}} \tilde{\phi}_{\rm{oct}}    &=&\frac{(1-e_{\rm{in}}^2)^{1/2}}{(1-e_{\rm{out}}^2)^{3/2}}\left \{
({\bf \hat{h}}_{\rm{in}}\cdot {\bf \hat{h}}_{\rm{out}})  
- 2 \epsilon_{\rm{oct}} \left[
({\bf e}_{\rm{in}}\cdot {\bf \hat{e}}_{\rm{out}})  ({\bf \hat{h}}_{\rm{in}}\cdot {\bf \hat{h}}_{\rm{out}})  +
({\bf e}_{\rm{in}}\cdot {\bf \hat{h}}_{\rm{out}})  ({\bf \hat{h}}_{\rm{in}}\cdot {\bf \hat{e}}_{\rm{out}})  
\right]
\right\} {\bf \hat{h}}_{\rm{out}} ~~~~~~~~~~~~~~~~\nonumber \\
&&-
 \frac{2 \epsilon_{\rm{oct}} (1-e_{\rm{in}}^2)^{1/2}}{(1-e_{\rm{out}}^2)^{3/2}}
({\bf \hat{h}}_{\rm{in}}\cdot {\bf \hat{h}}_{\rm{out}})  ({\bf e}_{\rm{in}}\cdot {\bf \hat{h}}_{\rm{out}})  
{\bf \hat{e}}_{\rm{out}},
\ea 
\ba
\nabla_{e_{\rm{out}}} \tilde{\phi}_{\rm{oct}}    &=&
\frac{5}{(1-e_{\rm{out}}^2)^{5/2}} 
\left[ \frac{1}{2}(1-e_{\rm{in}}^2) ({\bf \hat{h}}_{\rm{in}}\cdot {\bf \hat{h}}_{\rm{out}})^2  
+ \frac{3}{5}\right(e_{\rm{in}}^2 -\frac{1}{6}\left)-
\frac{5}{2}  ({\bf e}_{\rm{in}}\cdot {\bf \hat{h}}_{\rm{out}})^2   \right]
{\bf \hat{e}}_{\rm{out}}~~~~~~~~~~~~~~~~~~~~~
~~~~~~~~~~~~~~~~~	\nonumber\\
&&+
\frac{7\epsilon_{\rm{oct}}}{(1-e_{\rm{out}}^2)^{5/2}} 
\bigg\{  
 ({\bf e}_{\rm{in}}\cdot {\bf \hat{e}}_{\rm{out}})\left[ 
 \left(\frac{1}{7}-\frac{8}{7}e_{\rm{in}}^2\right)
- (1-e_{\rm{in}}^2)({\bf \hat{h}}_{\rm{in}}\cdot {\bf \hat{h}}_{\rm{out}})^2+
 7  ({\bf e}_{\rm{in}}\cdot {\bf \hat{h}}_{\rm{out}})^2
\right]
	 \nonumber\\
&&-
2(1-e_{\rm{in}}^2)({\bf \hat{h}}_{\rm{in}}\cdot {\bf \hat{h}}_{\rm{out}})({\bf e}_{\rm{in}}\cdot {\bf \hat{h}}_{\rm{out}})
 ({\bf \hat{h}}_{\rm{in}}\cdot {\bf \hat{e}}_{\rm{out}})
 \bigg\} {\bf \hat{e}}_{\rm{out}}	\nonumber \\
 &&+
 \frac{\epsilon_{\rm{oct}}}{e_{\rm{out}}(1-e_{\rm{out}}^2)^{3/2}} 
\left[\left(\frac{1}{5}-\frac{8}{5}e_{\rm{in}}^2\right)
 - (1-e_{\rm{in}}^2) ({\bf \hat{h}}_{\rm{in}}\cdot {\bf \hat{h}}_{\rm{out}})^2+
 7  ({\bf e}_{\rm{in}}\cdot {\bf \hat{h}}_{\rm{out}})^2\right]
{\bf e}_{\rm{in}}		\nonumber\\
&&-
 \frac{2 \epsilon_{\rm{oct}}(1-e_{\rm{in}}^2)}{e_{\rm out}(1-e_{\rm{out}}^2)^{3/2}}
({\bf \hat{h}}_{\rm{in}}\cdot {\bf \hat{h}}_{\rm{out}})  ({\bf e}_{\rm{in}}\cdot {\bf \hat{h}}_{\rm{out}})  
{\bf \hat{h}}_{\rm{in}},
\ea 
\ba
\nabla_{h_{\rm{out}}} \tilde{\phi}_{\rm{oct}}    &=&\frac{(1-e_{\rm{in}}^2)}{(1-e_{\rm{out}}^2)^{2}}\left \{
({\bf \hat{h}}_{\rm{in}}\cdot {\bf \hat{h}}_{\rm{out}})  
- 2 \epsilon_{\rm{oct}} \left[
({\bf e}_{\rm{in}}\cdot {\bf \hat{e}}_{\rm{out}})  ({\bf \hat{h}}_{\rm{in}}\cdot {\bf \hat{h}}_{\rm{out}})  +
({\bf e}_{\rm{in}}\cdot {\bf \hat{h}}_{\rm{out}})  ({\bf \hat{h}}_{\rm{in}}\cdot {\bf \hat{e}}_{\rm{out}})  
\right]
\right\} {\bf \hat{h}}_{\rm{in}} ~~~~~~~~~~
~~~~~~~~~\nonumber \\
&&-
\frac{1}{(1-e_{\rm{out}}^2)^{2}}\left \{
5 ({\bf e}_{\rm{in}}\cdot {\bf \hat{h}}_{\rm{out}})  
- 2 \epsilon_{\rm{oct}} \left[
7({\bf e}_{\rm{in}}\cdot {\bf \hat{e}}_{\rm{out}})  ({\bf e}_{\rm{in}}\cdot {\bf \hat{h}}_{\rm{out}})  -
(1-e_{\rm{in}}^2)({\bf \hat{h}}_{\rm{in}}\cdot {\bf \hat{h}}_{\rm{out}})  ({\bf \hat{h}}_{\rm{in}}\cdot {\bf \hat{e}}_{\rm{out}})  
\right] \right\} 
{\bf e}_{\rm{in}}. \nonumber\\
\ea

\section{Time-averaged eccentricity distribution over a Kozai-Lidov cycle}

The averaged Hamiltonian that represents the gravitational interaction 
up to the quadrupole approximation (Eq. A3 with $\epsilon_{\rm oct}=0$) 
can be  written in terms of the orbital elements as
\ba
H_{q}=\frac{\phi_0}{6(1-e_{\rm{out}}^2)^{3/2}} 
\left[2+3e_{\rm{in}}^2-(3-3e_{\rm{in}}^2+15e_{\rm{in}}^2\sin^2 \omega_{\rm{in}} ) \sin^2i_{\rm{tot}} \right],
\ea
where $i_{\rm tot}$ is the angle between ${\bf h}_{\rm{in}}$ and 
${\bf h}_{\rm{out}}$ (Eqs. A1-A2) and $\omega_{\rm{in}}$ is the 
argument of pericenter of the inner orbit.

In the limit of the outer body having all the angular momentum 
(i.e., $h_{\rm{in}}\ll h_{\rm{out}}$) the vector ${\bf h}_{\rm{out}}$ remains fixed and
we can use it to define a reference frame in which $i_{\rm{tot}}\equiv i_{\rm{in}}$.

Let us fix the energy by setting $e_{\rm{in}}=e_0,i_{\rm{in}}=i_0,$ 
and $ \omega_{\rm{in}}=\omega_0$:
\ba
H_{q,0}&=&\frac{\phi_0}{6(1-e_{\rm{out}}^2)^{3/2}} 
\left(2+3\theta_0\right), \mbox{ where }
\theta_0\equiv e_{0}^2-(1-e_{0}^2+5e_{0}^2\sin^2 \omega_{0} ) \sin^2i_{0}.
\ea

We can now write the distribution function as
\ba
f(\omega_{\rm{in}},\Omega_{\rm{in}},G_{\rm{in}},H_{\rm{in}})&=&
\delta\left(H_{q}-H_{q,0}\right) \nonumber\\
&\propto&\delta\left[ e_{\rm{in}}^2-(1-e_{\rm{in}}^2+5e_{\rm{in}}^2\sin^2 \omega_{\rm{in}} ) \sin^2i_{\rm{in}} -\theta_0\right].
\ea

Let us define the dimensionless momenta 
$\tilde{G}_{\rm{in}}= G_{\rm{in}}/L_{\rm{in}}=\sqrt{1-e_{\rm{in}}^2}$ and 
$\tilde{H}_{\rm{in}}= H_{\rm{in}}/L_{\rm{in}}=\sqrt{1-e_{\rm{in}}^2}\cos i_{\rm{in}}$,
where the latter is a constant of motion 
(i.e., $\tilde{H}_{\rm{in}}=\sqrt{1-e_{0}^2}\cos i_{0}$)
because the Hamiltonian 
is independent of the longitude of the ascending node $\Omega_{\rm{in}}$.
This allows us to express the time-averaged eccentricity distribution as
\ba
n_e(e_{\rm{in}}|\theta_0,\tilde{H}_{\rm{in}})\propto\int\int d\tilde{G}_{\rm{in}}d\omega_{\rm{in}}
\delta\left[ e_{\rm{in}}^2 - (1-e_{\rm{in}}^2+5e_{\rm{in}}^2\sin^2 \omega_{\rm{in}} ) 
\left(1-\frac{\tilde{H}_{\rm{in}}^2}{\tilde{G}_{\rm{in}}^2}\right)  -\theta_0\right]
\delta\left( e_{\rm{in}}-\sqrt{1-\tilde{G}_{\rm{in}}^2}\right),
\ea
and integrating over $\tilde{G}_{\rm{in}}$ we get
\ba
n_e(e_{\rm{in}}|\theta_0,\tilde{H}_{\rm{in}})&\propto&\int d\omega_{\rm{in}}
\frac{e_{\rm{in}}}{(1-e_{\rm{in}}^2)^{1/2}}
\delta\left[ e_{\rm{in}}^2 - (1-e_{\rm{in}}^2+5e_{\rm{in}}^2\sin^2 \omega_{\rm{in}} ) 
\left(1-\frac{\tilde{H}_{\rm{in}}^2}{1-e_{\rm{in}}^2}\right)  -\theta_0\right]\\
&\propto&
\frac{e_{\rm{in}}}{
\left\{\left[2e_{\rm{in}}^2+\tilde{H}_{\rm{in}}^2-\theta_0-1 \right]
\left[\left(1+4e_{\rm{in}}^2\right)\left(1-e_{\rm{in}}^2-\tilde{H}_{\rm{in}}^2 \right)
-\left(e_{\rm{in}}^2-\theta_0\right)\left(1-e_{\rm{in}}^2\right)
\right]\right\}^{1/2}  
  },
  \label{eq:n_e}
\ea
which is defined for $e_{\rm in}\in[0,1]$ 
such that $n_e(e_{\rm{in}}|\theta_0,\tilde{H}_{\rm{in}})$
is real and can be normalized to 1 as:
$\int_{0}^{1} de_{\rm{in}}n(e_{\rm{in}}|\theta_0,\tilde{H}_{\rm{in}})=1$.
We checked that this distribution matches the time-averaged
eccentricity distribution obtained from
solving the Equations (A4)-(A5) when $\epsilon_{\rm oct}=0$
and no extra forces are included.

A relevant limit for this work is to start with a planet in a low-eccentricity
orbit ($e_0^2\ll1$) and a stellar companion with inclination
$i_0$, so  $\theta_0\to-\sin^2 i_0$ and 
$\tilde{H}_{\rm{in}}\to \cos i_0$.
This limit simplifies the Equation (\ref{eq:n_e}) to
\ba
n_e(e_{\rm{in}}|i_0)\propto \frac{1}{e_{\rm{in}}
\left[\frac{3}{5}(1-e_{\rm{in}}^2)-\cos^2 i_0\right]^{1/2}},
\label{eq:n_e_i}
\ea
where the eccentricity distribution is defined for
a minimum eccentricity, say $e\sim0.01$,
and $\cos^2 i_0<\frac{3}{5}(1-e_{\rm{in}}^2)$.
The latter condition implies that
KL oscillations are restricted to $|\cos i_0|<\sqrt{3/5}$
and the maximum eccentricity is given by $\sqrt{1-5/3\cos^2i_0}$.

In contrast, by restricting to the initial high-inclination case 
$i_0\simeq \pi/2$
and setting $\omega_0=\pi/2$ (or $3\pi/2$) we can define the 
minimum eccentricity $e_{\rm{min}}=e_0$ and the maximum 
eccentricity $e_{\rm max}=\sqrt{1-5/3\cos^2i_0}$. 
Conveniently, this distribution only depends only on 
$e_{\rm{min}}$ and $e_{\rm{max}}$: 
\ba
n_e(e_{\rm{in}}|e_{\rm min},e_{\rm max})&\equiv& 
n_e(e_{\rm{in}}|\theta_0(e_{\rm min},e_{\rm max}),
\tilde{H}_{\rm{in}}(e_{\rm min},e_{\rm max})),\mbox{ where} \nonumber\\
\theta_0(e_{\rm min},e_{\rm max})&=&e_{\rm min}^2
-\frac{1}{5}\left(1+4e_{\rm min}^2\right)\left(2+3e_{\rm max}^2\right), ~~
\tilde{H}_{\rm{in}}(e_{\rm min},e_{\rm max})=
\sqrt{\frac{3}{5}\left(1-e_{\rm min}^2\right)\left(1-e_{\rm max}^2\right)},
\ea
which simplifies our analysis in \S\ref{sec:tides}.


\end{document}